\documentclass{IEEEoj}
\def\BibTeX{{\rm B\kern-.05em{\sc i\kern-.025em b}\kern-.08em
    T\kern-.1667em\lower.7ex\hbox{E}\kern-.125emX}}
\AtBeginDocument{\definecolor{ojcolor}{cmyk}{0.93,0.59,0.15,0.02}}
\def\OJlogo{\vspace{-4pt}\hskip-4pt\includegraphics[height=18pt]{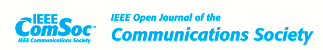}}

\usepackage{array,xcolor}

\usepackage{booktabs}
\usepackage{multirow}
\usepackage{comment}
\usepackage{setspace}

\renewenvironment{thebibliography}[1]{
  \begin{oldthebibliography}{#1}
    \setlength{\itemsep}{0em}
    \setlength{\parskip}{0em}
}
{
  \end{oldthebibliography}
}

\usepackage{lipsum}
\usepackage{bm}
\usepackage{tikz}
\usetikzlibrary{positioning,calc,arrows.meta,fit,backgrounds}
\usepackage{pgfplots}
\usepgfplotslibrary{colormaps}
\usepackage{mathrsfs}
\usepackage{mathtools}
\usetikzlibrary{shapes.geometric, arrows}

\usepackage{slashbox}

\newcommand{\tikzcircle}[2][red,fill=red]{\tikz[baseline=-0.7ex]\draw[#1,radius=#2] (0,0) circle ;}%
\usepackage{pifont}
\pgfplotsset{grid style={dashed,gray}}
\pgfplotsset{minor grid style={dotted,gray}}
\pgfplotsset{major grid style={dashed,gray}}

\usepackage[font=footnotesize]{caption}

\usepackage{amsmath, tabularx} 

\usepackage[noend]{algpseudocode}
\usepackage{ifpdf}
\usepackage{graphicx,amssymb,lineno}
\usepackage{relsize}
\usepackage{filecontents}
\usepackage{subcaption}
\usepackage{lipsum}
\usepackage{color,soul}
\usepackage{algorithm}
\usepackage{color,soul}
\usepackage{cite}
\usepackage{balance}
\usepackage[noend]{algpseudocode}
\makeatletter
\def\BState{\State\hskip-\ALG@thistlm}
\makeatother
\makeatletter
\newcommand*{\rom}[1]{\expandafter\@slowromancap\romannumeral #1@}
\makeatother

\graphicspath{ {Images/} }

\DeclareMathOperator{\EX}{\mathbb{E}}
\usepackage[T1]{fontenc}
\usepackage{float}

\makeatletter
\newcommand{\multiline}[1]{%
  \begin{tabularx}{\dimexpr\linewidth-\ALG@thistlm}[t]{@{}X@{}}
    #1
  \end{tabularx}
}
\makeatother

\renewcommand{\arraystretch}{1.5}
\usepackage{changepage}
\usepackage{adjustbox}
\DeclarePairedDelimiter\abs{\lvert}{\rvert}%
\DeclarePairedDelimiter\norm{\lVert}{\rVert}%

\makeatother
\DeclareMathOperator*{\argmax}{argmax} 
\makeatother
\DeclareMathOperator*{\argmin}{argmin} 
\usepackage{hyperref}
\usepackage{cleveref}

\author{
Amir Mansourian\IEEEauthorrefmark{1}, Alireza Fadakar\IEEEauthorrefmark{2}\IEEEmembership{(Graduate Student Member, IEEE)},
Saeed Akhavan\IEEEauthorrefmark{1}, Behrouz Maham\IEEEauthorrefmark{3}\IEEEmembership{(Senior Member, IEEE)}
\affil{School of Electrical and Computer Engineering, University of Tehran, Tehran 1439957131, Iran}
\affil{Ming Hsieh Department of Electrical and Computer Engineering, University of Southern California, Los Angeles, CA 90089, USA}
\affil{Department of Electrical and Computer Engineering, Nazarbayev University, 010000 Astana, Kazakhstan}
\authornote{This work was supported by the Faculty Development Competitive Research Grant Program of Nazarbayev University under Grant 20122022FD4125.}
}

\usepackage{eso-pic}
\usepackage{lipsum} 

\begin{document}
\markboth{This article has been accepted for publication in IEEE Open Journal of the Communications Society. DOI 10.1109/OJCOMS.2025.3558476}{ }
\title{Robust 3D Multi-Source Localization with a Movable Antenna Array via Sparse Signal Processing}
\begin{abstract}
Accurately localizing multiple sources is a critical task with various applications in wireless communications, such as emergency services, including natural post-disaster search and rescue operations.
However, scenarios where the receiver is moving have not been sufficiently addressed in recent studies.
This paper tackles the angle of arrival (AOA) 3D-localization problem for multiple sparse signal sources with a moving receiver, which has a limited number of antennas that may be outnumbered by the sources. 
First, an energy detector algorithm is proposed to leverage signal sparsity for eliminating noisy samples. 
Subsequently, an iterative algorithm is developed to refine and estimate
the AOAs accurately, initialized with previously estimated source locations and coarse elevation and azimuth AOAs obtained via the two-dimensional multiple signal classification (2D-MUSIC) method.
To this end, we introduce a sparse recovery algorithm to exploit signal sparsity, followed by a phase smoothing algorithm to refine the estimates. 
The K-SVD algorithm is then applied to the smoothed output to accurately determine the elevation and azimuth AOAs of the sources. 
For localization, a new multi-source 3D-localization algorithm is proposed to estimate source positions using the refined AOA estimates over a sequence of time windows.
Extensive simulations are carried out to demonstrate the effectiveness of the proposed framework.
\end{abstract}

\maketitle

\section{Introduction}
\label{sec:Int}
In recent decades, localization technologies have been widely studied in many applications such as communications, internet of things, and emergency services \cite{bochem2022robustness, fadakar2024ris, Trevlakis2023loc, Rekkas2023Artificial}. 
Consequently, there is an increasing interest in improving localization systems using advanced technologies \cite{fadakar2024ris, zou2023convergent, zheng2018exploiting}.
In environments with noise-afflicted sensor data, a variety of methodologies have been adopted for source positioning, including time-of-arrival (TOA) \cite{rao2021toa,xiong2022message,shamaei2021receiver, Xu2025Optimal}, time-difference-of-arrival (TDOA) \cite{radnosrati2020localization,jin2023differential, Tu2025Parameterized}, angle-of-arrival (AOA) \cite{zou2023convergent, wang2015asymptotically, zhang2023autoloc}, angle-of-departure (AOD) \cite{fascista2020low}, received-signal-strength (RSS) \cite{chu2022rss, li2023rss}, and fingerprint \cite{gong2023deep,li2023radio,liu2024fingerprint, Jia2022Extrinsic}, along with combinations of these methods \cite{liu2023hybrid_toa_aoa, fadakar2024ris, kang2023hybrid,zhang2024user, Lin2025Hybrid}. 
While time-based techniques depend on precise synchronization \cite{zou2023convergent}, RSS methods are influenced by signal propagation characteristics such as shadow-fading \cite{chu2022rss}.
Notably, \cite{chu2022rss} proposes an RSS-based method to enhance robustness in co-channel multi-source localization under shadow fading conditions.
Moreover, the considerable time and effort required to construct the radio map, restrict the practical applications of fingerprint-based localization \cite{li2023radio}.

AOA localization has garnered significant interest over the years due to its ability to passively estimate the position of a signal source \cite{wang2015asymptotically}. 
In contrast to TOA and TDOA techniques, AOA localization does not necessitate synchronization with the signal source or coordination among multiple receivers. 
Additionally, AOA localization is less influenced by environmental conditions and their varying parameters compared to RSS-aided techniques.
As a result, it has broader application scenarios \cite{wang2015asymptotically,zou2023convergent}.
Most of the recent 2D or 3D AOA localization studies such as \cite{wang2015asymptotically,zou2023convergent} assume that the estimates of source AOAs are available at the receiver. 
In particular, they begin their formulation of the elevation and azimuth AOAs as their true values plus the measurement error. 
Moreover, most of the recent studies only investigate the localization problem when there is only one source.
However, in this study, we consider a more general and challenging case, i.e., we assume that there are multiple sources and we also investigate how to accurately estimate the elevation and azimuth AOAs. 
It is noteworthy that the number of sources may be more than the number of antennas.
In \cite{badriasl2014three}, a closed-form improved pseudolinear estimator (IPLE) is proposed for positioning, which is based on a linear approximation of the maximum likelihood (ML) cost function. 
Subsequently, in \cite{wang2015asymptotically}, a closed-form solution based on weighted least squares (WLS) is introduced for 3D localization of a single source using AOA measurements. This solution addresses the challenge of sensor position errors.
Expanding upon these advancements, \cite{zou2023convergent} presents an iterative method that avoids matrix operations. 

Many localization studies assume stationary receiver sensors, often requiring multiple static beacons for effective localization. Nevertheless, it is noteworthy that source localization can be achieved with a single moving receiver \cite{fadakar2024passive, mansourian2024enhanced, fadakar2024localization_conv, liu2016source, zou2017emitter, Tzoreff2014single, Dogancay2008maximum, Witzgall2010single, Hmam2010passive}. 
This approach offers several advantages over conventional static multi-anchor localization methods, including economic efficiency by utilizing a single receiver, obviating the need for synchronization between stations since only one beacon is employed, facilitating real-time operation, and eliminating the necessity for data transmission among beacons \cite{Tzoreff2014single}. 
However, employing a moving array introduces several challenges, including variations in received signal power as well as continuous elevation and azimuth AOA changes from each source due to changes in the array's position, and potential inaccuracies in array position and mobility direction information. Our simulations validate the effectiveness of the proposed method in addressing these challenges.
In \cite{junbo2011design, tang2019study}, the localization problem for searching and rescuing survivors after a natural disaster, such as an earthquake, is investigated.
In \cite{junbo2011design}, a two-stage localization method is proposed, leveraging RSS measurements, measured by a mobile beacon node moving on an equilateral triangle path.
The first stage utilizes a trilateration technique to estimate the x and y coordinates of the survivors, assuming they possess mobile phones capable of transmitting signals post-disaster.
Subsequently, in the second stage, the 3D location is determined using recorded RSS values collected by the moving beacon node, exploiting the geometric properties of its trajectory.
In \cite{tang2019study}, a hybrid AOA/RSS-based 2D localization method is proposed to estimate survivor positions using both RSS and AOA measurements. 
In \cite{liu2016source, zou2017emitter}, ML based approaches is proposed for 2D localization of a single source using the recorded TOA samples. 
More recently, \cite{fadakar2024passive} introduced a novel approach for 3D multi-source localization utilizing a moving receiver and deep learning techniques, under the assumption of Gaussian signal transmission by the sources. However, despite its low computational complexity, this method is constrained by the number of antennas, thus limiting the number of sources that can be localized. 

The existing solutions for search and rescue operations and
passive localization often suffer from limited accuracy, high complexity, and scalability issues, especially in large areas with many survivors. 
To address these challenges, this paper introduces a novel 3D AOA multi-source localization approach with a moving multi-antenna array. 
We assume that mobile phones emit sparse signals, allowing for efficient localization of multiple sources even when the number of sources exceeds the number of antennas. 
Under this assumption, neither synchronization nor bandwidth scheduling is required. Consequently, the sparsity feature provides a cost-effective method for separating multiple sources at the receiver. 
The main contributions of this study are as follows:
\begin{itemize}
\item
\textbf{System level design for multi-source 3D localization with a moving array:}
We address the 3D AOA localization problem using a moving array with a limited number of antennas, starting from signal reception. Our system sequentially performs three main tasks: noise filtering with an energy detector, 2D-AOA estimation, and multi-source localization.
Unlike previous works \cite{fadakar2024passive, liu2016source, zou2017emitter}, our approach enables accurate localization of multiple stationary sources using sparse signals,
even when the sources outnumber the array elements.
\item
\textbf{Novel iterative energy detection algorithm for filtering noisy samples:}
We propose a novel iterative energy detection algorithm to filter noisy samples in the received signal by leveraging the sparsity of the transmitted signals. The algorithm iteratively removes noisy samples by comparing their energy with a threshold and exploiting the continuity features of each pulse. The filtered signal is then used for AOA estimation.
\item
\textbf{Novel precise 2D-AOA estimation algorithm:}
The sample covariance matrix (SCM)-based AOA estimation lacks accuracy in multi-source scenarios, as it does not exploit signal sparsity, phase characteristics, or each pulse continuity. To address this, we propose a novel iterative algorithm that integrates a new sparse recovery technique, phase smoothing, and the K-SVD algorithm to refine the initial 2D-AOA estimates. The initial values are derived from previously estimated source locations, along with the rough AOAs of potential undetected sources, obtained using an SCM-based method such as two-dimensional multiple signal classification (2D-MUSIC). The proposed approach significantly enhances estimation accuracy and enables detecting more sources than the number of antennas.
\item
\textbf{Novel multi-source localization algorithm:}
We propose an iterative gradient projection algorithm to estimate source positions based on the refined AOA estimates obtained at each time window.
Unlike existing methods \cite{zou2023convergent, badriasl2014three, wang2015asymptotically}, our algorithm
utilizes the receiver’s prior knowledge of the environment’s surface uncertainty map, a practical assumption, to improve the accuracy of source localization.
\item
\textbf{Extensive simulations under various imperfections and complexity analysis:}
Numerous simulations validate the effectiveness and robustness of the proposed method against various imperfections, including inaccuracies in receiver position and orientation, as well as the absence of prior environmental knowledge. Additionally, a comprehensive complexity analysis is conducted.
\end{itemize}

We use bold capital letters to show matrices $\bm{X}$ and bold small letters to show vectors $\bm{x}$.
$[\bm{X}]_{\bm{u}, \bm{v}}$ denotes the submatrix of $\bm{X}$, formed by extracting the rows indicated by the indices in $\bm{u}$ and the columns specified by the indices in $\bm{v}$,
where $\bm{u}$ and $\bm{v}$ can also be represented by binary vectors. 
The symbol $:$ in place of $\bm{u}$ or $\bm{v}$, signifies that all rows or all columns of $\bm{X}$ are being selected, respectively. 
For integers $i$ and $j$ with $i<j$, the expression $i:j$ represents a vector consisting of the integers from $i$ to $j$ in sequential order. 
The superscript $(\cdot)^{T},(\cdot)^{H}$ and $(\cdot)^\dagger$ represent the operations of transposing a vector or matrix, taking its Hermitian transpose, and computing its pseudo-inverse, respectively. 
$[\bm{x}_1,\dots, \bm{x}_n]$ shows the horizontal concatenation of the vectors $\bm{x}_1, \dots,\bm{x}_n$. 
The identity matrix of size $n\times n$ is denoted by $\bm{I}_n$. 
$\norm{\bm{X}}_F$ and $\norm{\bm{x}}_2$  show the Frobenius and Euclidean norms, respectively. 
The zero-norm $\norm{\bm{x}}_0$ denotes the number of nonzero elements in $\bm{x}$.
The floor $\lfloor a \rfloor$ function
rounds $a\in \mathbb{R}$ to the nearest smaller integer. $\binom{n}{m} = \frac{n!}{m!(n-m)!}$ represents the combinatorial number.
Finally, big-O notation, $O(\cdot)$, is employed to express the computational complexity of the proposed method modules.

The subsequent sections of this paper are arranged as follows.
Section~\ref{sec:sys} explains the system model and formulates the signal model. 
Section~\ref{sec:Energy-Detector}, Section~\ref{sec:2D-AOA-algorithm} and Section~\ref{sec:loc-estimator} present the proposed energy detector, 2D-AOA estimation, and localization approaches, respectively, as the three main tasks of the proposed method.
Section~\ref{sec:complexity-analysis} analyses the computational complexity of the proposed method, while
Section~\ref{sec:simul} conducts simulations to showcase the method's effectiveness. 
Finally, Section~\ref{sec:con} provides a summary and conclusion of the paper.
\section{System and Signal Model}\label{sec:sys}
\begin{figure}[ht!] 
\centering
\includegraphics[width=\columnwidth]{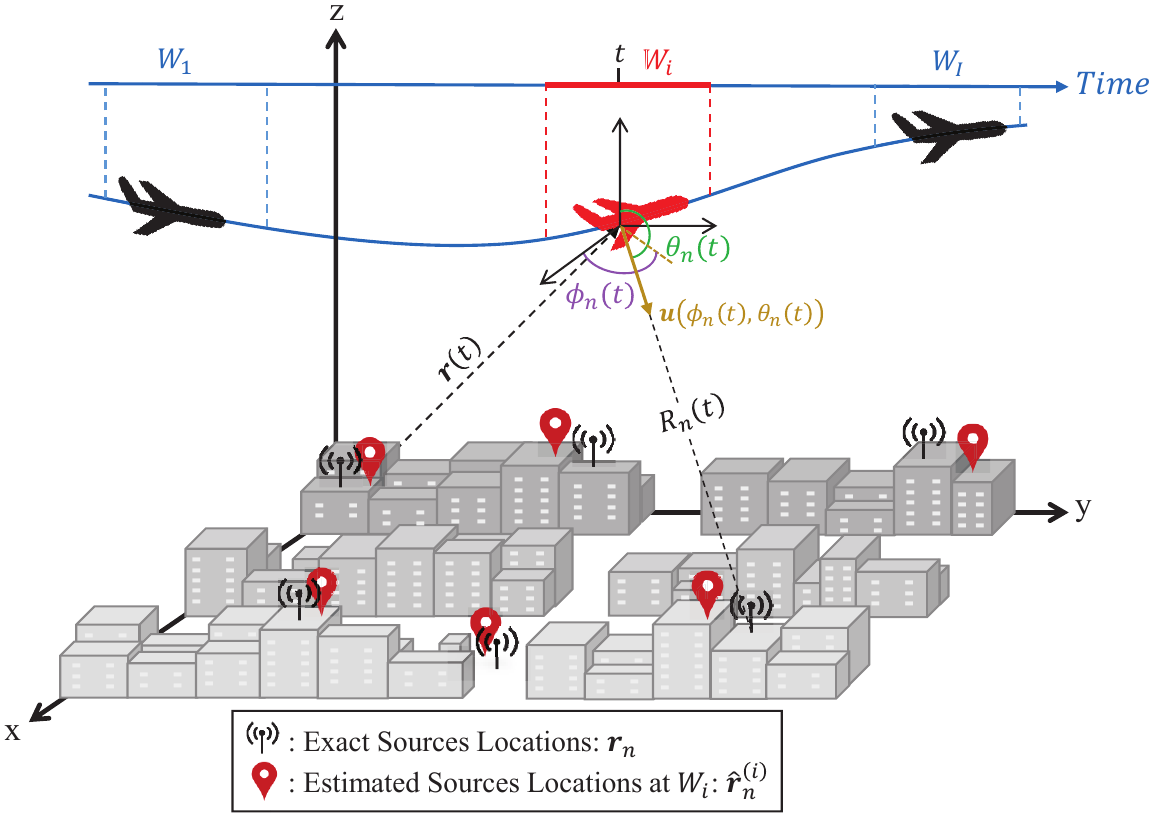}
\caption{System model.
\label{fig:system-model}
}
\end{figure}

As illustrated in Fig.~\ref{fig:system-model}, we consider a mobile receiver equipped with an arbitrary 2D or 3D array comprising $M$ elements positioned at $\bm{D}\in\mathbb{R}^{3\times M}$ whose $m$-th column represents the Cartesian coordinates of the $m$-th array element relative to the array center. 
The mobile receiver captures the received signal over a sequence of $I\ge 1$ consecutive time windows denoted as $\{W_i\}_{i=1}^{I}$, each with a duration $T$ consistent with Fig.~\ref{fig:system-model}.
Given the assumption that the initial time window begins at $t_0$, then
\begin{equation}\label{def:time-window}
W_i=[t_0+(i-1)T,\ t_0+iT],\ \ \forall i\in\{1,\dots, I\}.
\end{equation}

Let $\bm{r}(t)$ denote the location of the array center at time instant $t$. Additionally, suppose there are $N$ stationary sources randomly distributed in 3D space, denoted by Cartesian positions $\{\bm{r}_n\}_{n=1}^{N}$, where $\bm{r}_n=[r_{n_x},r_{n_y},r_{n_z}]^T$. 
During each time window $W_i$, our objective is twofold: to detect new sources among $\{\bm{r}_n\}_{n=1}^{N}$ and concurrently refine the localization of sources previously detected within the preceding $i-1$ time windows $\{W_j\}_{j=1}^{i-1}$.

Let $s_n(t)$ represent the baseband signal transmitted by the $n$-th source 
\begin{equation}\label{eq:sum_p_n}
s_n(t)=\sum_{j} p_n(t-t_{n_j}),
\end{equation}
where $p_n(t)\in\mathbb{R}$ denotes the transmitted continuous spike from the $n$-th source, lasting for a duration of $T_{p_n}$. 
Furthermore, $t_{n_j}$ represents the starting time of the $j$-th pulse within the spike $p_n(t)$ emitted by the $n$-th source.
The pulses are assumed to be uniformly distributed over any time interval, with an average inter-pulse duration
$T_{\text{avg}}$ such that $T_{\text{avg}}=\EX\{\abs{t_{n_{j+1}}-t_{n_j}}\}\gg T_{p_n}$, imparting sparsity to the signals $\{s_n(t)\}_{n=1}^{N}$. Therefore, the number of $n$-th source pulses in a given time span is a Poisson random process with parameter $\lambda=1/T_{\text{avg}}$ and distribution as follows:
\begin{equation}
Pr\left\{N_{p_n}(\Delta t)=\eta\right\} = \frac{(\lambda\Delta t)^\eta}{\eta!}e^{-\lambda\Delta t},
\end{equation}
where $N_{p_n}(\Delta t)$ is the number of $n$-th source pulses in an arbitrary time duration $\Delta t$.
Additionally, we assume that the receiver moves at an altitude of several hundred meters above ground level, while the sources are positioned on the surface of the uncertainty map, such as building rooftops.  
Thus, assuming operation in the UHF band, which provides favorable propagation characteristics, this setup ensures a line-of-sight (LoS) channel between the mobile receiver and the sources.
As a result, the baseband signal received from the $n$-th source at the receiver can be expressed as \cite{theodoridis2013academic}:
\begin{equation} \label{eq:noise-free-xi}
\bm{x}_n(t)\approx
\beta_n(t)
s_n\left(t-\tau_n(t)\right)
\bm{a}\left(\theta_n(t),\phi_n(t)\right),
\end{equation}
where $\tau_n(t)=R_n(t)/c$ represents the time delay between the $n$-th source and the array center, with $R_n(t)=\norm{\bm{r}(t)-\bm{r}_n}_2$ denoting their distance and $c$ the speed of light. Additionally, $\beta_n(t)$ incorporates the complex path loss and receiver antenna gain, calculated as follows:
\begin{equation}\label{def:beta}
\beta_n(t)
=
\frac{\mathcal{G}(\theta_n(t),\phi_n(t))}{\sqrt{4\pi}R_n(t)}
e^{-jKR_n(t)},
\end{equation} 
where $K=2\pi f_c/c$ represents the wave number, with $f_c$ indicating the carrier frequency,
$\theta_n(t)$ and $\phi_n(t)$ are the elevation and azimuth AOAs for the $n$-th source at time instant $t$, respectively,
and $\mathcal{G}(\theta,\phi)$ denotes the receiver antenna gain for 2D-AOA pair $(\theta,\phi)$.
Here, the elevation AOA $\theta\in[0,\pi]$ is the angle between the unit direction vector towards the source and the $z$-axis, while the azimuth AOA $\phi\in[0,2\pi]$ is the angle between the projection of this vector on the $xy$ plane and the $x$-axis.
In \eqref{eq:noise-free-xi}, $\bm{a}(\theta,\phi)\in \mathbb{C}^{M}$ represents the array steering vector expressed as
\begin{equation} \label{eq:streeing_vec}
\bm{a}(\theta,\phi)
=
e^{jK\bm{D}^T\bm{u}(\theta,\phi)},
\end{equation}
where $\bm{u}(\theta,\phi)\in\mathbb{R}^{3}$ is the unit direction vector pointing towards the 2D-AOA pair $(\theta,\phi)$:
\begin{equation}\label{eq:unit-dir}
\bm{u}(\theta,\phi)
=
[
\sin(\theta)\cos(\phi),
\sin(\theta)\sin(\phi),
\cos(\theta)
]^T.
\end{equation}

Let $t_s$ denote the sampling time, and $G$ be the number of samples in each time window. Consider the vector $\bm{s}_{n}^{(i)}\in\mathbb{C}^{G}$ as the received baseband signal samples from the $n$-th source at the array center during the $i$-th time window:
\begin{equation} \label{eq:s_i}
[\bm{s}_{n}^{(i)}]_g=
\beta_n(t_g^{(i)})
s_n(t_g^{(i)}-\tau_n(t_g^{(i)})),
\end{equation}
for $g=1,\dots, G$, and $i=1,\dots, I$, where 
$
t_g^{(i)}=(i-1)T+t_0+gt_s
$ 
represents the time instant corresponding to the $g$-th sample within the $i$-th time window. 
If the changes in the orientation of the array and its traveled distance compared to $R_n(t)$ are negligible within an arbitrary time window $W_i$,
$\theta_n(t)$ and $\phi_n(t)$ remain approximately constant throughout the time window. Thus, they can be approximated by their values at the midpoint of the time window. Subsequently, after collecting all $G$ received samples, a compact expression is derived utilizing \eqref{eq:s_i} and \eqref{eq:noise-free-xi}:
\begin{equation} \label{eq:Xi-approx}
\bm{X}_n^{(i)}\approx
\bm{a}
(
\theta_{n}^{(i)},
\phi_{n}^{(i)}
)
\bm{s}_n^{(i)^T},
\end{equation}
where $\bm{X}_n^{(i)}\in \mathbb{C}^{M\times G}$ represents the collected samples from the $n$-th source during the $i$-th time window.
In addition, the elevation and azimuth AOA for this source at the midpoint of the $i$-th time window, $t_{m}^{(i)}$, are given by $\theta^{(i)}_n=\theta_n(t_{m}^{(i)})$ and $\phi^{(i)}_n= \phi_n(t_{m}^{(i)})$, respectively.
By applying the superposition principle, the noise-free recorded signals from all $N$ sources during the $i$-th time window can be expressed as
\begin{equation} \label{eq:entire-rec-sig}
\bm{X}^{(i)}
=
\sum_{n=1}^{N}\bm{X}_n^{(i)}
\stackrel{\eqref{eq:Xi-approx}}{\approx}
\bm{A}^{(i)}
\bm{S}^{(i)},
\end{equation}
where $\bm{A}^{(i)}\in\mathbb{C}^{M\times N}$ 
represents the array manifold matrix at the midpoint of the  $i$-th time window 
\begin{equation} \label{eq:array-manifold}
\bm{A}^{(i)}
=
\left[
\bm{a}(\theta^{(i)}_1,\phi^{(i)}_1),\dots,
\bm{a}(\theta^{(i)}_N,\phi^{(i)}_N)
\right].
\end{equation}
Moreover,
$\bm{S}^{(i)}\in\mathbb{C}^{N\times G}$ 
is a matrix which $\bm{s}_n^{(i)^T}$ represents its $n$-th row. 
Finally, after noise addition at the receiver, the recorded signal at the $i$-th time window is formulated as
\begin{equation} \label{eq:X_fact}
\bm{Y}^{(i)} =\bm{X}^{(i)}+\bm{V}^{(i)}
\stackrel{\eqref{eq:entire-rec-sig}}{\approx}
\bm{A}^{(i)}
\bm{S}^{(i)}
+
\bm{V}^{(i)},
\end{equation}
where the matrix $\bm{V}^{(i)}\in\mathbb{C}^{M\times G}$ comprises columns that represent samples of time dependent noise $\bm{v}(t)\in\mathbb{C}^{M}$. This noise follows a complex normal distribution, specifically 
$
\bm{v}(t)\sim \mathcal{CN}(0,\sigma_v^2\bm{I}_M)
$, and is considered to be uncorrelated with sources.
It is important to highlight that the factorization presented in \eqref{eq:X_fact} serves as a key assumption in most AOA estimation methods.

Given the array's limited velocity, the changes of $KR_n(t)$ in \eqref{def:beta}, and consequently the phase variations of $\bm{s}_{n}^{(i)}$ as written in \eqref{eq:s_i} are negligible over the sampling period $t_s$. As a result, according to \eqref{eq:sum_p_n}, $\bm{S}^{(i)}$ is a complex sparse matrix with row elements demonstrating smooth phase variations.

\begin{figure*}[ht!]
\centering
\begin{subfigure}[b]{2\columnwidth}
\includegraphics[width=\textwidth]{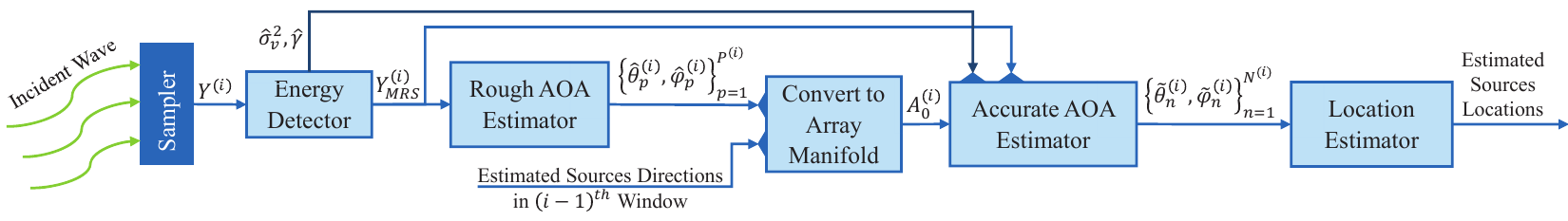}
\end{subfigure}%
\caption{The structure of the proposed method.}
\label{fig:Proposed-Method}
\end{figure*}

Fig.~\ref{fig:Proposed-Method} illustrates the overall structure of the proposed method, which is divided into several blocks detailed in the following sections.
The method involves three main tasks. 
First, Section~\ref{sec:Energy-Detector} introduces an energy detector block to filter noisy samples from the received signal, ensuring that these samples do not impair localization performance. The second task, detailed in Section~\ref{sec:2D-AOA-algorithm}, covers AOA estimation, which includes Rough AOA Estimator, Array Manifold Initializer, and AOA Refiner blocks to determine the elevation and azimuth AOAs of sources in the current time window. Finally, Section~\ref{sec:loc-estimator} describes the Location Estimator block, which uses these AOAs and data from previous time windows to estimate sources positions.

\section{Energy Detector for Noise Filtering} \label{sec:Energy-Detector}
The main objective of this block is 
noise reduction through exploiting the sparsity of the received signal and the continuity of each transmitted pulse.
First, due to the signal's sparsity, the estimation of $\sigma^2_v$ begins with the average energy of samples of $\bm{Y}^{(i)}$, denoted as $\hat{\sigma}^2_{v,i}$. Next, for an arbitrary element $\upsilon$ from the noise matrix $\bm{V}^{(i)}$ and a given probability  $P_0$ for false detection (identifying noise as a signal) defined as $P_0=\text{Pr} \{ \abs{\upsilon} > V_{\text{th}} \}$,
the threshold value ($V_{\text{th}}$), will be obtained using \cite{theodoridis2013academic} as:
\begin{equation}
V_{\text{th}} = \sqrt{-(\log P_0)\hat{\sigma}^2_{v,i}}.
\end{equation}

Subsequently, $\bm{q}_{\text{MRS}}^{(i)}\in \mathbb{Z}^{G_{\text{MRS}}^{(i)}}$ is determined as a vector consisted of the column indices of the largest subset of columns in $\bm{Y}^{(i)}$, each containing at least one element with an absolute value greater than $V_{\text{th}}$, where MRS stands for Modified Received Signal. As a result, $G_{\text{MRS}}^{(i)}\leq G$ where $G_{\text{MRS}}^{(i)}$ indicates the number of such columns.
%
Next, to exploit the continuity of pulses, we compute the first-order discrete derivative of $\bm{q}_{\text{MRS}}^{(i)}$ as $\text{diff}_{1}(\bm{q}_{\text{MRS}}^{(i)})\in \mathbb{Z}^{G_{\text{MRS}}^{(i)}-1}$, obtained by subtracting $\bm{q}_{\text{MRS}}^{(i)}$ from its one-sample-shifted vector:
\begin{equation}
\text{diff}_{1}(\bm{q}_{\text{MRS}}^{(i)})
=
[\bm{q}_{\text{MRS}}^{(i)}]_{2:G_{\text{MRS}}^{(i)}}
-
[\bm{q}_{\text{MRS}}^{(i)}]_{1:(G_{\text{MRS}}^{(i)}-1)}.
\end{equation}
Then, we obtain the binary vector $\bm{q}_{\text{adj}}^{(i)}\in \{0,1\}^{G_{\text{MRS}}^{(i)}-1}$ by comparing the resulting vector with the threshold $\text{diff}_{\text{max}}$:
\begin{equation}
\bm{q}^{(i)}_{\text{adj}} 
=
\text{diff}_{1}(\bm{q}^{(i)}_{\text{MRS}})\leq \text{diff}_{\text{max}}.
\end{equation} 
Each nonzero component of $\bm{q}^{(i)}_{\text{adj}}$ signifies that its corresponding element in $\bm{q}^{(i)}_{\text{MRS}}$ and its downward neighbor, refer to two columns of $\bm{Y}^{(i)}$ associated with the same pulse transmitted by one of the sources\footnote{It is important to note that despite potential pulse interleaving due to nearly simultaneous transmissions by multiple sources, the continuity of the received signal remains unaffected, ensuring no impact on the methodology.},
due to the continuity of pulses. It is important to emphasize that according to the definition of $\bm{q}^{(i)}_{\text{MRS}}$, each of these columns has at least one sample of recorded signal with an amplitude grater than $V_\text{th}$.

Therefore, to exploit the continuity feature of the pulses, we selectively retain the elements of $\bm{q}_\text{MRS}^{(i)}$ that correspond to consecutive ones in $\bm{q}_\text{adj}^{(i)}$, with at least $L_\text{adj}-1$ elements. Additionally, we keep the element immediately below each of these sequences. This selective retention is then used to update $\bm{q}_\text{MRS}^{(i)}$ and hence $G_\text{MRS}^{(i)}$. Consequently, pure noise samples with absolute values greater than $V_{\text{th}}$, but not contributing to any pulses, are eliminated as much as possible.

By assuming that there is only noise on the subset of columns of $\bm{Y}^{(i)}$ with indices other than $\bm{q}_\text{MRS}^{(i)}$, the estimation of the noise variance will be updated as
\begin{equation}
\hat{\sigma}_{v,i}^2 = 
\tfrac{1}{M(G-G_{\text{MRS}}^{(i)})}
\norm{
[\bm{Y}^{(i)}]_{:,\overline{\bm{q}_{\text{MRS}}^{(i)}}}
}_F^2,
\end{equation}
where $\overline{\bm{q}_{\text{MRS}}^{(i)}}$ denotes the column indices of $\bm{Y}^{(i)}$ other than $\bm{q}_{\text{MRS}}^{(i)}$.
This procedure repeats until convergence, which is reached when $\bm{q}^{(i)}_{\text{MRS}}$ remains unchanged for two consecutive iterations or when the number of iterations surpasses the maximum limit, $N_\text{max}^{\text{Eng}}$.
Hence, after reaching convergence, the output of Energy Detector block at the $i$-th time window, can be represented as
\begin{equation}\label{eq:ed-out}
\bm{Y}_{\text{MRS}}^{(i)}=
\bm{A}^{(i)}\bm{S}_{\text{MRS}}^{(i)}
+
\bm{V}^{(i)}_\text{MRS},
\end{equation}
where 
$
\bm{Y}_{\text{MRS}}^{(i)}=
\big[
\bm{Y}^{(i)}
\big]_{:,\bm{q}_{\text{MRS}}^{(i)}}$, 
$
\bm{S}_{\text{MRS}}^{(i)}=
\big[
\bm{S}^{(i)}
\big]_{:,\bm{q}_{\text{MRS}}^{(i)}}$
, 
$
\bm{V}_{\text{MRS}}^{(i)}=
\big[
\bm{V}^{(i)}
\big]_{:,\bm{q}_{\text{MRS}}^{(i)}}
$. 
Let $\gamma^{(i)}$ be the exact $\text{SNR}$ in the output of Energy Detector block at the $i$-th time window \eqref{eq:ed-out}, which is given by
\begin{equation}\label{def:gamma}
\gamma^{(i)}=
\dfrac{\norm{\bm{A}^{(i)}\bm{S}_{\text{MRS}}^{(i)}}_F^2}{\norm{\bm{V}^{(i)}_\text{MRS}}_F^2}.
\end{equation}
The estimate of $\gamma^{(i)}$, denoted by $\hat{\gamma}^{(i)}$, can be obtained as
\begin{equation}\label{eq:gamma-hat}
\hat{\gamma}^{(i)}=\dfrac{
\norm{\bm{Y}^{(i)}_{\text{MRS}}}_F^2/(MG_\text{MRS}^{(i)})-\hat{\sigma}^2_{v,i}
}
{
\hat{\sigma}^2_{v,i}
}.
\end{equation}

Since $\gamma^{(i)}$ changes in each time window, it is called instantaneous $\text{SNR}$ in the rest of the paper. 
The aforementioned steps, are summarized in Algorithm~\ref{alg:energy-detector}. 

This section concludes with the definition of $\text{SNR}^*$.
Note that after filtering the noisy samples by the energy detector, the output of this block should ideally consist only of pulses $\{p_n(t)\}_{n=1}^{N}$.
Motivated by this, we define $\mathrm{SNR}^*$ as the ratio of the maximum achievable received pulse power at time $t_0$ (assuming simultaneous transmission by all sources) to the noise power at the receiver as follows:
\begin{equation}\label{def:SNR-star}
\text{SNR}^*=
\sum_{n=1}^{N}\dfrac{P_n}{4\pi \norm{\bm{r}(t_0)-\bm{r}_n}^2_2\sigma^2_\nu},
\end{equation}
where $P_n=\frac{1}{T_{p_n}}\EX\{|p_n(t)|^2\}$ represents the power of the pulse $p_n(t)$.
Given $\text{SNR}^*$ value, the noise power level is determined.
\begin{algorithm}[ht]
\caption{Energy Detector}\label{alg:energy-detector}
 \hspace*{\algorithmicindent} \textbf{Input}:
 $\bm{Y}^{(i)}$
  \\
  \hspace*{\algorithmicindent} \textbf{Design Parameters}:
$P_{0}$, $\text{diff}_{\text{max}}$, $L_\text{adj}$, $N_\text{max}^{\text{Eng}}$
  \\
 \hspace*{\algorithmicindent} \textbf{Outputs}: 
 $\bm{q}^{(i)}_{\text{MRS}}$, $\bm{Y}^{(i)}_{\text{MRS}}$, $\hat{\sigma}_{v,i}^2$, $\hat{\gamma}^{(i)}$
\begin{algorithmic}[1]
\State 
\multiline{%
$M$, $G$: Number of rows and columns in $\bm{Y}^{(i)}$.

Initialize $\hat{\sigma}_{v,i}^2$ with $\dfrac{1}{MG}\norm{\bm{Y}^{(i)}}_F^2$,
$\bm{q}^{(i)}_{\text{MRS}}=0$,
$\kappa=0$
}
\Repeat
\State $V_{\text{th}} \gets \sqrt{-(\log P_{0})\hat{\sigma}_{v,i}^2}$
\State $\bm{q}_{\text{old}} \gets \bm{q}^{(i)}_{\text{MRS}}, \quad \kappa \gets \kappa+1$
\State 
\multiline{%
$\bm{q}^{(i)}_{\text{MRS}}$ is obtained by selecting the column indices of $\bm{Y}^{(i)}$ that have at least one element with an absolute value greater than $V_\text{th}$.
}
\State 
\multiline{%
$\bm{q}^{(i)}_{\text{adj}} \gets \text{diff}_{1}(\bm{q}^{(i)}_{\text{MRS}})\leq \text{diff}_{\text{max}}$ 
}
\State 
\multiline{%
$\bm{q}_{\text{MRS}}^{(i)}$ is updated by selecting the elements corresponding to all sequences of $L_\text{adj}-1$ or more consecutive 1-valued elements in $\bm{q}_{\text{adj}}^{(i)}$, along with the element directly below each sequence.

$G_{\text{MRS}}^{(i)}\gets $ length of $\bm{q}_{\text{MRS}}^{(i)}$
}
\State 
\multiline{%
$\hat{\sigma}_{v,i}^2 \gets 
\frac{1}{M(G-G_{\text{MRS}}^{(i)})}
\norm{
[\bm{Y}^{(i)}]_{:,\overline{\bm{q}_{\text{MRS}}^{(i)}}}
}_F^2$
}
\Until{$\bm{q}_{\text{old}}=\bm{q}_{\text{MRS}}^{(i)}$ or $\kappa=N_\text{max}^{\text{Eng}}$}
\State
\multiline{%
$
\bm{Y}_{\text{MRS}}^{(i)}=
\big[
\bm{Y}^{(i)}
\big]_{:,\bm{q}_{\text{MRS}}^{(i)}}$
,
$
\hat{\gamma}^{(i)}=\frac{
\norm{\bm{Y}_{\text{MRS}}^{(i)}}_F^2\big/\left(MG_\text{MRS}^{(i)}\right)-\hat{\sigma}^2_{v,i}
}
{
\hat{\sigma}^2_{v,i}
}
$
}
\end{algorithmic}
\end{algorithm}

\section{Multi-Source 2D-AOA Estimation}\label{sec:2D-AOA-algorithm}
AOA estimation in the proposed method begins with the Rough AOA Estimator, which obtains the initial AOAs of the potential undetected sources using the signal filtered by the Energy Detector block. Next, the Array Manifold Initializer uses these initial AOAs and the previous estimated locations to compute an initial array manifold. Finally, the AOA Refiner block accurately estimates the AOAs using this initial value and the output of the Energy Detector block.
The mentioned modules are explained in the following subsections.

\subsection{Rough AOA Estimator}\label{sec:AOA-Estimator}
During array movement, the received energy from an undetected source may be increased enough to be detected. The purpose of the Rough AOA Estimator block is to obtain initial AOAs of possibly undetected sources.
Various methods exist for 2D-AOA estimation \cite{FADAKAR2024106743, fadakar2024multi_AOA}. In this paper, we utilize the well-known 2D-MUSIC algorithm \cite{theodoridis2013academic} for initial 2D-AOA estimation of sources, although alternative techniques may also be employed.
It is important to note that this step relies solely on second-order statistics, i.e., the SCM, and does not exploit the sparsity structure and other characteristics of the signal, leading to less accurate estimations.
In the subsequent subsections, we leverage these estimations as an initial point in our proposed novel algorithm for achieving more precise 2D-AOA estimation.

As shown in Fig.~\ref{fig:Proposed-Method}, at each time window $W_i$, Rough AOA Estimator block first receives the output of Energy Detector module $\bm{Y}_{\text{MRS}}^{(i)}$. 
Next, it computes the corresponding SCM $\bm{R}^{(i)}\in\mathbb{C}^{M\times M}$ as follows:
\begin{equation}\label{eq:scm-computation}
\bm{R}^{(i)}=
\frac{1}{
G_{\text{MRS}}^{(i)}
}
\bm{Y}_{\text{MRS}}^{(i)}
\bm{Y}_{\text{MRS}}^{{(i)}^H}
.
\end{equation} 

In the next step, the minimum description length (MDL) algorithm \cite{theodoridis2013academic} is employed to estimate the number of incoming signals $P^{(i)}$
\footnote{
Let $\hat{\lambda}_1, \hat{\lambda}_2, \dots, \hat{\lambda}_M$ denote the eigenvalues of the SCM $\bm{R}^{(i)}$ in \eqref{eq:scm-computation}. The MDL criterion is defined as $\mathrm{MDL}(m) = -2 G_{\mathrm{MRS}}^{(i)} (M - m) \ln \varrho(m) + 2m(2M - m) \ln(G_{\mathrm{MRS}}^{(i)}) / 2$, where $\varrho(m) = \frac{1}{\hat{\sigma}^2_m} \left(\prod_{i=m+1}^{M} \hat{\lambda}_{i} \right)^{\frac{1}{M-m}}$ and $\hat{\sigma}^2_m = \frac{1}{M - m} \sum_{i=m+1}^{M} \hat{\lambda}_{i}$. The estimated number of sources $P^{(i)}$ is obtained by selecting $m \in \{0, \dots, M-1\}$ that minimizes $\mathrm{MDL}(m)$.  
}. 
Finally, the 2D-MUSIC spectrum $\bm{S}_{\text{MUSIC}}\in\mathbb{R}_{+}^{K_\theta\times K_\phi}$ is calculated as
\begin{equation}\label{eq:2d-music-map}
[\bm{S}_{\text{MUSIC}}]_{u,v}
=
\dfrac{1}{\bm{a}^H(\theta_u,\phi_v)\bm{U}_{\text{null}}\bm{U}_{\text{null}}^H\bm{a}(\theta_u,\phi_v)},
\end{equation}
where $\theta_u$ and $\phi_v$ are the $u$-th and $v$-th element of elevation and azimuth scope ranges  
$
\bm{\Theta}=\{\theta_1,\dots,\theta_{K_\theta}\}
$ 
and  
$\bm{\Phi}=\{\phi_1,\dots,\phi_{K_\phi}\}
$, respectively, and $\bm{U}_{\text{null}}$ denotes the noise subspace of $\bm{R}^{(i)}$, obtained via eigenvalue decomposition (EVD) by selecting the $P^{(i)}$ eigenvectors corresponding to the smallest eigenvalues.
Finally, the $P^{(i)}$ AOAs $\{\hat{\theta}_p^{(i)},\hat{\phi}_p^{(i)}\}_{p=1}^{P^{(i)}}$ are determined by choosing $P^{(i)}$ highest peaks of the 2D spectrum $\bm{S}_{\text{MUSIC}}$. 

\subsection{Array Manifold Initializer}\label{sec:arr-manifold-init} 
Assuming the proposed method has been applied to the first $i-1$ time windows $\{W_j\}_{j=1}^{i-1}$, the following text provides a detailed explanation of calculating the initial array manifold for the $i$-th time window.

Define $\bm{U}_1\in\mathbb{R}^{3\times N_{\text{ESL}}^{(i-1)}}$ to represent a matrix where each column determines the unit direction from the receiver center position in the midpoint of the $i$-th time window, denoted by $\bm{r}^{(i)}$, to the formerly estimated positions of sources
$\{\hat{\bm{r}}_s^{(i-1)}\}_{s=1}^{N_{\text{ESL}}^{(i-1)}}$. Here, $N_{\text{ESL}}^{(i-1)}$ is the number of estimated source locations (ESL) and $\hat{\bm{r}}_s^{(i-1)}$ represents the estimated location of source $s$ until the $(i-1)$-th time window. $\bm{U}_1$ is obtained as
\begin{equation}
[\bm{U}_1]_{:,s}= \dfrac{\hat{\bm{r}}_s^{(i-1)}-\bm{r}^{(i)}}{\norm{\hat{\bm{r}}_s^{(i-1)}-\bm{r}^{(i)}}_2}\\
,\ s=1,\dots,N_{\text{ESL}}^{(i-1)}.
\end{equation} 
Furthermore, $\bm{U}_2\in\mathbb{R}^{3\times N_{U_2}^{(i-1)}}$ represents a matrix whose columns contain the unit direction vectors allocated to sources, each having exactly one assigned vector during the initial $i-1$ time windows. Here, $N_{U_2}^{(i-1)}$ is the number of such directions.
According to the proposed localization algorithm in Section~\ref{sec:loc-estimator}, no locations are estimated for these sources yet.
Then, we define $\bm{U}_D^{(i-1)}
\in\mathbb{R}^{3\times (N_{\text{ESL}}^{(i-1)}+N_{U_2}^{(i-1)})}$
as the matrix containing the last directions of the detected sources up to the $(i-1)$-th time window, given by:
\begin{equation}\label{eq:U_D}
\bm{U}_D^{(i-1)}=[\bm{U}_1, \bm{U}_2].
\end{equation}

Recall from Section~\ref{sec:AOA-Estimator} that
$\{\hat{\theta}_p^{(i)}\}_{p=1}^{P^{(i)}},\ \{\hat{\phi}_p^{(i)}\}_{p=1}^{P^{(i)}}$ are the estimated elevation and azimuth AOAs 
at the $i$-th time window by the rough AOA estimator algorithm. 
These AOAs are then used to define the corresponding unit directions $\{\bm{u}_p^{\text{new}}\}_{p=1}^{P^{(i)}}$ according to \eqref{eq:unit-dir}.
Among them, only the directions related to possibly undetected sources are used to calculate the initial array manifold.
To achieve this, we calculate the following for each obtained direction $\bm{u}_p^{\text{new}}$:
\begin{equation}
m_p= \max\abs{\bm{u}_p^{{\text{new}}^T} 
\bm{U}_D^{(i-1)}},
\ p=1,\dots ,P^{(i)}.
\end{equation}
Let $\bm{\mathcal{P}}$ denote the largest subset of indices $\{1,2,\dots ,P^{(i)}\}$ such that $m_{p'}<\xi$ for each $p'\in \bm{\mathcal{P}}$,
where $\xi$ is a threshold value for assigning new estimated directions to the previously detected sources. The subset $\bm{\mathcal{P}}$ specifies new estimated unit directions that are not attributed to any of previously detected sources. Considering $\bm{\mathcal{P}}=\{p_1,\dots,p_{\abs{\bm{\mathcal{P}}}}\}$, where $\abs{\bm{\mathcal{P}}}$ is the size of the subset $\bm{\mathcal{P}}$, we define $\bm{U}_3$ to store the corresponding unit directions as follow:
\begin{equation}
\bm{U}_3=\left[
\bm{u}_{p_1}^{\text{new}},\dots,\bm{u}_{p_{\abs{\bm{\mathcal{P}}}}}^{\text{new}}
\right].
\end{equation}

Finally, the estimation of the array manifold, $\hat{\bm{A}}^{(i)}$, is initialized as 
\begin{equation}\label{eq:A-init}
\bm{A}^{(i)}_{0}=\text{exp}(jK\bm{D}^T[\bm{U}_D^{(i-1)}, \bm{U}_3]).
\end{equation}

\subsection{AOA Refiner}\label{sec:Alg-AOA-ref}
\begin{figure*}[ht!]
\centering
\begin{subfigure}[b]{1.8\columnwidth}
\includegraphics[width=\textwidth]{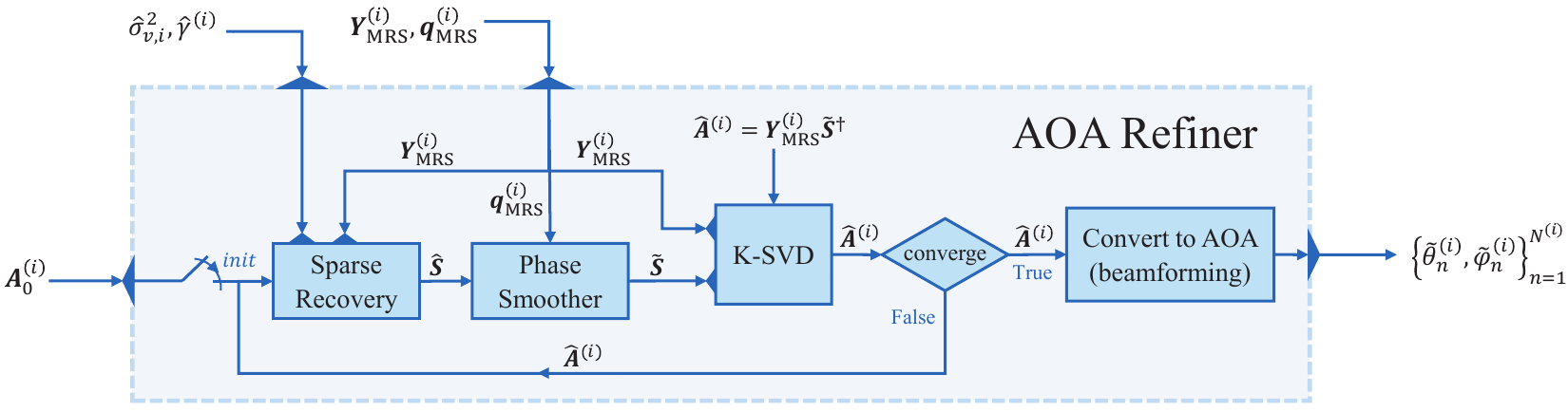}
\end{subfigure}%
\caption{Structure of AOA Refiner.}
\label{fig:block1}
\end{figure*}

AOA Refiner block aims to enhance AOA estimation by exploiting signal sparsity, phase characteristics, and each pulse continuity. It includes several sub-blocks and can estimate more AOAs than the number of receiver antennas.
Its architecture is illustrated in Fig.~\ref{fig:block1}, with detailed descriptions of the sub-blocks provided in this subsection.

In the following, we propose an algorithm to obtain an accurate estimation of the array manifold matrix $\bm{A}^{(i)}$ defined in \eqref{eq:array-manifold}, denoted by $\hat{\bm{A}}^{(i)}$. 
To this end, we first initialize $\hat{\bm{A}}^{(i)}$ with $\bm{A}^{(i)}_{0}$, which is derived in \eqref{eq:A-init}. 
Next, we estimate the signal matrix $\bm{S}^{(i)}_{\text{MRS}}$\eqref{eq:ed-out}.
To this end, we propose Sparse Recovery and Phase Smoothing algorithms, which are explained in the next subsections. 
Subsequently, we perform least squares (LS) method to update $\hat{\bm{A}}^{(i)}$ using the obtained estimation of 
$\bm{S}^{(i)}_{\text{MRS}}$ in the output of the Phase Smoothing algorithm ($\tilde{\bm{S}}$), 
as $\hat{\bm{A}}^{(i)}=\bm{Y}_{\text{MRS}}^{(i)}\tilde{\bm{S}}^{\dagger}$.
Finally, we employ the K-SVD algorithm to refine the estimated array manifold $\hat{\bm{A}}^{(i)}$ as follows:

First, we define the matrices $\bm{A}_j$ and $\bm{S}_j$ by excluding the $j$-th column of $\hat{\bm{A}}^{(i)}$ and the $j$-th row of $\tilde{\bm{S}}$, respectively.
Next, we construct
\begin{equation}
\bm{Y}_r = \bm{Y}_\text{MRS}^{(i)}-\bm{A}_j\bm{S}_j,
\end{equation}
which effectively suppresses the contribution of all sources except the $j$-th source in the received signal.
On the other hand, the columns of $\bm{Y}_r$ corresponding to zero elements of $[\tilde{\bm{S}}]_{j,:}$ are noisy and have no information about $j$-th source. Therefore, we delete these columns from $\bm{Y}_r$.

Then, according to \eqref{eq:Xi-approx}, we apply singular value decomposition (SVD) to approximate $\bm{Y}_r$ with the closest rank-1 matrix in Frobenius norm.
This matrix is given by $\bm{u}\sigma\bm{v}^H$, where $\sigma$ is the largest singular value and $\bm{u}$ and $\bm{v}$ are the corresponding left and right singular vectors, respectively.
Finally, we update $[\hat{\bm{A}}^{(i)}]_{:,j}$ and nonzero elements of $[\tilde{\bm{S}}]_{j,:}$ using $\bm{u}$ and $\sigma \bm{v}^H$, respectively.

We apply this procedure for $j = 1, 2, \dots, N^{(i)}$, where $N^{(i)}$ is the number of columns in $\hat{\bm{A}}^{(i)}$.
Thus, all columns of $\hat{\bm{A}}^{(i)}$ are updated once.
The K-SVD algorithm is described in lines 7-12 of Algorithm~\ref{alg:AOA-Refiner}.

The aforementioned steps are repeated until the convergence condition is met or the iteration count exceeds the maximum limit, $N_{\text{max}}^{\text{AOA}}$. Convergence is determined by
\begin{equation}
\frac{\|\hat{\bm{A}}^{(i)}-\hat{\bm{A}}^{(i)}_{\text{old}}\|_F}{N^{(i)}} \leq \varepsilon_{\text{AOA}},
\end{equation}
where $\hat{\bm{A}}^{(i)}_{\text{old}}$ represents the estimate from the previous iteration, and $\varepsilon_{\text{AOA}}$ is the convergence threshold for AOA Refiner.
The details of the proposed algorithm for array manifold refinement are shown in Algorithm~\ref{alg:AOA-Refiner}. 

\begin{algorithm}[ht]
\caption{Array Manifold Refining}\label{alg:AOA-Refiner}
 \hspace*{\algorithmicindent} \textbf{Inputs}:
$\bm{A}_{0}^{(i)}$, $\bm{Y}_{\text{MRS}}^{(i)}$, $\bm{q}_{\text{MRS}}^{(i)}$, $\hat{\sigma}_{v,i}^2$, $\hat{\gamma}^{(i)}$
  \\
  \hspace*{\algorithmicindent} \textbf{Design Parameters}:
$\varepsilon_{\text{AOA}}$, $N_\text{max}^{\text{AOA}}$
  \\
 \hspace*{\algorithmicindent} \textbf{Output}: 
$\hat{\bm{A}}^{(i)}$
\begin{algorithmic}[1]
\State 
\multiline{%
$\hat{\bm{A}}^{(i)}\gets \bm{A}_{0}^{(i)}$, 
$N^{(i)}$: number of $\hat{\bm{A}}^{(i)}$ columns, $\kappa=0$
}
\Repeat
\State
$\hat{\bm{A}}^{(i)}_{\text{old}} \gets \hat{\bm{A}}^{(i)}, \quad \kappa \gets \kappa+1$
\State 
\multiline{%
$\widehat{\bm{S}}$ is obtained by the Sparse Recovery block (Algorithm~\ref{alg:sparse-recovery}) using $\bm{Y}_{\text{MRS}}^{(i)}$, $\hat{\bm{A}}^{(i)}$, $\hat{\sigma}_{v,i}^2$ and $\hat{\gamma}^{(i)}$ as inputs.
}
\State 
\multiline{%
$\widetilde{\bm{S}}$ is obtained as the output of the Phase Smoother block (Algorithm~\ref{alg:phase-smoother}) using $\widehat{\bm{S}}$ and $\bm{q}_{\text{MRS}}^{(i)}$ as inputs.
}
%
\State 
\multiline{%
$\hat{\bm{A}}^{(i)}\gets \bm{Y}_{\text{MRS}}^{(i)}\tilde{\bm{S}}^{\dagger}$, update $N^{(i)}$
}
\For{$j=1:N^{(i)}$}
\State
\multiline{%
$\bm{S}_j\gets$ $\tilde{\bm{S}}$ without its $j$-th row

$\bm{A}_j\gets$ $\hat{\bm{A}}^{(i)}$ without its $j$-th column
}
\State 
$\bm{Y}_r\gets \bm{Y}_{\text{MRS}}^{(i)}-\bm{A}_j\bm{S}_j$
\State 
\multiline{%
Remove columns from $\bm{Y}_r$ which correspond to zero elements in $[\tilde{\bm{S}}]_{j,:}$.
}
\State 
\multiline{%
Apply the SVD algorithm to $\bm{Y}_r$ to obtain the largest singular value ($\sigma$) and its right and left singular vectors ($\bm{u}$, $\bm{v}$).
}
\State
\multiline{%
Replace
$[\hat{\bm{A}}^{(i)}]_{:,j}$
with $\bm{u}$, and also replace the nonzero elements of
$[\tilde{\bm{S}}]_{j,:}$
with $\sigma\bm{v}^H$.
}
\EndFor
\Until{$\frac{\|\hat{\bm{A}}^{(i)}-\hat{\bm{A}}^{(i)}_{\text{old}}\|_F}{N^{(i)}} \leq \varepsilon_{\text{AOA}}$ or $\kappa=N_\text{max}^{\text{AOA}}$}
\end{algorithmic}
\end{algorithm}

Finally, after obtaining an accurate estimation of the array manifold (i.e., $\hat{\bm{A}}^{(i)}$) using Algorithm~\ref{alg:AOA-Refiner}, we employ beamforming technique to precisely estimate the 2D-AOAs, as outlined below:
\begin{equation}\label{eq:refined-AOAs}
\{\tilde{\theta}_n^{(i)}, \tilde{\phi}_n^{(i)}\}=
\argmax_{\theta\in\bm{\Theta},\phi\in\bm{\Phi}}\bm{a}(\theta,\phi)^H[\hat{\bm{A}}^{(i)}]_{:,n},
n=1,\dots ,N^{(i)},
\end{equation}
where $\bm{a}(\theta,\phi)$ indicates the steering vector, defined in \eqref{eq:streeing_vec}, 
$N^{(i)}$ is the number of columns in $\hat{\bm{A}}^{(i)}$, and $\bm{\Theta}$, $\bm{\Phi}$ are the search ranges associated with the elevation and azimuth AOAs, respectively, as defined in Section~\ref{sec:AOA-Estimator}. 
\vspace{-6mm}
\subsubsection{Sparse Recovery}\label{sec:sparse_recovery}
This block focuses on estimating
$\bm{S}_{\text{MRS}}^{(i)}$ by leveraging its sparsity, with the result denoted as $\widehat{\bm{S}}$.
Let $\bm{y}_g$, $\bm{s}_g$ and $\bm{v}_g$ represent the $g$-th column of $\bm{Y}_{\text{MRS}}^{(i)}$, $\bm{S}_{\text{MRS}}^{(i)}$ and $\bm{V}_{\text{MRS}}^{(i)}$, respectively, where $g\in\{1,\dots, G_{\text{MRS}}^{(i)}\}$. 
From \eqref{eq:ed-out}, we have
\begin{equation}\label{eq:en-vec-out}
\bm{y}_g=
\bm{A}^{(i)}
\bm{s}_g
+
\bm{v}_g.
\end{equation}

We assume that the array manifold $\bm{A}^{(i)}$ has already been estimated as $\hat{\bm{A}}^{(i)}$.  
Hence, the sparse recovery optimization problem for estimating $\bm{s}_g$ can be written as follows:
\begin{equation}\label{prob:sparse_recovery}
\begin{aligned}
\min \quad & \norm{\bm{s}_g}_{0}\\
\textrm{s.t.} \quad & \norm{\hat{\bm{A}}^{(i)}\bm{s}_g-\bm{y}_g}_2\le \epsilon,
\end{aligned}
\end{equation}
where $\epsilon$ is a design threshold parameter. 
In the following, first we explain how to estimate and choose the suboptimal threshold value $\epsilon$ for solving \eqref{prob:sparse_recovery}.

To establish a theoretical model for the optimal value of $\epsilon$, we begin by assuming that the precise value of $\bm{A}^{(i)}$ is available. Thus, under this assumption, the constraint of optimization problem \eqref{prob:sparse_recovery} for its solution modifies to:
\begin{equation}\label{eq:best_case_sparse_prob}
\norm{\bm{A}^{(i)}(\hat{\bm{s}}_g-\bm{s}_g)-\bm{v}_g}_2^2\le \epsilon^2,
\end{equation}
where $\bm{A}^{(i)}\bm{s}_g$ denotes the noise-free received signal, whose elements have average energy $E_{s,\text{avg}}$. 
Clearly, the best estimation is happened when $\hat{\bm{s}}_g=\bm{s}_g$. 
Hence, according to \eqref{eq:best_case_sparse_prob}, the minimum value of $\epsilon^2$ equals to $\norm{\bm{v}_g}_2^2$. 
Motivated by this fact, we model the optimal $\epsilon^2$ as follows
\begin{equation}\label{eq:epsilon-opt}
\begin{split}
\hat{\epsilon}^2_{\text{opt}}
& \approx 
f(N,\gamma^{(i)})\EX\left\{\norm{\bm{A}^{(i)}\bm{s}_g}_2^2\right\}+
\EX\left\{
\norm{\bm{v}_g}^2_2
\right\}\\
& =
f(N,\gamma^{(i)})M E_{s,\text{avg}}+M\sigma^2_v,
\end{split}
\end{equation}
where, $f(\cdot)$ indicates a nonlinear function of the number of sources ($N$) and the instantaneous SNR at $i$-th time window ($\gamma^{(i)}$). 
Intuitively, with smaller values of $\gamma^{(i)}$, the difference between $\hat{\bm{s}}_g$ and $\bm{s}_g$ gets larger and as a result, $f(N,\gamma^{(i)})$ increases.
Additionally, as $N$ grows, the correlation between the columns of $\bm{A}^{(i)}$ and the probability of simultaneous signal transmission from sources both increase.
This amplifies ambiguity in sparse recovery, resulting in a greater disparity between $\hat{\bm{s}}_g$ and $\bm{s}_g$, consequently leading to an increase in $f(N,\gamma^{(i)})$.
Please refer to Appendix~\ref{app:f-est} for the details on how to choose and obtain the function $f(N,\gamma^{(i)})$. 

Notably, since the exact values for $N$, $\gamma^{(i)}$, $E_{s,\text{avg}}$, and $\sigma^2_v$ are unavailable, the proposed system relies on their estimated counterparts, $N^{(i)}$, $\hat{\gamma}^{(i)}$, $\hat{\sigma}^2_{v,i}$ and $\hat{E}_{s,\text{avg}} = \hat{\gamma}^{(i)}\hat{\sigma}^2_{v,i}$, to compute $\hat{\epsilon}_\text{opt}$ using \eqref{eq:epsilon-opt}. 
%
Once $\hat{\epsilon}_{\text{opt}}$ is obtained, each column vector $\bm{s}_g = [\bm{S}_{\text{MRS}}^{(i)}]_{:,g}$ is estimated by solving the optimization problem \eqref{prob:sparse_recovery}:

If $\norm{\bm{y}_g}_2\le \hat{\epsilon}_{\text{opt}}$ then $\hat{\bm{s}}_g=\bm{0}$, 
otherwise let $L_{\text{max}}$ denote the maximum sparsity level of the columns in $\bm{S}_{\text{MRS}}^{(i)}$. 
To solve \eqref{prob:sparse_recovery}, for each $j\in\{1,\dots ,L_\text{max}\}$,
we first define the vector $\bm{q}_k\in \mathbb{Z}^j$, which represents the $k$-th combination of selecting $j$ elements from the set $\{1,\dots,N^{(i)}\}$. Based on $\bm{q}_k$, the binary vector $\bm{b}_k \in \{0,1\}^{N^{(i)}}$ constructed as follows:
\begin{equation}
[\bm{b}_k]_l=
\begin{cases*}
1,&\text{if}  $l\in \bm{q}_k$,  \\
0,&\text{otherwise},
\end{cases*}
\quad l=1,\dots,N^{(i)}.
\end{equation}
Next, we define the set $\bm{\mathcal{B}}^{(j)}=\{\bm{b}_k\}_{k=1}^{\binom{N^{(i)}}{j}}$, which contains all distinct binary vectors of length $N^{(i)}$ with exactly $j$ ones.

Subsequently, for each $\bm{b}_k\in\bm{\mathcal{B}}^{(j)}$, we define $\bm{A}_k=[\hat{\bm{A}}^{(i)}]_{:,\bm{b}_k}$ as the submatrix of $\hat{\bm{A}}^{(i)}$ consisting of the selected $j$ columns. Similarly, let $\bm{s}_g^{(k)}\in\mathbb{C}^{j}$ be the vector containing the elements of $\bm{s}_g$ indexed by the nonzero positions in $\bm{b}_k$. 
Hence, for a given sparsity level $j$ and the corresponding nonzero indices specified by $\bm{b}_k$, problem  \eqref{prob:sparse_recovery} reduces to finding a feasible solution under the constraint
\begin{equation}\label{prob:sparse-recovery-simple}
\norm{\bm{A}_k\bm{s}_g^{(k)}-\bm{y}_g}_2\le \hat{\epsilon}_{\text{\text{opt}}}.
\end{equation}

It is noteworthy that the problem of minimizing $\norm{\bm{A}_k\bm{s}_g^{(k)}-\bm{y}_g}_2$ is a least squares (LS) problem, whose optimal solution can be obtained as 
$
\hat{\bm{s}}_g^{(k)}=\bm{A}_k^{\dagger}\bm{y}_g
$. 
The corresponding LS estimation error can be calculated as
\begin{equation}\label{eq:sr-est-err}
\norm{\bm{y}_g-\bm{A}_k\hat{\bm{s}}_g^{(k)}}_2
=
\left\Vert\left(\bm{I}_M-\bm{A}_k\bm{A}_k^{\dagger}\right)\bm{y}_g\right\Vert_2.
\end{equation}

Next, assuming $j=1$, we determine the index $\hat{k}$ that minimizes the LS error given in \eqref{eq:sr-est-err}. If the constraint \eqref{prob:sparse-recovery-simple} is satisfied with $\hat{\bm{s}}_g^{(\hat{k})}$, the algorithm concludes. Otherwise, we gradually increment $j$ and repeat the previous steps until convergence is achieved. 
The convergence is happened either the constraint \eqref{prob:sparse-recovery-simple} is satisfied or the sparsity level $j$ reaches the maximum level $L_{\text{max}}$. 
As a result, the estimated nonzero elements of $\bm{s}_{g}$, given by $\hat{\bm{s}}_g^{(\hat{k})}$, are placed at positions indicated by $\bm{b}_{\hat{k}}$ in $\hat{\bm{s}}_g$, with all other elements set to zero.

Finally, $\widehat{\bm{S}}$ is obtained by concatenating $\{\hat{\bm{s}}_g\}_{g=1}^{G_{\text{MRS}}^{(i)}}$, yielding $\widehat{\bm{S}} = [\hat{\bm{s}}_1, \dots, \hat{\bm{s}}_{G_{\text{MRS}}^{(i)}}]$.
The details are given in Algorithm~\ref{alg:sparse-recovery}.
\begin{algorithm}[ht]
\caption{Sparse Recovery Algorithm}\label{alg:sparse-recovery}
\begin{algorithmic}[0]
\State
\multiline{
\textbf{Inputs}:
$\hat{\gamma}^{(i)}$, $\hat{\sigma}^2_{v,i}$, $\bm{Y}_\text{MRS}^{(i)}$, $\hat{\bm{A}}^{(i)}$, $f(N,\gamma)$ defined in \eqref{eq:f}
}
\State \textbf{Design Parameter}: $L_{\text{max}}$
\State \textbf{Output}: $\widehat{\bm{S}}$
\end{algorithmic}
\begin{algorithmic}[1]
\State
\multiline{%
$M$, $N^{(i)}$: Number of rows and columns in $\hat{\bm{A}}^{(i)}$.

$G_{\text{MRS}}^{(i)}$: Number of columns in $\bm{Y}_\text{MRS}^{(i)}$.

$\widehat{\bm{S}} \gets \bm{0}_{N^{(i)}\times G_{\text{MRS}}}$
}
\State
\multiline{%
$\hat{E}_{s,\text{avg}}= \hat{\gamma}^{(i)}\hat{\sigma}^2_{v,i}$, 

$\hat{\epsilon}_{\text{opt}}=\sqrt{M\hat{E}_{s,\text{avg}}f(N^{(i)},\hat{\gamma}^{(i)})+M\hat{\sigma}^2_{v,i}}$
}
\For{$g=1:G_\text{MRS}^{(i)}$}
\State
\multiline{%
$m \gets \norm{\bm{y}_g}_2$,
$j\gets1$
\Comment{$\bm{y}_g=[\bm{Y}_{\text{MRS}}^{(i)}]_{:,g}$}
}
\While{$m>\hat{\epsilon}_{\text{opt}}$ and $j\le L_{\text{max}}$}
\State
\multiline{%
$\bm{\mathcal{B}}^{(j)} \gets \{\bm{b}_k\}_{k=1}^{\binom{N^{(i)}}{j}}$: All binary vector permutations of length $N^{(i)}$ containing $j$ ones.
}
\State
\multiline{%
$\bm{A}_k \gets [\hat{\bm{A}}^{(i)}]_{:,\bm{b}_k}$ 
}
\State
\multiline{%
$\hat{k} \gets \argmin_{k}\left\Vert\left(\bm{I}_M-\bm{A}_k\bm{A}_k^{\dagger}\right)\bm{y}_g\right\Vert_2$
}
\State
$\hat{\bm{s}}_g^{(\hat{k})} = \bm{A}_{\hat{k}}^{\dagger}\bm{y}_g$, $m = \norm{\bm{A}_{\hat{k}}\hat{\bm{s}}_g^{(\hat{k})}-\bm{y}_g}_2$,
$j\gets j+1$
\If{$m\le\hat{\epsilon}_{\text{opt}}$ or $j> L_{\text{max}}$}
\State
$\big[\widehat{\bm{S}}\big]_{\bm{b}_{\hat{k}},g}=\hat{\bm{s}}_g^{(\hat{k})}$
\EndIf
\EndWhile
\EndFor
\end{algorithmic}
\end{algorithm}

\vspace{-5mm}
\subsubsection{Phase Smoother}
\begin{figure*}[ht!] 
\centering
\begin{subfigure}[b]{\columnwidth}
\centering
\includegraphics[width=\columnwidth]{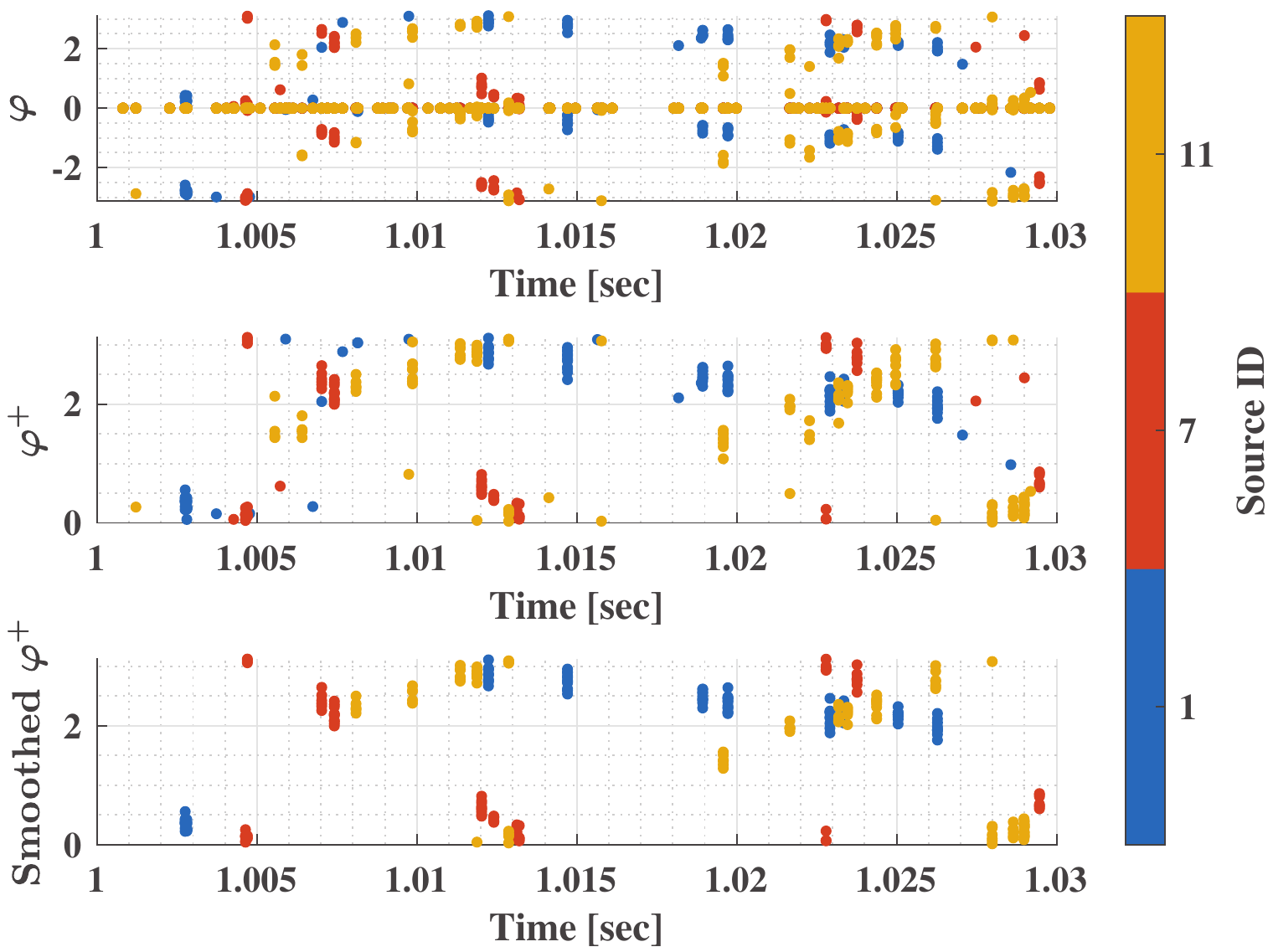}
\caption{$\text{SNR}^*=12\,\mathrm{dB}$}
\label{fig:11src_PhaseSmoother_CLsnr12_t1}
\end{subfigure}%
\hfill
\begin{subfigure}[b]{\columnwidth}
\centering
\includegraphics[width=\columnwidth]{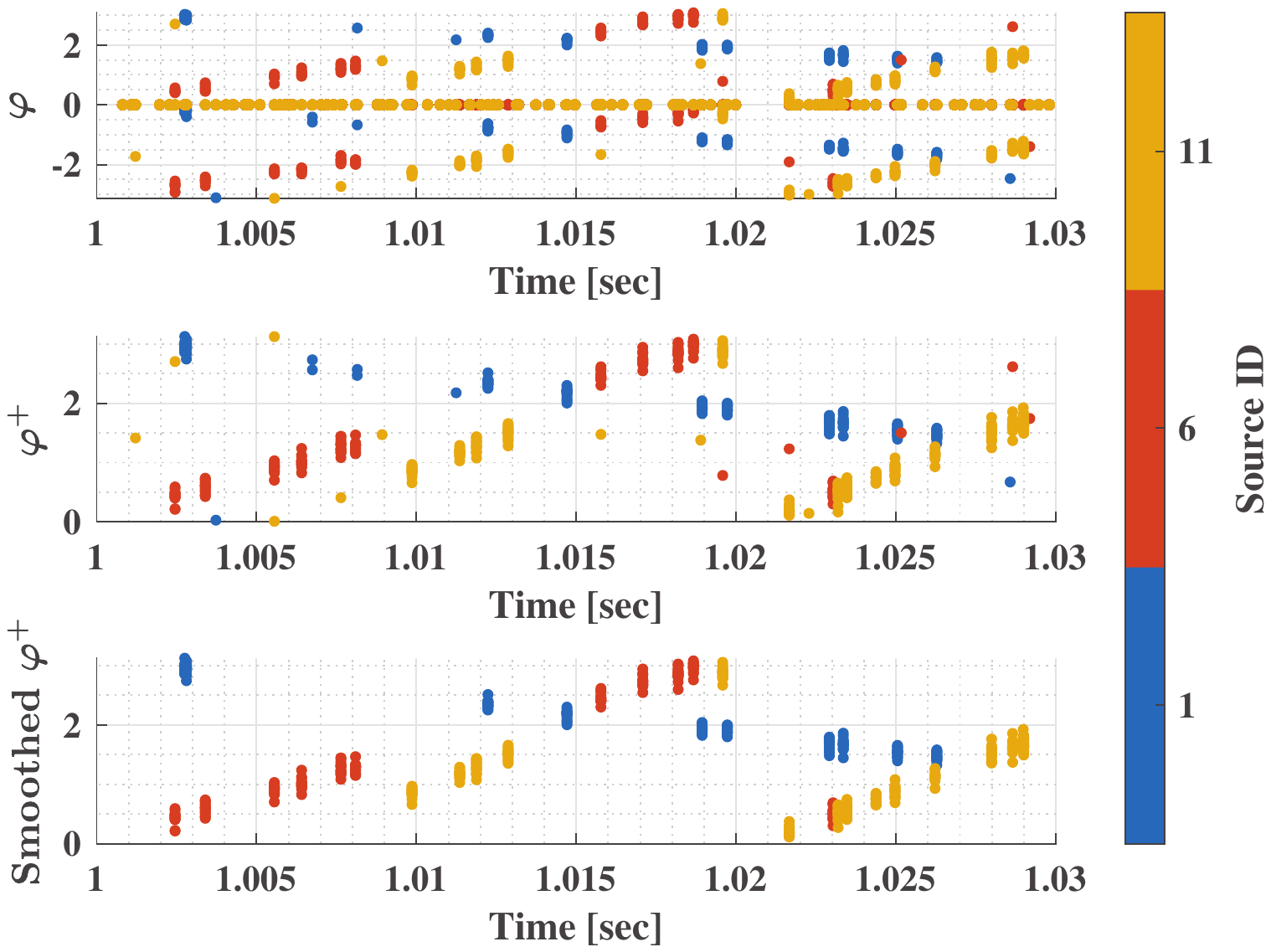}
\caption{$\text{SNR}^*=20\,\mathrm{dB}$
}
\label{fig:11src_PhaseSmoother_CLsnr20_t1}
\end{subfigure}%
\caption{\label{fig:phase_smoother}
Procedure of the Phase Smoothing algorithm (See Fig.~\ref{fig:inst-noise-snr} in Section~\ref{sec:performance_En} for corresponding instantaneous SNR).}
\end{figure*}

As mentioned in Section~\ref{sec:sys}, the phase of each row in $\bm{S}^{(i)}$ has approximately smooth variations over time. 
Considering this feature and also the continuity of waveforms of transmitted spikes $p_n(t)$, we propose a Phase Smoothing algorithm to refine the estimated signal matrix $\widehat{\bm{S}}$ in the output of Sparse Recovery block. 
For the sake of better exposition, the output of each main step, is sub-plotted for $\text{SNR}^*=12\,\mathrm{dB}$ and $\text{SNR}^*=20\,\mathrm{dB}$ in a specific time window $W = [1,1.03]\,\text{sec}$\footnote{The duration of time window $T$ is set to $0.03$ seconds.} as shown in Fig.~\ref{fig:phase_smoother}. 
The first subplot in Fig.~\ref{fig:phase_smoother}, illustrates the phase values of three rows in $\widehat{\bm{S}}$, with each row corresponding to a specific source.

The sign of $p_n(t)$ in \eqref{eq:sum_p_n} is either positive or negative. 
Thus, the phases of the elements of the $n$-th row of $\widehat{\bm{S}}$ are in two parallel lines $\ell_1$ and $\ell_2$ with distance $\pi$. 
Hence, in order to remove the distance between $\ell_1$ and $\ell_2$, at first, we negate the elements of $[\widehat{\bm{S}}]_{n,:}$, whose imaginary part is negative. After this step, the phase of all elements becomes a positive value. 
Next, we remove the elements with zero absolute value. 
The phase value of the output of this step is plotted in the second subplots of Fig.~\ref{fig:phase_smoother} (it is denoted by $\varphi^+$). 

Subsequently, similar to the Energy Detector algorithm, we exploit the continuous shape of the transmitted spikes by estimating the consecutive samples in the output of previous step.
However, at the same time, due to the smooth phase variations, we only keep those consecutive samples, whose phase difference is less than $\epsilon_\varphi$ or greater than $\pi-\epsilon_\varphi$\footnote{If the phase difference between two samples is less than $\epsilon_\varphi$, but their imaginary parts have opposite signs, adjusting their phases to be positive will result in a phase difference that exceeds $\pi-\epsilon_\varphi$.
Hence, the phase difference greater than $\pi-\epsilon_\varphi$ also indicates a smooth variation.}. 
The positive smoothed phase, are shown in the third subplots in Fig.~\ref{fig:phase_smoother}. 
Finally, we revert the remaining samples that were negated in the initial step to their original values, and then store all of them in their respective positions in the $n$-th row of $\widetilde{\bm{S}}$.
Other elements of this row remain zero. The aforementioned procedure is repeated for all rows of $\widehat{\bm{S}}$.
Please refer to Algorithm~\ref{alg:phase-smoother} for the details of the proposed Phase Smoother block.

\begin{algorithm}[ht]
\caption{Phase Smoothing Algorithm}\label{alg:phase-smoother}
 \hspace*{\algorithmicindent} \textbf{Inputs}:
$\widehat{\bm{S}}$, $\bm{q}_\text{MRS}^{(i)}$
  \\
 \hspace*{\algorithmicindent} \textbf{Design Parameters}:
$\text{diff}_\text{max}$, $L_{\text{adj}}$, $\epsilon_\varphi$
  \\
 \hspace*{\algorithmicindent} \textbf{Output}: 
$\widetilde{\bm{S}}$
\begin{algorithmic}[1]
\State 
\multiline{%
$N^{(i)}, G_\text{MRS}^{(i)}$: Number of rows and columns in $\widehat{\bm{S}}$.

$\widetilde{\bm{S}} \gets \bm{0}_{N^{(i)}\times G_{\text{MRS}}^{(i)}}$
}
\For{$n=1:N^{(i)}$}
\State
\multiline{%
$\hat{\bm{s}}_n^T\gets [\widehat{\bm{S}}]_{n,:}$, 
$\bm{q}_{\text{sel}}\gets[1,2,\dots ,G_\text{MRS}^{(i)}]$, 
$\bm{q}'_\text{MRS}\gets\bm{q}_\text{MRS}^{(i)}$
}
\State
\multiline{%
Negate elements of $\hat{\bm{s}}_n^T$ with negative imaginary part.
}
\State
\multiline{%
Remove elements from $\hat{\bm{s}}_n^T$, $\bm{q}'_\text{MRS}$, and $\bm{q}_{\text{sel}}$ corresponding to zero-valued entries in $\hat{\bm{s}}_n^T$.
}
\State 
\multiline{%
$\bm{\varphi} \gets$ Phase components of the elements of $\hat{\bm{s}}_n^T$.
}
\State
\multiline{%
$\bm{q}_\varphi\gets (\text{diff}_1(\bm{\varphi})<\epsilon_\varphi)|(\text{diff}_1(\bm{\varphi})>\pi-\epsilon_\varphi)$

$\bm{q}_{\text{adj}}\gets \text{diff}_1(\bm{q}'_\text{MRS})<\text{diff}_\text{max}$
}
\State 
\multiline{%
$\bm{q}_{\text{sel}}$ is updated by selecting the elements corresponding to all sequences of $L_\text{adj}-1$ or more consecutive 1-valued elements in $(\bm{q}_\text{adj}\&\bm{q}_\varphi)$, along with the element directly below each sequence.
}
\State 
\multiline{%
$\big[\widetilde{\bm{S}}\big]_{n,\bm{q}_{\text{sel}}}
=
\big[\widehat{\bm{S}}\big]_{n,\bm{q}_{\text{sel}}}
$
}
\EndFor
\State
\multiline{%
Remove rows of $\widetilde{\bm{S}}$ whose all elements are zero.
}
\end{algorithmic}
\end{algorithm}

\section{Location Estimation}\label{sec:loc-estimator}
This section explores the proposed approach for source localization, utilizing the estimated elevation and azimuth AOAs to determine source positions. The analysis starts with addressing the localization of a single source and subsequently introduces the method developed for multi-source localization.

\subsection{Proposed Method for Single Source Localization}
Let $N_a$ denote the number of time windows in which the 2D-AOAs of the source $s$ are estimated among the first $i-1$ such intervals.
Thus, we have $N_a$ anchors, each providing 2D-AOA estimations of this source.
Denote the positions of these anchors as $\bm{r}_1, \bm{r}_2, \ldots, \bm{r}_{N_a}$, where $\bm{r}_j = [r_{j,x}, r_{j,y}, r_{j,z}]^T$ for $j = 1, \ldots, N_a$. Each anchor point corresponds to a line $\ell_1, \ell_2, \ldots, \ell_{N_a}$, with associated unit vectors $\bm{u}_1, \bm{u}_2, \ldots, \bm{u}_{N_a}$ along these lines pointing to the location of source $s$. 
These unit vectors are derived by substituting the estimated azimuth and elevation AOAs into \eqref{eq:unit-dir}.  
However, vectors $\bm{u}_1, \ldots, \bm{u}_{N_a}$ may be subject to noise corruption.
Our objective is to estimate the location of source $s$ with coordinates $\bm{w}^* = [x^*, y^*, z^*]^T$ on the available map defined by $z = \mathcal{M}(x, y)$. This is achieved by minimizing the sum of squared distances from $\bm{w}^*$ to the lines $\ell_1, \ldots, \ell_{N_a}$. The mapping function $\mathcal{M}: \mathbb{R}^2 \longrightarrow \mathbb{R}$ describes the relationship between the height $z$ of each point and its $(x, y)$ coordinates in the $xy$ plane.

The line $\ell_j$ can be expressed as $ \ell_j = \alpha \bm{u}_j + \bm{r}_j, \alpha \in \mathbb{R}$. Let $\bm{w} \in \mathbb{R}^3$ represent an arbitrary point. 
The distance from $\bm{w}$ to the line $\ell_j$ can be calculated as:
\begin{equation} \label{eq:dist-point}
d_j^2=
\norm{\bm{w}-\bm{r}_j}^2_2-\left((\bm{w}-\bm{r}_j)^T\bm{u}_j\right)^2.
\end{equation}
Therefore, the sum of squared distances from the point $\bm{w}$ to the lines $\ell_1, \dots, \ell_{N_a}$ is computed as follows:
\begin{equation} \label{eq:sum-square}
S(\bm{w}) =
\sum_{j=1}^{N_a}
d_j^2 
= 
\sum_{j=1}^{N_a}
(\bm{w}-\bm{r}_j)^T(\bm{I}_3-\bm{u}_j\bm{u}_j^T)(\bm{w}-\bm{r}_j).
\end{equation}

As a result, the optimization problem can be expressed as
\begin{equation} \label{opt:lemma}
\begin{aligned}
\bm{w}^*
=
\argmin_{\bm{w}} \quad & S(\bm{w})\quad
\textrm{s.t.} \quad & z=\mathcal{M}(x,y),
\end{aligned}
\end{equation}
where $\bm{w}=[x,y,z]^T$. 
In addressing the optimization problem outlined in \eqref{opt:lemma}, we start by leveraging the constraint to reduce dimensionality. This is accomplished by defining the variables 
$\bm{w'}\in\mathbb{R}^2, \bm{r}'_j\in\mathbb{R}^2,
\bm{C}_j\in\mathbb{R}^{2\times 2},\bm{b}_j\in\mathbb{R}^2,d_j\in\mathbb{R}$ for $j = 1,\dots, N_a$ such that
\begin{equation}
\bm{w'} =
\renewcommand{\arraystretch}{0.8} 
\begin{bmatrix}
\bm{x}\\
\bm{y}
\end{bmatrix}
,\bm{r}'_j =
\renewcommand{\arraystretch}{0.8} 
\begin{bmatrix}
\bm{r}_{j,x}\\
\bm{r}_{j,y}
\end{bmatrix}
,\bm{I}_3-\bm{u}_j\bm{u}_j^T =
\renewcommand{\arraystretch}{0.8} 
\begin{bmatrix}
\bm{C}_j & \bm{b}_j\\
\bm{b}_j^T & d_j
\end{bmatrix}.
\end{equation}
Consequently, \eqref{opt:lemma} can be rewritten as follows:
\begin{equation} \label{opt:lemma-map}
\begin{aligned}
\bm{w}^* = &
\argmin_{\bm{w}', z} \:
\sum_{i=1}^{N_a}
\Big[(\bm{w}'-\bm{r}'_i)^T\bm{C}_i(\bm{w}'_i-\bm{r}'_i)
\\
& +2(\bm{w}'-\bm{r}'_i)^T\bm{b}_i(z-r_{i,z})
+ (z-r_{i,z})^2d_i\Big],\\
&\textrm{s.t.} \quad z=\mathcal{M}(\bm{w}').\\
\end{aligned}
\end{equation}
It is important to note that when $z$ remains constant, \eqref{opt:lemma-map} exhibits convexity concerning $\bm{w}'$. In this scenario, the optimal solution can be efficiently determined through a closed-form formula by equating the gradient of the objective function to zero. The solution is expressed as follows:
\begin{equation}\label{eq:v_prime_opt}
\bm{w}'(z)
=
\bigg[
\sum_{j=1}^{N_a}
\bm{C}_j
\bigg]^{\dagger}
\sum_{j=1}^{N_a}
\left(
\bm{C}_j\bm{r}'_i
+
\bm{b}_j(r_{j,z}-z)
\right),
\end{equation} 
where $\bm{w}'(z)$ represents the optimal solution corresponding to a specified value of $z$. 
Driven by this observation, we apply the gradient projection (GP) method \cite{boyd2004convex} to address the optimization problem in \eqref{opt:lemma-map}:
 
Initially, we compute $\bm{w}'$ using equation \eqref{eq:v_prime_opt} with a given initial value of $z$, subsequently updating $z$ via $z=\mathcal{M}(\bm{w}')$. 
This iterative process continues until convergence, determined by the distance between position estimates in successive iterations dropping below $\varepsilon_{\text{Loc}}$, or until the iteration count surpasses the predefined limit $N_{\text{max}}^{\text{Loc}}$. The convergence condition is expressed as
\begin{equation}
\sqrt{\|\bm{w}' - \bm{w}'_{\text{old}}\|_2^2 + (z - z_{\text{old}})^2} \leq \varepsilon_{\text{Loc}},
\end{equation}
where $\bm{w}'_{\text{old}}$ and $z_{\text{old}}$ represent the estimated values from the previous iteration.

Furthermore, based on \eqref{eq:v_prime_opt}, the positioning algorithm can be expressed recursively. 
To this end, we define the matrix $\bm{C}\in\mathbb{R}^{2\times 2}$ and the vectors $\bm{h}\in \mathbb{R}^2$ and $\bm{b}\in \mathbb{R}^2$ as:
\begin{equation}\label{eq:defs}
\bm{C}
=
\sum_{j=1}^{N_a}
\bm{C}_j
,\ 
\bm{h}
=
\sum_{j=1}^{N_a}
\bm{h}_j
,\ 
\bm{b}
=
\sum_{j=1}^{N_a}
\bm{b}_j,
\end{equation}
where $\bm{h}_j = [\bm{C}_j, \bm{b}_j] \bm{r}_j$.
By incorporating \eqref{eq:defs} in \eqref{eq:v_prime_opt}, the solution $\bm{w}'(z)$ can be reformulated as
\begin{equation}\label{eq:v_prime_opt_z}
\bm{w}'(z)
=
\bm{C}^{\dagger}
\left(
\bm{h}
-
\bm{b}z
\right).
\end{equation}
Therefore, by introducing an additional line $\ell_{N_a+1}$ in the $i$-th time window, the optimal solution \eqref{eq:v_prime_opt_z} can be readily determined by updating $\bm{C}$, $\bm{h}$, and $\bm{b}$ as follows:
\begin{equation}
\bm{C}\gets \bm{C}+\bm{C}_{N_a+1},\ 
\bm{h}\gets \bm{h}+\bm{h}_{N_a+1},\ 
\bm{b}\gets \bm{b}+\bm{b}_{N_a+1},
\end{equation}
and then reinserting the updated values into \eqref{eq:v_prime_opt_z}. 
The specifics of the low-complexity recursive GP-based algorithm are outlined in Algorithm~\ref{alg:opt-GP}.

\begin{algorithm}[h]
\caption{Recursive GP Localization Algorithm}\label{alg:opt-GP}
\begin{algorithmic}[0]
\State
\multiline{%
\textbf{Inputs:}
$\bm{C}$, $\bm{h}$, $\bm{b}$, mapping function $\mathcal{M}(x,y)$, and 
a new line directed along $\bm{u}$ that passes through $\bm{r}$
}
\State \textbf{Design Parameters}:
$\varepsilon_{\text{Loc}}$, $N_\text{max}^{\text{Loc}}$
\State \textbf{Output:} 
\multiline{%
$\bm{w}^*$ and updated $\bm{C},\ \bm{h},\ \bm{b}$
}
\end{algorithmic}
\begin{algorithmic}[1]
\State 
\multiline{%
$\bm{F}= \bm{I}_3-\bm{u}\bm{u}^T$,
$\bm{h}\gets \bm{h}+[\bm{F}]_{1:2,:}\bm{r}$,
$\bm{C}\gets \bm{C}+[\bm{F}]_{1:2,1:2}$,  
$\bm{b}\gets \bm{b}+[\bm{F}]_{1:2,3}$
}
\State 
\multiline{%
$z=0,\; \bm{w}'=[0,0]^T, \; \kappa=0$
\Comment{initialization}
}
\Repeat
\State 
\multiline{%
$z_{\text{old}}\gets z,\; \bm{w}'_{\text{old}}\gets \bm{w}', \; \kappa \gets \kappa+1$
}
\State 
\multiline{%
$\bm{w}'\gets \bm{C}^{\dagger}
\left(
\bm{h}
-
\bm{b}z
\right)$
}
\State $z\gets \mathcal{M}(\bm{w}')$
\Until{${\scriptstyle \sqrt{\|\bm{w}' - \bm{w}'_{\text{old}}\|_2^2 + (z - z_{\text{old}})^2}} \leq \varepsilon_{\text{Loc}}$ or $\kappa=N_{\text{max}}^{\text{Loc}}$}
\State
$\bm{w}^*=[\bm{w}'^T,z]^T$
\end{algorithmic}
\end{algorithm}
\subsection{Proposed Localization Method for Multiple Sources}\label{sec:proposed-loc-method}
Recall from Section~\ref{sec:Alg-AOA-ref} that $\{\tilde{\theta}_n^{(i)}\}_{n=1}^{N^{(i)}},\ \{\tilde{\phi}_n^{(i)}\}_{n=1}^{N^{(i)}}$ are the estimated elevation and azimuth AOAs 
at the $i$-th time window by the AOA Refiner. We define the corresponding unit directions $\{\tilde{\bm{u}}_n^{\mathrm{new}}\}_{n=1}^{N^{(i)}}$ according to \eqref{eq:unit-dir} using these AOAs. 
Next, the inner product of each obtained direction $\tilde{\bm{u}}_n^{\text{new}}$ with the unit vectors in columns of $\bm{U}_D^{(i-1)}$, defined in \eqref{eq:U_D}, is calculated. Let $m_n$ be the greatest obtained value,
\begin{equation}
\begin{split}
& \hat{s}_n= \argmax_{s}\abs{\tilde{\bm{u}}_n^{{\text{new}}^T} 
[\bm{U}_D^{(i-1)}]_{:,s}},
\ n=1,\dots ,N^{(i)}, \\
& m_n = \abs{\tilde{\bm{u}}_n^{{\text{new}}^T}[\bm{U}_D^{(i-1)}]_{:,\hat{s}_n}}.
\end{split}
\end{equation}
If $m_n$ exceeds the threshold $\xi$, the unit direction $\tilde{\bm{u}}_n^{\text{new}}$ is assigned to the $\hat{s}_n$-th source among the previously estimated locations. Conversely, if $m_n < \xi$, it suggests that this direction may correspond to a new, undetected source.

In the first scenario, where $m_n \geq \xi$, the matrix $\bm{C}$ and vectors $\bm{h}$ and $\bm{b}$, as specified in \eqref{eq:defs}, are assumed to correspond to the $\hat{s}_n$-th source. 
The position of this source, as well as the associated $\bm{C},\ \bm{h}$, and $\bm{b}$, are updated by applying Algorithm~\ref{alg:opt-GP}.
In this algorithm ,$\tilde{\bm{u}}_n^{\mathrm{new}}$ and the array position $\bm{r}^{(i)}$ are used as the unit direction $\bm{u}$ and the anchor position $\bm{r}$, respectively.

In the scenario $m_n < \xi$, as previously indicated, the direction $\tilde{\bm{u}}_n^{\text{new}}$ may be oriented towards a new, undetected source $s'$. 
Then, we define the matrix $\bm{F} = \bm{I}_3 - \tilde{\bm{u}}_n^{\text{new}} \tilde{\bm{u}}_n^{\text{new}^T}$ to initialize the associated $\bm{C}, \bm{h},$ and $\bm{b}$ \eqref{eq:defs} as:
\begin{equation}
\bm{C}=[\bm{F}]_{1:2,1:2},\ \ \ 
\bm{h}= [\bm{F}]_{1:2,:}\bm{r}^{(i)},\ \ \ 
\bm{b}=[\bm{F}]_{1:2,3}.
\end{equation}

The number of time windows in which a source $s$ is detected and its position updated may vary. To evaluate the reliability of the estimated sources, we propose a metric using a counter, ${hist}_s$, for each source $s$. Initially set to $1$ upon detection, ${hist}_s$ increments with subsequent detections in the following time windows.
The source $s$ probability of reliability is defined as $P_{r,s}={hist}_s/\mathcal{I}_p$, where $\mathcal{I}_p$ is the number of processed time windows.
The range of $P_{r,s}$ varies from 0 to 1, with values closer to 1 indicating higher reliability of the estimated source.
Additionally, a timer $T_s$ measures the time interval between successive observations of source $s$. 
If $T_s$ surpasses the threshold $T_{\text{death}}$, all information pertaining to source $s$ is removed. This suggests that source $s$ is either idle or its emitted signal has experienced significant attenuation, making it ineffective at the receiver. Every $T_{\text{death}}$ seconds, the counters ${hist}_s$ and $\mathcal{I}_p$ are reset for all detected sources, enabling newly identified sources to enhance their reliability.

\section{Complexity Analysis}
\label{sec:complexity-analysis}
This section is dedicated to evaluating the complexity of the algorithms presented in earlier sections.

The dominant computational complexity of the energy detector in Algorithm~\ref{alg:energy-detector} is related to the calculation of $\hat{\sigma}_{v,i}^2$ and the search for all 1-valued sequences of minimum length $L_\text{adj}-1$ in $\bm{q}_\text{adj}^{(i)}$, which are performed with complexities $O(MG')$ and $O(L_\text{adj}G_\text{MRS}^{(i)})$, respectively. 
Here, $G'=G$ for initialization and
$G' = G_\text{MRS}^{(i)}$\footnote{Because $\scriptstyle G_\text{MRS}^{(i)}<G/2$ is common due to signal sparsity, and since $\scriptstyle\norm{\bm{Y}^{(i)}}_F^2$ is available from the initial step, calculating $\scriptstyle\hat{\sigma}^2_{v,i}$ using
$\scriptstyle\norm{\bm{Y}^{(i)}}_F^2-\norm{[\bm{Y}^{(i)}]_{:,\bm{q}_{\text{MRS}}^{(i)}}}_F^2$
is more cost effective than direct computation.}
in other iterations of the algorithm. Let $N_{\text{itt}}^{\text{Eng}}$ denote the number of iterations until convergence. Then the overall complexity is obtained as 
\begin{equation}
\mathcal{O}_1
=
O\left(MG+N_{\text{itt}}^{\text{Eng}}(L_{\text{adj}}+M)G_{\text{MRS}}^{(i)}\right).
\end{equation}
Although the value of $G_\text{MRS}^{(i)}$ may vary in each iteration of this algorithm, its order of magnitude does not change.

Regarding Rough AOA estimation method presented in Section~\ref{sec:AOA-Estimator}, the SCM calculation in \eqref{eq:scm-computation} requires a complexity of $O(M^2G_\text{MRS}^{(i)})$. 
Moreover, the EVD is performed with complexity $O(M^3)$ and
the 2D-MUSIC spectra in \eqref{eq:2d-music-map} is obtained with complexity $O(K_\theta K_\phi M^2)$. 
Hence, the overall complexity of the Rough AOA estimation stage is
\begin{equation}
\mathcal{O}_2
=
O(M^2(G_\text{MRS}^{(i)}+K_\theta K_\phi+M)).
\end{equation}

Given that subspace based methods such as MUSIC can estimate at most $M-1$ sources \cite{FADAKAR2024106743}, the overall complexity of initializing the array manifold in \eqref{eq:A-init} is
\begin{equation}
\mathcal{O}_3 = O(M(M+N_{U_2}^{(i-1)}+N_{ESL}^{(i-1)})).
\end{equation}

Regarding AOA Refiner algorithm explained in Section~\ref{sec:Alg-AOA-ref}, the K-SVD algorithm described in lines 7-12 of Algorithm~\ref{alg:AOA-Refiner} has a computational complexity of $O(MN^{(i)}(M+N^{(i)})G_{\text{MRS}}^{(i)})$. 
Given the search for the LS estimation error across all members of the set $\mathcal{\bm{B}}$ and the repetition of this operation for every column of $\bm{S}$, the sparse recovery process detailed in Algorithm~\ref{alg:sparse-recovery} exhibits a complexity of 
$
O
(
G_{\text{MRS}}
\sum_{j=1}^{J}
\binom{N^{(i)}}{j}
M^2j
),
$
where $J=\text{min}(L_{\text{max}},N^{(i)})$. It can be simplified as follows:
\begin{equation}
\mathcal{O}_4
=
O\left(G_{\text{MRS}}^{(i)}M^2N^{(i)}
\binom{N^{(i)}-1}{\text{min}\{{\scriptstyle L_\text{max}-1}, \lfloor {\scriptstyle\tfrac{N^{(i)}-1}{2}}\rfloor\}}\right).
\end{equation}
The dominant computational part of the phase smoother block in Algorithm~\ref{alg:phase-smoother} is stemmed from exploiting continuity feature of spikes (line 8 in Algorithm~\ref{alg:phase-smoother}). This block requires a complexity of 
\begin{equation}
\mathcal{O}_5
=
O(
N^{(i)}G_\text{MRS}^{(i)}L_\text{adj}
).
\end{equation}
Finally, the 2D-AOA search in \eqref{eq:refined-AOAs} is performed with a complexity of $O(K_\theta K_\phi M N^{(i)})$.
Hence, the overall complexity of the proposed AOA Refiner block, which includes Algorithm~\ref{alg:AOA-Refiner} and the 2D-AOA search, can be obtained as
\begin{equation}
\begin{aligned}
\mathcal{O}_6
=
&O(
MN^{(i)}
(N_{\text{itt}}^{\text{AOA}}
(M+N^{(i)})
G_\text{MRS}^{(i)}
+
K_\theta K_\phi)
)
\\
&+N_{\text{itt}}^{\text{AOA}}
(\mathcal{O}_4+\mathcal{O}_5),
\end{aligned}
\end{equation}
where $N_{\text{itt}}^{\text{AOA}}$ represents the number of iterations needed for convergence. It is noteworthy that although the value of $N^{(i)}$ may be different in each iteration of this algorithm, its orders of magnitude remains relatively stable.

In the end, considering the proposed localization algorithm in Section~\ref{sec:loc-estimator}, each step in Algorithm~\ref{alg:opt-GP} has a complexity of $O(1)$, leading to an overall complexity of
$
\mathcal{O}_7=O(N_\text{itt}^{\text{Loc}}),
$
where $N_\text{itt}^{\text{Loc}}$ is the number of iterations until convergence.

The aforementioned complexities are summarized in Table~\ref{tab:blocks-complexity}. 
The average number of iterations for the algorithms is reported in the third column.

\begin{table}
\caption{Complexities of the Proposed Blocks.}
\centering
\fontsize{14}{14}\selectfont 
\resizebox{1\columnwidth}{!}{
\begin{tabular}{|l|l|l|}
\hline
\textbf{Block} & \textbf{Order of Complexity} & \textbf{Description}\\
\hline
Energy Detector & 
$\mathcal{O}_1
=
O\left(MG + N_{\text{itt}}^{\text{Eng}}(M+L_{\text{adj}})G_{\text{MRS}}^{(i)}\right)$ & 
$N_{\text{itt}}^{\text{Eng}}\approx 3$
\\
\hline
Rough AOA Estimator & 
$\mathcal{O}_2
=
O(M^2(G_\text{MRS}^{(i)}+K_\theta K_\phi+M))
$ & 
-
\\
\hline
Array Manifold Initializer& 
$
\mathcal{O}_3 = O(M(M+N_{U_2}^{(i-1)}+N_{ESL}^{(i-1)}))
$ & 
-
\\
\hline
Sparse Recovery & 
$
\mathcal{O}_4
=
O\left(G_{\text{MRS}}^{(i)}M^2N^{(i)}
\binom{N^{(i)}-1}{\text{min}\{{\scriptstyle L_\text{max}-1}, \lfloor {\scriptstyle\tfrac{N^{(i)}-1}{2}}\rfloor\}}\right)$ & 
-
\\
\hline
Phase Smoother & 
$\mathcal{O}_5
=
O(
N^{(i)}G_\text{MRS}^{(i)}L_\text{adj}
)$
& 
-
\\
\hline
AOA Refiner & 
$\begin{aligned}[t]
\mathcal{O}_6
=&
O(
MN^{(i)}
(N_{\text{itt}}^{\text{AOA}}
(M+N^{(i)})
G_\text{MRS}^{(i)}
\\
&+
K_\theta K_\phi)
)
+N_{\text{itt}}^{\text{AOA}}
(\mathcal{O}_4+\mathcal{O}_5)
\end{aligned}$ & 
$N_{\text{itt}}^{\text{AOA}}\approx 4$
\\
\hline
Location Estimator & 
$\mathcal{O}_7=O(N_\text{itt}^{\text{Loc}})$ & 
$N_\text{itt}^{\text{Loc}}\approx 3.5$
\\
\hline
\end{tabular}
}
\label{tab:blocks-complexity}
\end{table} 
\section{Simulation and Discussion}\label{sec:simul}
\subsection{Simulation Setup}\label{sec:setup}
The environment is a city district with a size of $2\,\mathrm{km}\times 2\,\mathrm{km}$. Buildings have a fixed area of $10\,\mathrm{m}\times 20\,\mathrm{m}$ and varying heights between $3.5\,\mathrm{m}$ and $20\,\mathrm{m}$, with small random offsets. Streets, alleys, and other urban features have randomly chosen dimensions, adding a small random number for realism. 
The carrier and sampling frequencies are $f_c=0.5\,\mathrm{GHz}$ and $f_s=10\,\mathrm{MHz}$, respectively. 
The moving receiver uses a uniform circular array (UCA) with $M=6$ isotropic antennas ($\mathcal{G}(\theta,\phi)=1$) and a radius of $0.2\,\mathrm{m}$.
Thus, the position of the $m$-th antenna relative to the array center is expressed as $[\bm{D}]_{:,m}=[0.2\cos(2\frac{m-1}{M}\pi), 0.2\sin(2\frac{m-1}{M}\pi),0]^T$.
Simulations involve $N=11$ sources with positions shown in Table.~\ref{tab:source-pos}. All sources emit the same pulse shape 
$p_n(t)=\sqrt{6}\sin(2\pi t/T_{p_n})$ with pulse duration $T_{p_n}=3\,\mu \text{sec}$ and power $3\,\text{Watt}$. 
The average period of transmitted signals for each source is $T_{\text{avg}}=3\,\text{msec}$, which creates sparsity in the signals. The duration of each time window is $T=0.03\,\text{sec}$. 
\begin{table}[ht] 
\captionsetup{font=scriptsize}
\caption{\label{tab:source-pos}
Positions of Sources (in meter)}
\centering
\resizebox{\columnwidth}{!}{
\Huge
\begin{tabular}{|c|c|c|c|c|c|c|c|c|c|c|c|}
\hline
Position & $\bm{r}_1$ & $\bm{r}_2$ & $\bm{r}_3$ &
$\bm{r}_4$ & $\bm{r}_5$ & $\bm{r}_6$ &
$\bm{r}_7$ & $\bm{r}_8$ & $\bm{r}_9$ &
$\bm{r}_{10}$ & $\bm{r}_{11}$\\
\hline
$r_{n_x}$ & $0$ & $-13$ & $66$ &
$53.33$ & $-240.22$ & $600$ & $-300$ & $520$ &
$-250$ & $200$ & $406$\\
\hline
$r_{n_y}$ & $50$ & $-233$ & $-85$ &
$-611.87$ & $357.43$ & $-300$ & $-100$ & $159.17$ &
$-550$ & $-300$ & $-36$ \\
\hline
$r_{n_z}$ & $4.46$ & $3.47$ & $3.2$ &
$2.85$ & $10.69$ & $4.1$ & $5.98$ & $20.46$ &
$10.27$ & $13.71$ & $4.09$\\
\hline
\end{tabular}
}
\end{table}

In the Energy Detector block, the false detection probability is set to $P_{0}=0.001$, and the maximum number of iterations is limited to $N_{\text{max}}^{\text{Eng}}=10$.
$\text{diff}_{\text{max}}=20$ and 
$L_{\text{adj}}=5$ are common parameters for both Energy Detector and Phase Smoother blocks. 
Moreover, in Phase Smoother we use $\epsilon_\varphi=\frac{\pi}{10}$. The maximum sparsity level for the Sparse Recovery algorithm is set to $ L_{\text{max}} = 3$. 
The AOA Refiner and Location Estimator blocks employ convergence thresholds $\varepsilon_{\text{AOA}}=10^{-4}$ and $\varepsilon_{\text{Loc}}=10^{-2}$, with maximum iterations set to $N_{\text{max}}^{\text{AOA}}=20$ and $N_{\text{max}}^{\text{Loc}}=15$, respectively. Additionally, both blocks utilize a common threshold of $\xi=\cos(10^\circ)=0.9848$ for direction assignment. The death time is set to $T_{\text{death}}=0.3\,\text{sec}$. 
The time origin is $t=0$, with the initial array position at $\bm{r}_0=[27,\,11,\,500]^T\,\mathrm{m}$. Processing algorithms start at $t_0=0.1\,\text{sec}$, and the array velocity is $\bm{v}=[44,\,33,\,0]^T\,\frac{\mathrm{m}}{\mathrm{s}}$. 
The results in the next subsections (excluding Section~\ref{sec:Performance_Loc_HeatMap}) are based on averaging 100 independent Monte Carlo trials.
\subsection{Performance of Energy Detector Block}
\label{sec:performance_En}
\begin{figure}[h] 
\centering
\begin{subfigure}{\columnwidth}
\centering
\includegraphics[width=\columnwidth]{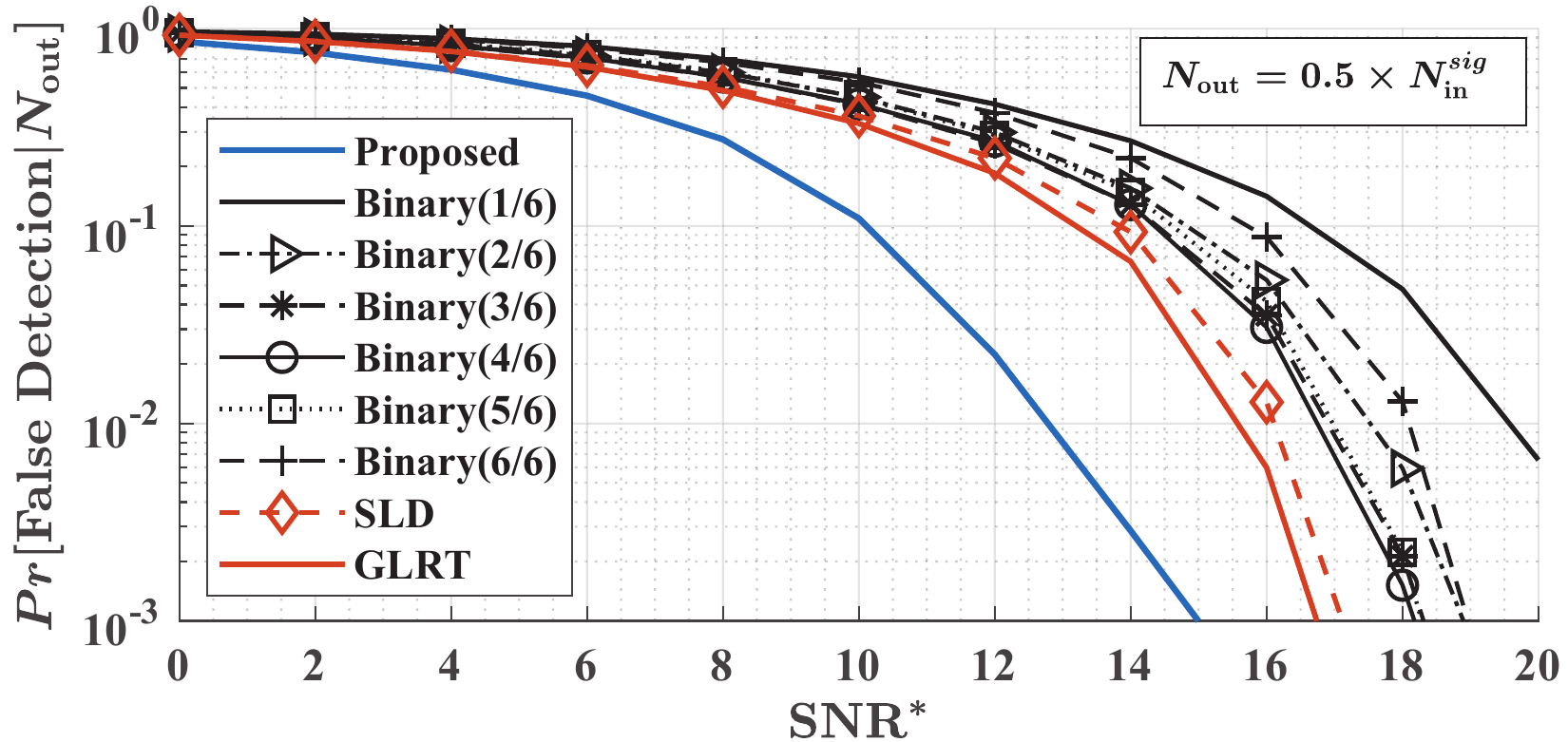}
\caption{}
\end{subfigure}%
\hfill
\begin{subfigure}{\columnwidth}
\centering
\includegraphics[width=\columnwidth]{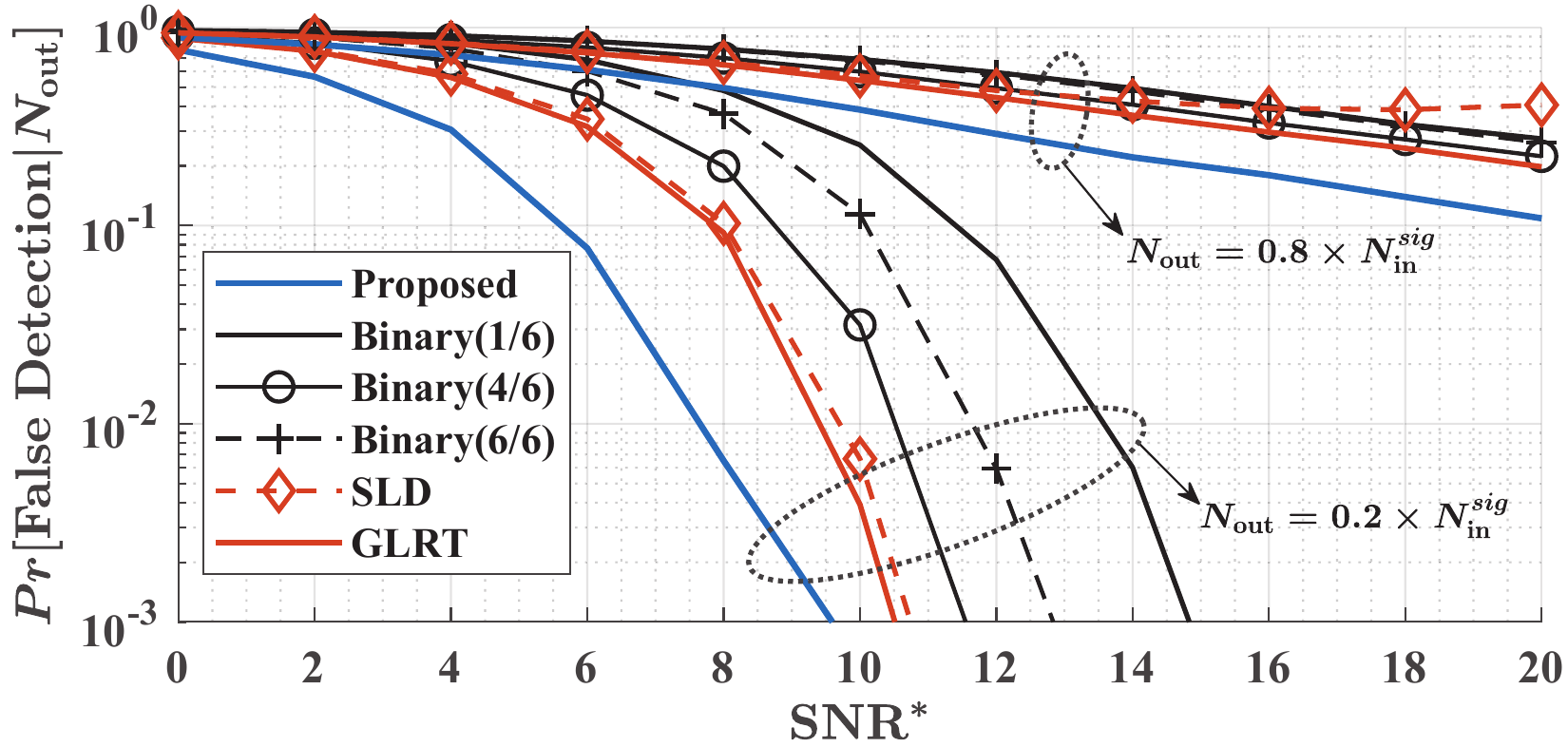}
\caption{}
\end{subfigure}%
\caption{\label{fig:Ent_Comp}False Detection Probability. $N_{\text{in}}^{sig}$ represents the number of signal-containing samples per time window.}
\end{figure}

Fig.~\ref{fig:Ent_Comp} presents a performance comparison between the proposed Energy Detector and baseline methods in \cite{richards2005fundamentals}:
\begin{itemize}
\item \textbf{Binary $\bm{(n/M)}$ detector:} A signal is detected if
\begin{equation}
\sum_{k=1}^{M} u(|y_k| - \mathcal{T}) \geq n,
\end{equation}
where  $y_k = [Y^{(i)}]_{k,g}$ denotes the received sample at the $k$-th antenna at time $t_g^{(i)}$, $u(x)$ represents the Heaviside function and $\mathcal{T}$ is the detection threshold.
\item \textbf{Generalized Likelihood Ratio Test (GLRT):} Detection is determined by
\begin{equation}
\label{eq:GLRT}
\sum_{k=1}^{M} \ln \left[ I_0 \left( 2\sqrt{\frac{\gamma^{(i)}}{\sigma_v^2}} |y_k| \right) \right] \geq \mathcal{T},
\end{equation}
where $I_0(\cdot)$ is the zeroth-order modified Bessel function. Unlike the proposed approach, GLRT assumes exact knowledge of the SNR and noise power.
\item \textbf{Square Law Detector (SLD):} Approximating $I_0(\cdot)$ in \eqref{eq:GLRT} via a second-order Taylor expansion simplifies the detection condition to
$
\sum_{k=1}^{M} |y_k|^2 \geq \mathcal{T}.
$
\end{itemize}
For a fair comparison, the detection threshold $\mathcal{T}$ is calibrated such that all detectors yield the same number of output samples, $N_{\text{out}}$. The false detection probability is computed as the ratio of number of noise-only outputs to $N_{\text{out}}$.  
In Fig.~\ref{fig:Ent_Comp}, the proposed method is evaluated using $L_{\text{adj}} = 10$ and $\text{diff}_{\text{max}} = 5$. Results are averaged over 100 independent configurations with $N = 11$ sources, where $R_n$, $\theta_n$, and $\phi_n$ are uniformly distributed in $[500,2000]$ meters, $[130^\circ,180^\circ]$, and $[0^\circ,360^\circ]$, respectively, for $n=1,\dots,N$. Each configuration is averaged over 10 noise realizations.
As observed, the proposed Energy Detector algorithm outperforms others by leveraging signal sparsity and pulse shape continuity.

\begin{figure}[h] 
\centering
\begin{subfigure}{\columnwidth}
\centering
\includegraphics[width=\columnwidth]{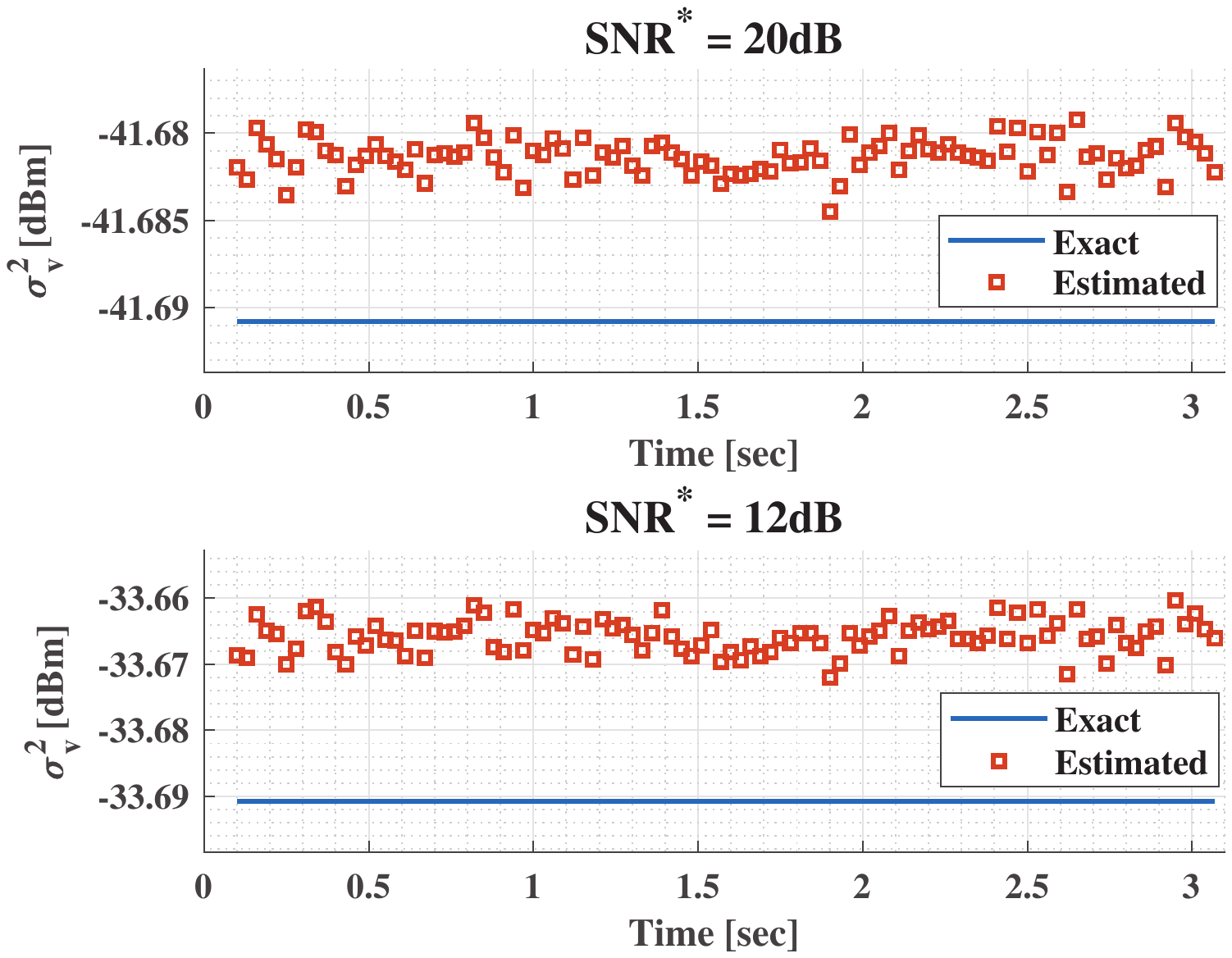}
\caption{Instantaneous noise power ($\sigma^2_v$) and its estimation $\hat{\sigma}^2_{v,i}$}
\label{fig:11src_noisePowExtEst}
\end{subfigure}%
\hfill
\begin{subfigure}{\columnwidth}
\centering
\includegraphics[width=\columnwidth]{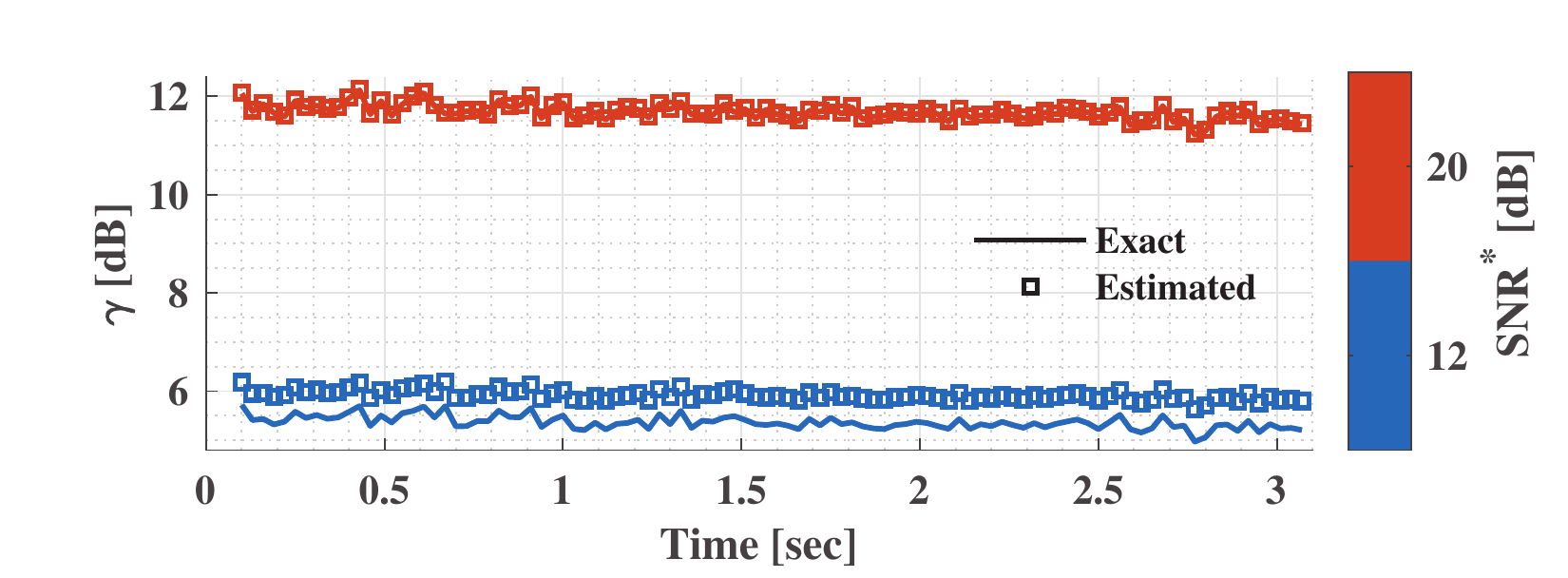}
\caption{Instantaneous SNR ($\gamma^{(i)}$) and its estimation $\hat{\gamma}^{(i)}$
}
\label{fig:11src_snrExtEst}
\end{subfigure}%
\caption{\label{fig:inst-noise-snr}Performance of Energy Detector block.}
\end{figure}

Fig.~\ref{fig:inst-noise-snr} compares the estimated noise energy $\hat{\sigma}^2_{v,i}$ and instantaneous SNR $\hat{\gamma}^{(i)}$ from the Energy Detector block with their true values, $\sigma^2_v$ and $\gamma^{(i)}$, over time.
Throughout this and the following sections, the system configuration adheres to the specifications outlined in Section~\ref{sec:setup}.
The results confirm the accurate estimation of noise power and instantaneous SNR for $\text{SNR}^*$ levels of $12\,\mathrm{dB}$ and $20\,\mathrm{dB}$.
It is noteworthy that $\gamma^{(i)}$ and $\hat{\gamma}^{(i)}$ depicted in Fig.~\ref{fig:11src_snrExtEst} represents the SNR incorporating all $11$ sources. However, if all sources equally contribute to the received power, the SNR for each individual source is $10\log(N)\,\mathrm{dB}$ ($10.41\,\mathrm{dB}$ for $N=11$) lower than this aggregate value.

\subsection{Performance of Detection and AOA Estimation}
\begin{figure}[h] 
\centering
\begin{subfigure}{\columnwidth}
\centering
\includegraphics[width=\columnwidth]{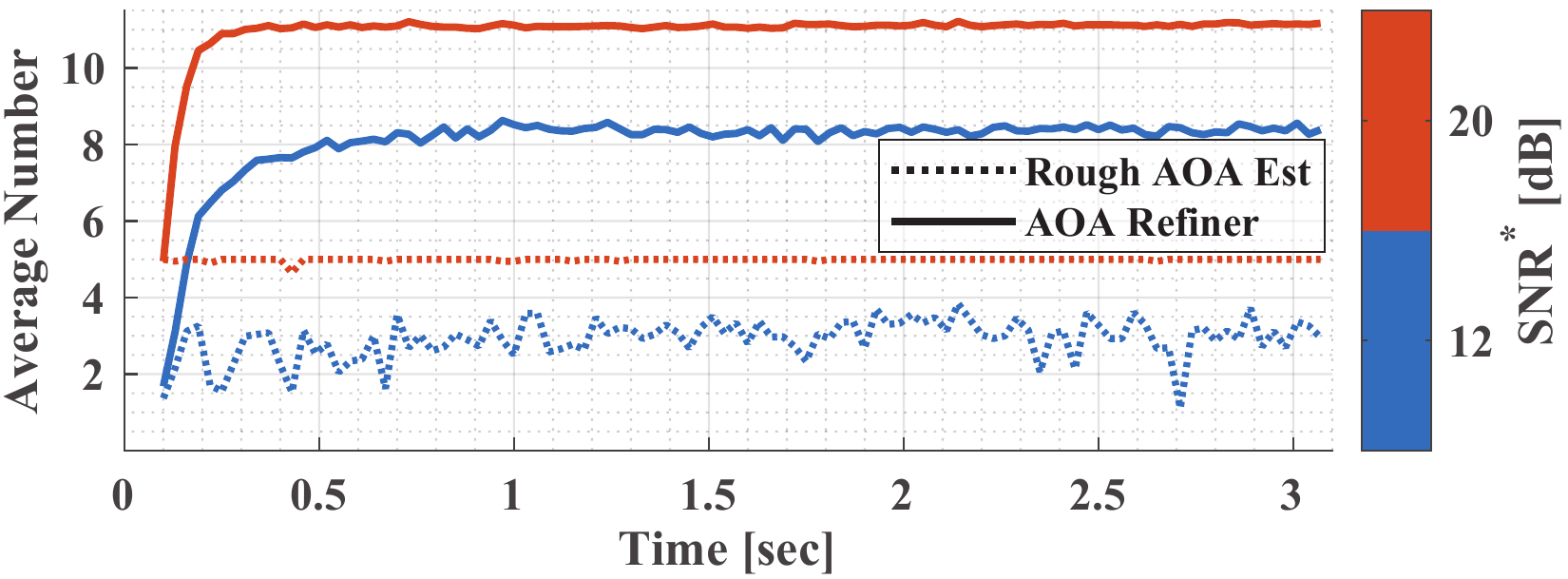}
\caption{Number of Detected AOAs}
\label{fig:11src_AvgNumAOAinaWin}
\end{subfigure}%
\hfill
\begin{subfigure}{\columnwidth}
\centering
\includegraphics[width=\columnwidth]{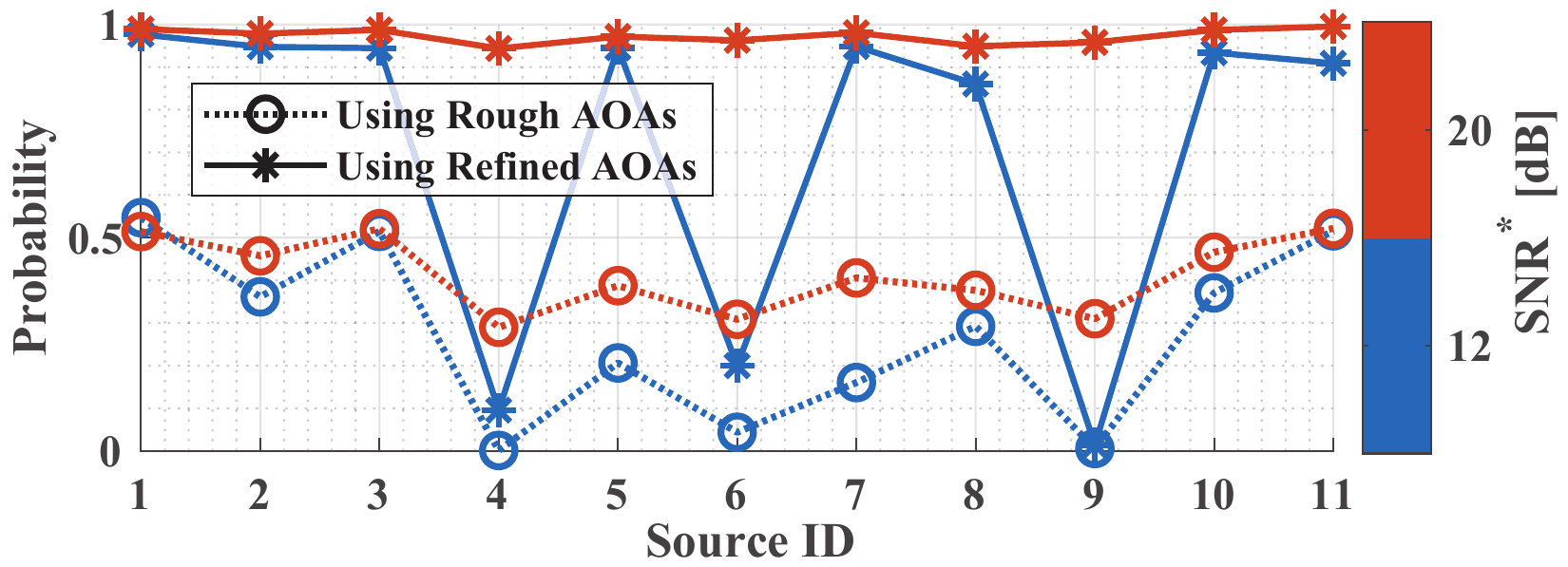}
\caption{The Probability of Reliability
}
\label{fig:11src_ReliabilityProb}
\end{subfigure}%
\hfill
\begin{subfigure}{\columnwidth}
\centering
\includegraphics[width=\columnwidth]{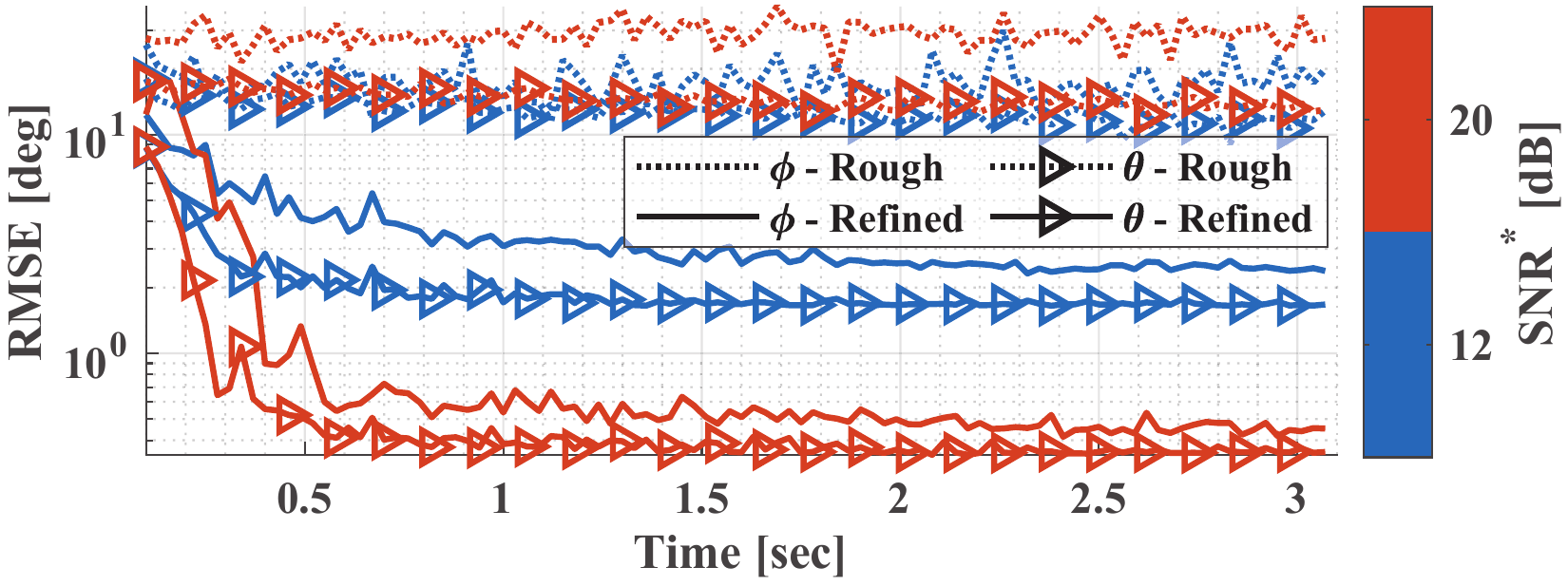}
\caption{Elevation and azimuth estimation RMSE}
\label{fig:11src_AvgNumSrcLoc}
\end{subfigure}%
\caption{Performance of source detection and AOA estimation.\label{fig:step2_Prob}}
\end{figure}

Fig.~\ref{fig:11src_AvgNumAOAinaWin} displays the number of detected AOAs by Rough AOA Estimator and AOA Refiner blocks.
Since 2D-MUSIC is a subspace-based method, the Rough AOA Estimator module can estimate the AOAs of at most $M-1=5$ sources in each time window, even at infinite SNR.
However, with the AOA Refiner module, it is possible to estimate more sources than the number of the array elements.
Additionally, in the initial time window, both blocks estimate a similar number of AOAs due to the initial iteration of the AOA Refiner algorithm. However, as time progresses, the performance of the AOA Refiner improves until it achieves convergence.

Fig.~\ref{fig:11src_ReliabilityProb} depicts the localization reliability probability ($P_{r,s}$) for each of the 11 sources. At an $\text{SNR}^*$ of $12\,\mathrm{dB}$, the proposed system reliably estimates $8$ sources with high probability by disregarding the 3 weakest sources. However, when relying solely on the roughly estimated AOAs, only $3$ sources achieve a reliability probability above $0.5$.
At an $\text{SNR}^*$ of $20\,\mathrm{dB}$, the proposed system reliably estimates all $11$ sources. However, when localization is based solely on the roughly estimated AOAs, there is only a slight improvement compared to the previous lower $\text{SNR}^*$ condition of $12\,\mathrm{dB}$.
It should be noted, as detailed at the end of Section~\ref{sec:proposed-loc-method}, that the reliability probability for each source closely approximates the probability of estimating its AOAs in each time window. Consequently, the sum of probabilities for each curve in Fig.~\ref{fig:11src_ReliabilityProb} aligns approximately with the average number of detected AOAs across all time windows, as shown in Fig.~\ref{fig:11src_AvgNumAOAinaWin}.

Fig.~\ref{fig:11src_AvgNumSrcLoc} displays the azimuth and elevation RMSE for both rough and refined estimated AOAs. RMSE is calculated only for sources with a sufficient number of detections relative to the most commonly detected source among them. Thus, at $\text{SNR}^*=12\,\mathrm{dB}$, only the 8 sources with the highest reliability probability (as shown in Fig.~\ref{fig:11src_ReliabilityProb}) contribute to the RMSE metrics,
while at $\text{SNR}^*=20\,\mathrm{dB}$, all 11 sources are included.

\begin{figure*}[h] 
\centering
\begin{subfigure}{\columnwidth}
\centering
\includegraphics[width=\columnwidth]{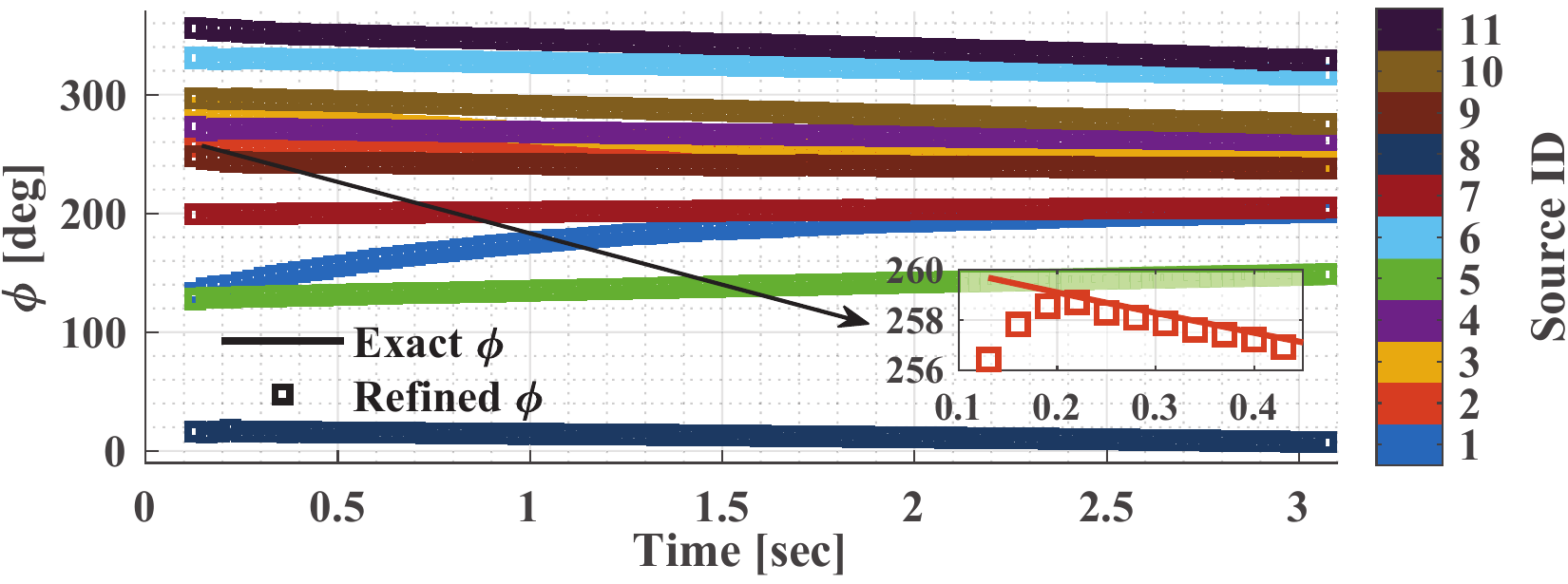}
\caption{}
\label{fig:11src_Azimuth_snr20_CL}
\end{subfigure}%
\hfill
\begin{subfigure}{\columnwidth}
\centering
\includegraphics[width=\columnwidth]{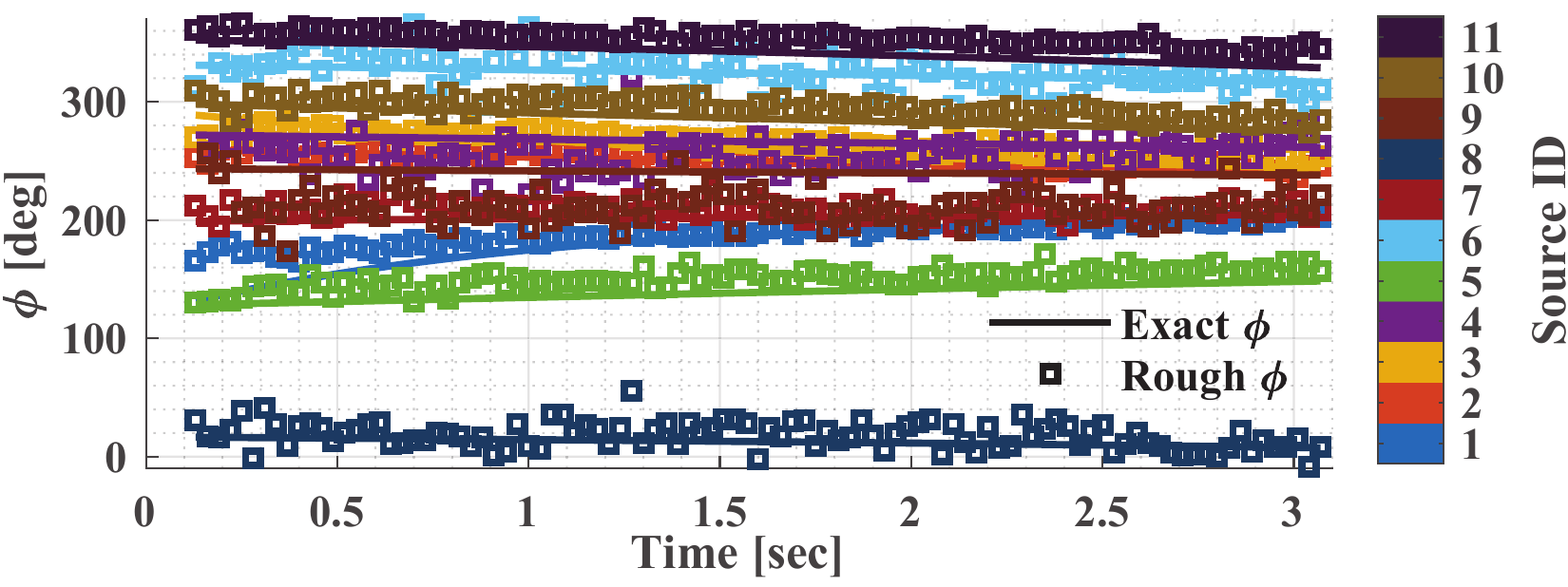}
\caption{}
\label{fig:11src_Azimuth_snr20_OL}
\end{subfigure}%
\hfill
\begin{subfigure}{\columnwidth}
\centering
\includegraphics[width=\columnwidth]{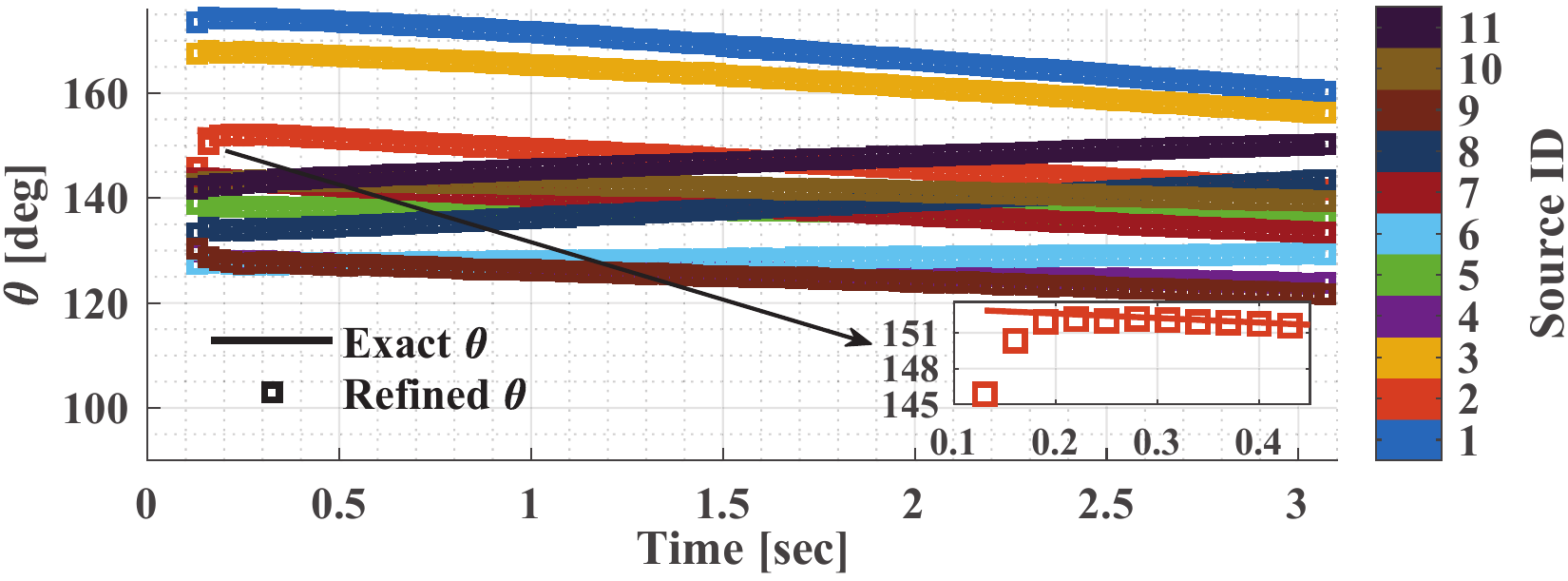}
\caption{}
\label{fig:11src_Elevation_snr20_CL}
\end{subfigure}%
\hfill
\begin{subfigure}{\columnwidth}
\centering
\includegraphics[width=\columnwidth]{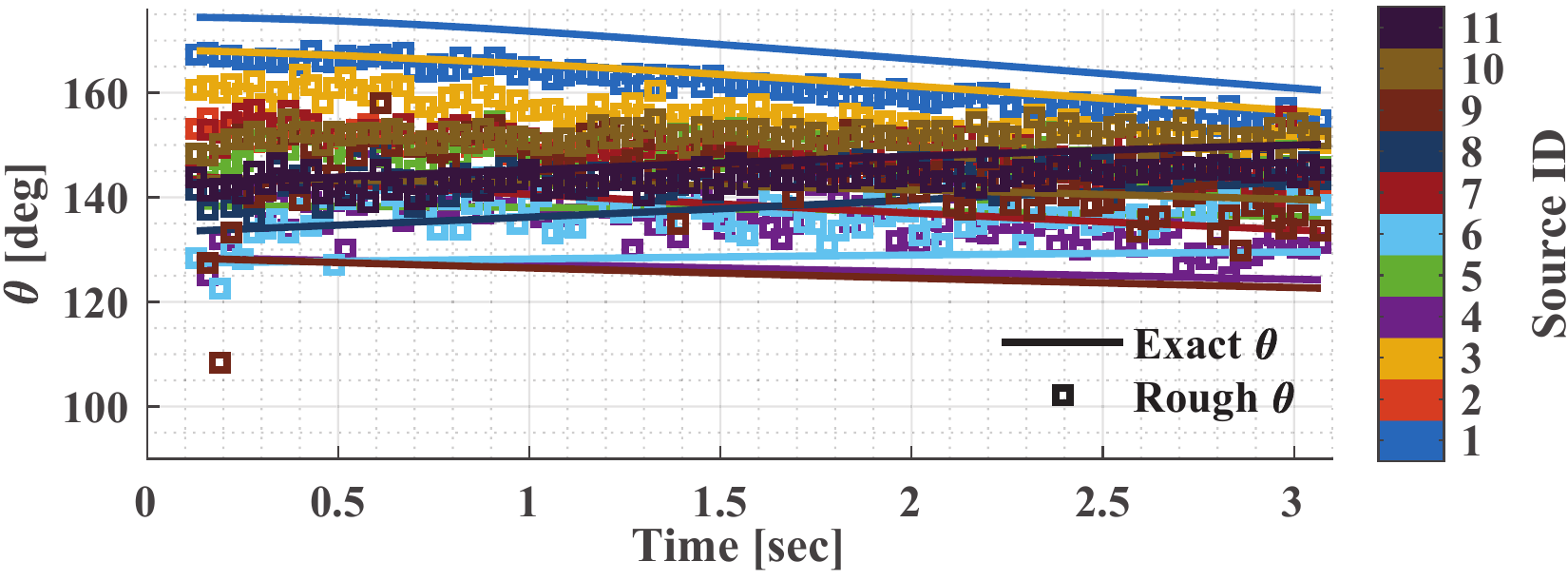}
\caption{}
\label{fig:11src_Elevation_snr20_OL}
\end{subfigure}%
\caption{Comparison of estimated azimuth and elevation angles with ground truth at $\text{SNR}^*=20\,\text{dB}$. Figures (a) and (b) illustrate refined and rough azimuth angle estimates over time, respectively, while figures (c) and (d) depict refined and rough elevation angle estimates over time.
}
\label{fig:step5_AOA}
\end{figure*}

Fig.~\ref{fig:step5_AOA} displays the estimated rough and refined azimuth and elevation AOAs for all $11$ sources, along with the exact AOAs (solid curves) at $\text{SNR}^*=20\,\mathrm{dB}$.
The estimated AOAs by the AOA Refiner module converge and closely follow their exact values throughout the simulation, as evident in the zoomed-in regions of Figs.~\ref{fig:11src_Azimuth_snr20_CL} and \ref{fig:11src_Elevation_snr20_CL} for a selected source. 

\subsection{Performance of Localization}
\label{sec:Performance_Loc_HeatMap}

\begin{figure*}[h]
\centering
\begin{subfigure}{0.67\columnwidth}
\centering
\includegraphics[width=\columnwidth]{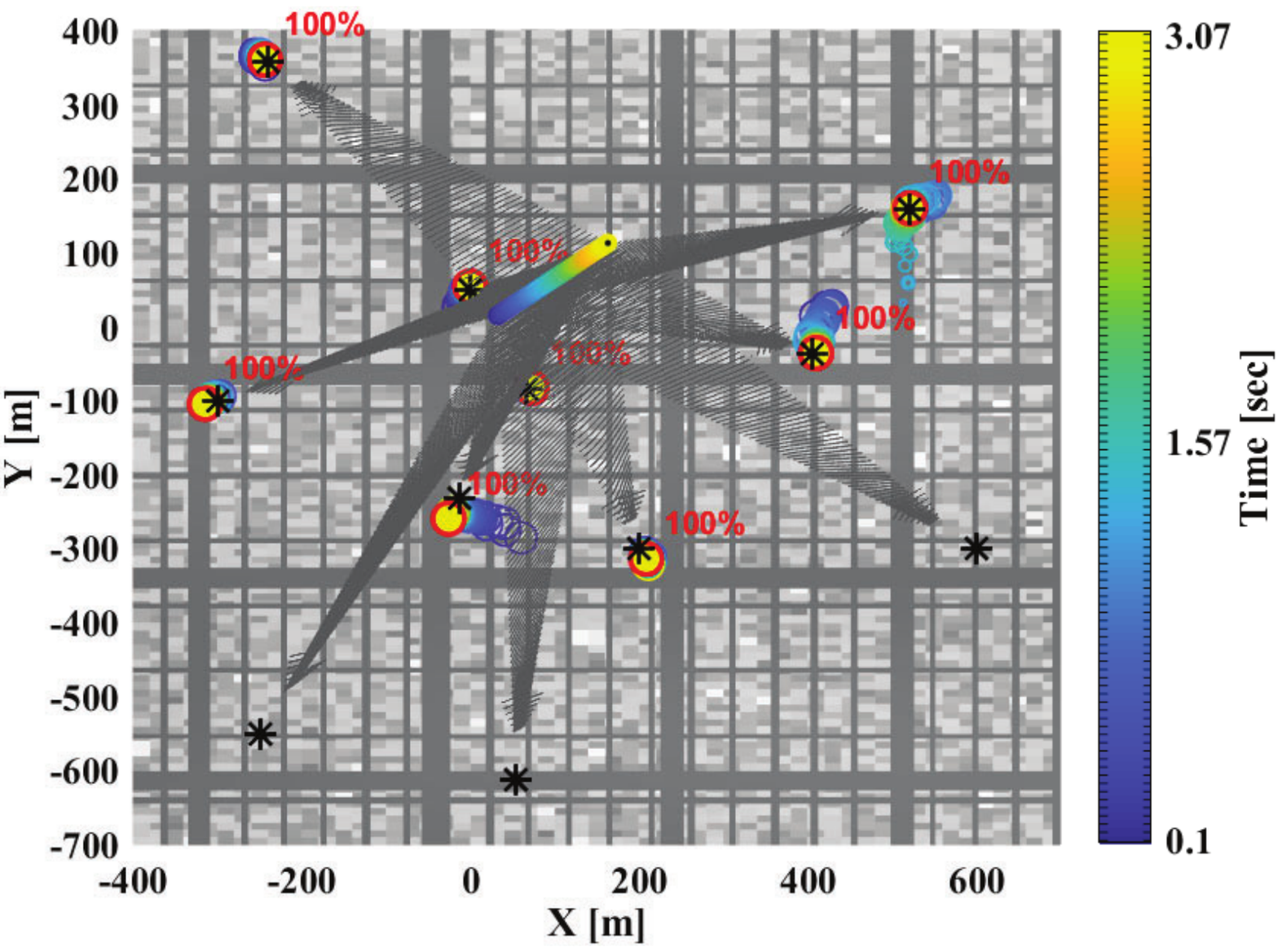}
\caption{Top View $\text{SNR}^*=12\,\mathrm{dB}$
}
\label{fig:11src_3D_CL_snr12_topView}
\end{subfigure}%
\hfill
\begin{subfigure}{0.67\columnwidth}
\centering
\includegraphics[width=\columnwidth]{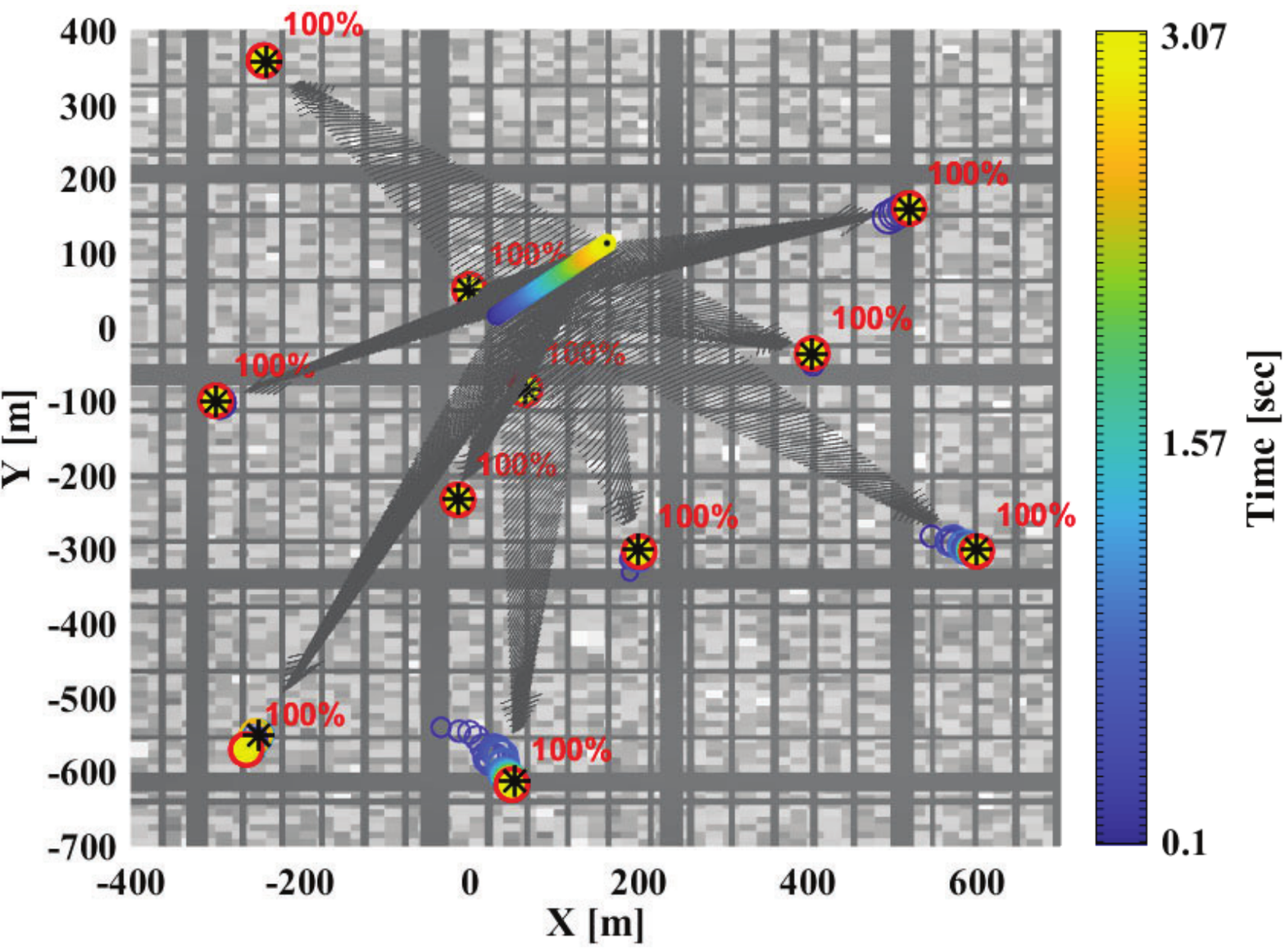}
\caption{Top View $\text{SNR}^*=20\,\mathrm{dB}$
}
\label{fig:11src_3D_CL_snr20_topView}
\end{subfigure}%
\hfill
\begin{subfigure}{0.67\columnwidth}
\centering
\includegraphics[width=\columnwidth]{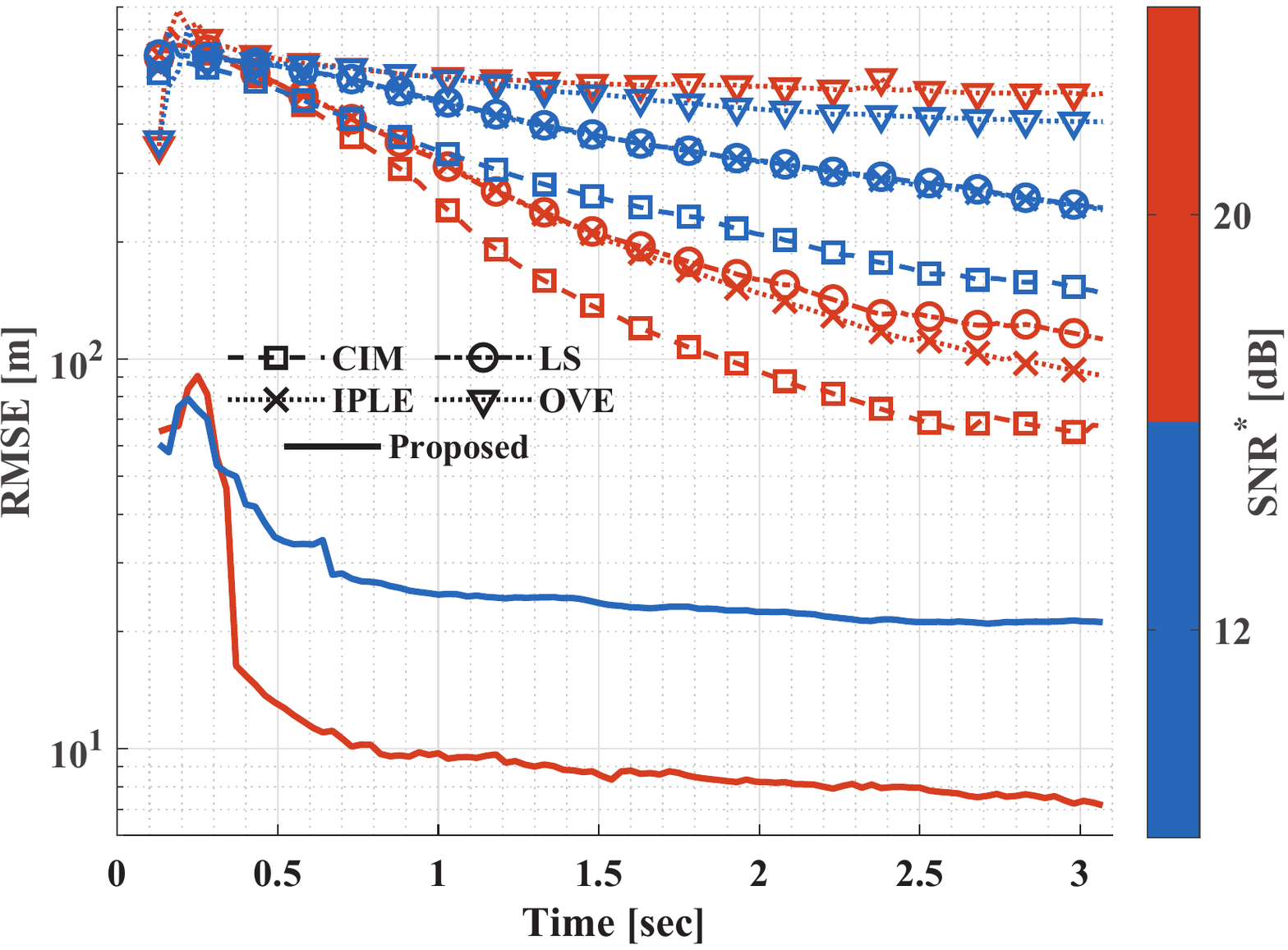}
\caption{
Localization RMSE
}
\label{fig:11src_TotAbsError}
\end{subfigure}%
\caption{Exact and estimated locations of sources, the array movement over time, and localization RMSE of detected sources.}
\label{fig:11src_loc}
\end{figure*}

Figs.~\ref{fig:11src_3D_CL_snr12_topView} and \ref{fig:11src_3D_CL_snr20_topView} illustrate graphical maps of the proposed system during a single performance evaluation. These two plots show the position of the array at each time instant, the direction of sources from the array, and the estimated source locations. The top view is shown for both $\text{SNR}^*=12\,\mathrm{dB}$ and $\text{SNR}^*=20\,\mathrm{dB}$ cases. 
Empty circles on the map indicate the estimated locations of the sources at each time window, with larger circles showing greater reliability.
The symbol \ding{107} represents the true sources' locations, while the yellow circles with red borders (\tikzcircle[red, fill=yellow]{3.5pt}) indicate the final estimates, with reliability of probabilities are displayed in red next to them.  
As the simulation progresses, the coloration of the path line and the circles transfer from blue to yellow. At $\text{SNR}^*=12\,\mathrm{dB}$, $8$ out of $11$ sources are accurately estimated, while at $\text{SNR}^*=20\,\mathrm{dB}$, all $11$ sources are precisely estimated. 

Fig.~\ref{fig:11src_TotAbsError} compares the positioning RMSE of the proposed system with leading state-of-the-art methods, averaged over 100 Monte Carlo trials. The evaluated baselines include:
\begin{itemize}  
\item \textbf{Least Squares (LS)} \cite{fadakar2024ris}:  
A direct AOA localization method utilizing the LS criterion to estimate the source position.  
\item \textbf{Convergent Iterative Method (CIM)} \cite{zou2023convergent}:  
An iterative approach for AOA localization that circumvents matrix inversions, improving numerical computations.  
\item \textbf{Orthogonal Vector Estimator (OVE)} \cite{badriasl2014three}:  
A closed-form technique that exploits orthogonal unit vectors to mitigate noise in bearing angle measurements.  
\item \textbf{Improved Pseudolinear Estimator (IPLE)} \cite{badriasl2014three}:  
A closed form localization approach that linearizes the ML cost function to enable efficient closed-form estimation.  
\end{itemize}
To ensure a fair comparison, all methods use refined estimated AOAs for localization.
Similar to Fig.~\ref{fig:11src_AvgNumSrcLoc}, RMSE at $\text{SNR}^*=12\,\mathrm{dB}$ is based on the 8 most reliable sources, while at $\text{SNR}^*=20\,\mathrm{dB}$, it incorporates all sources.

\begin{figure*}[h!]
\centering
\begin{subfigure}{0.67\columnwidth}
\centering
\includegraphics[width=\columnwidth]{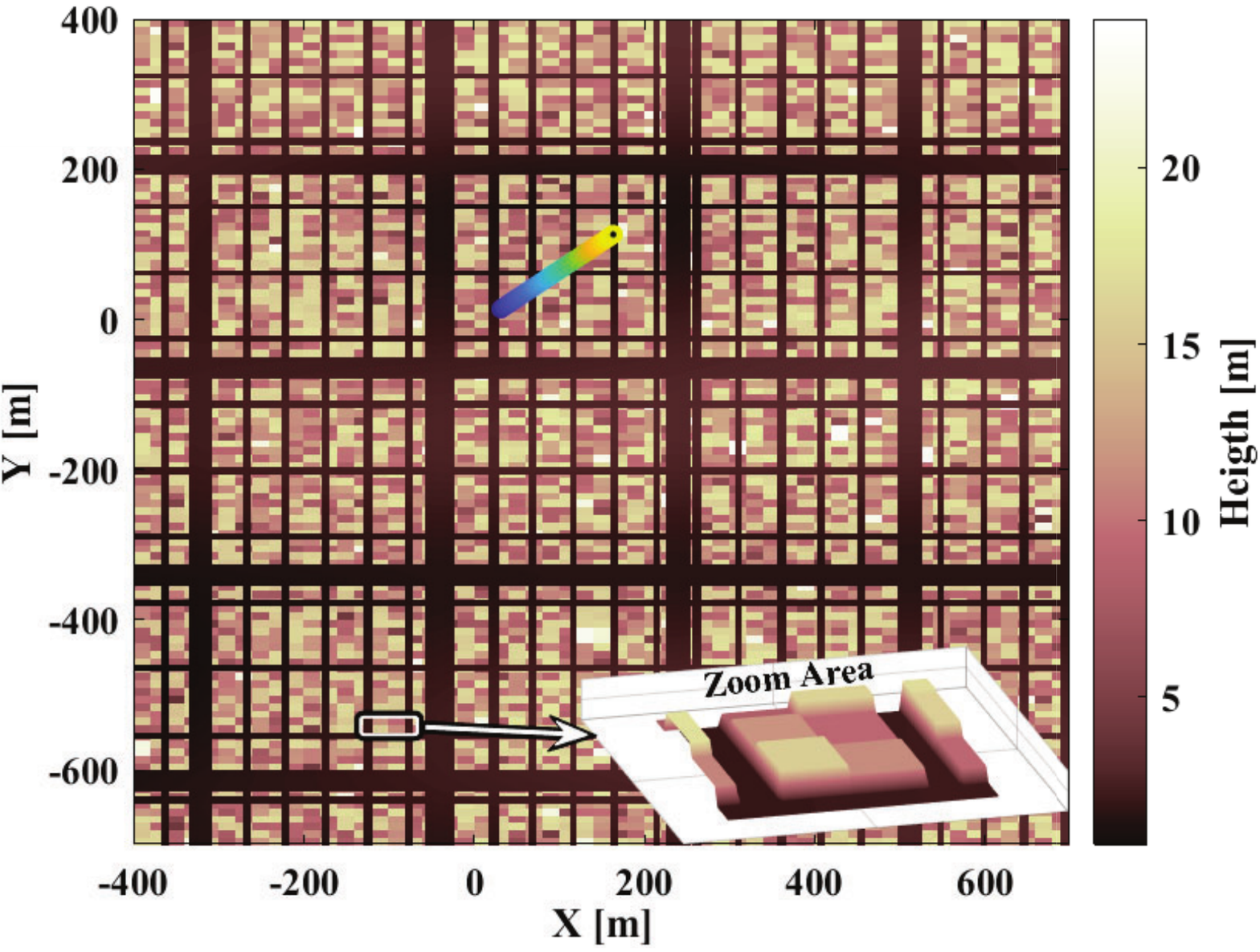}
\caption{City Height Map
}
\label{fig:Map_major}
\end{subfigure}%
\hfill
\begin{subfigure}{0.67\columnwidth}
\centering
\includegraphics[width=\columnwidth]{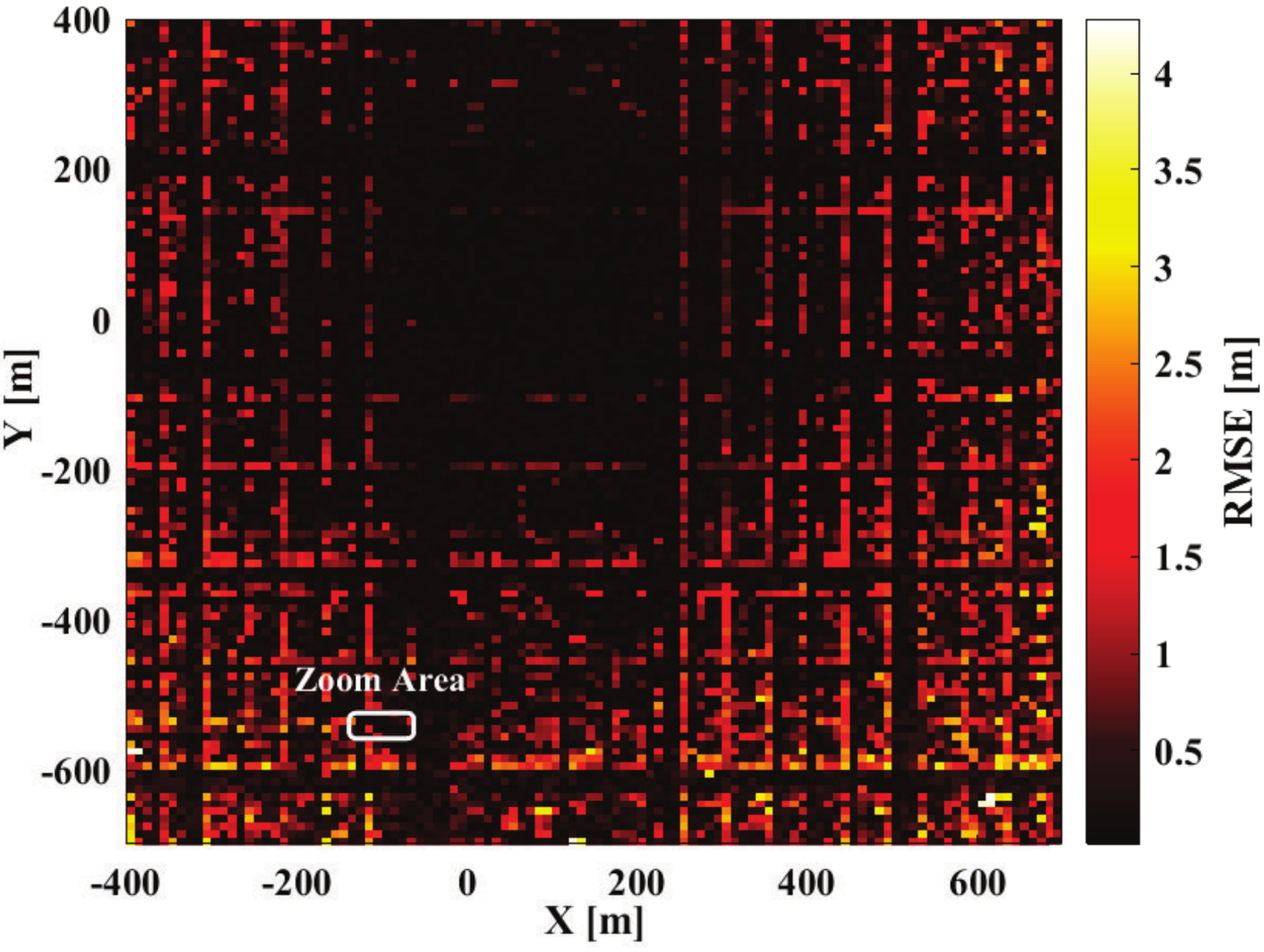}
\caption{Localization RMSE Heat-Map
}
\label{fig:HitMap_major}
\end{subfigure}%
\hfill
\begin{subfigure}{0.67\columnwidth}
\centering
\includegraphics[width=\columnwidth]{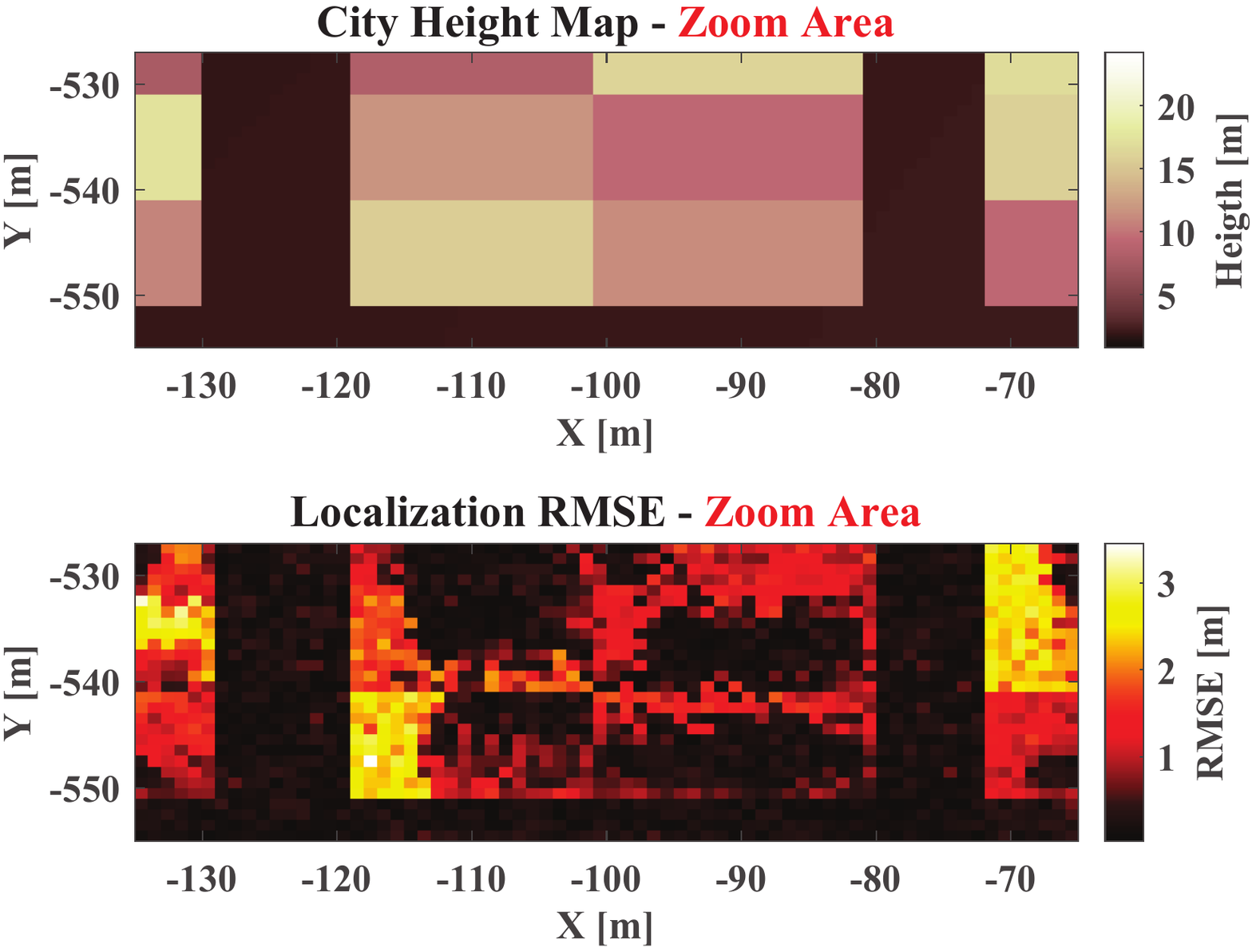}
\caption{
Zoom Area Map and Localization RMSE
}
\label{fig:HitMap_minor}
\end{subfigure}%
\caption{City height and localization RMSE heat-map for single source at $\text{SNR}^*=5\,\mathrm{dB}$.}
\end{figure*}

Fig.~\ref{fig:Map_major} displays the city height map from a top view and includes a side view of a zoomed-in region. It also depicts the receiver array's movement trajectory, similar to Fig.~\ref{fig:11src_loc}.

In addition, Fig.~\ref{fig:HitMap_major} shows the localization RMSE heat-map for a single source at the last time window. The area is divided into uniform grids with a $10\,\mathrm{m}$ step size in both $x$ and $y$ directions. The RMSE metrics in Figs.~\ref{fig:HitMap_major} and \ref{fig:HitMap_minor} are averaged over 10 independent Monte Carlo experiments at $\text{SNR}^*=5\,\mathrm{dB}$ for each grid.
As shown in the figure, despite a consistent $\text{SNR}^*$ across all map points, the RMSE increases as the distance from the array in the $xy$ plane grows. For a detailed mathematical explanation, refer to Appendix~\ref{app:Max_RMSE}.
It should be noted that in the single-source scenario, $\text{SNR}^*$ and instantaneous $\text{SNR}$ are nearly identical at the start, but the latter varies as the array moves.

Fig.~\ref{fig:HitMap_minor} shows the city height map and RMSE heat-map for the zoomed-in area with $1\,\mathrm{m}$ grid steps.
We can observe that RMSE increases at building edges, where greater height differences leading to larger RMSEs, as small $xy$ inaccuracies can lead to significant $z$-coordinate estimation errors.

\subsection{System Performance in Presence of Various Imperfections}
In previous subsections, we assumed perfect knowledge of the receiver's location, movement direction, and the environment map. However, these assumptions may not hold in practical scenarios. In this subsection, we evaluate the proposed system's performance under these imperfections using binary parameters $\delta_\text{Loc}$, $\delta_\text{Dir}$, and $\delta_\text{Map}$ to indicate the presence of errors in each aspect. 
If $\delta_\text{Loc}=1$, Gaussian random position errors (mean: $0$, SD\footnote{Standard Deviation}: $5 \, \text{m}$) are added to $\bm{r}^{(i)}$ in all time windows $\{W_i\}_{i=1}^{I}$. If $\delta_{Dir}=1$, Gaussian random angular noise (mean: $0$, SD: $5^\circ$) is introduced by rotating the array around the $z$-axis in each time window. 
For $\delta_\text{Map}=1$, the receiver assumes an unavailable map and sets $\mathcal{M}(x,y)=0$ (see \eqref{opt:lemma}).
\begin{figure*}[h]
\centering
\begin{subfigure}{\columnwidth}
\centering
\includegraphics[width=0.9\columnwidth]{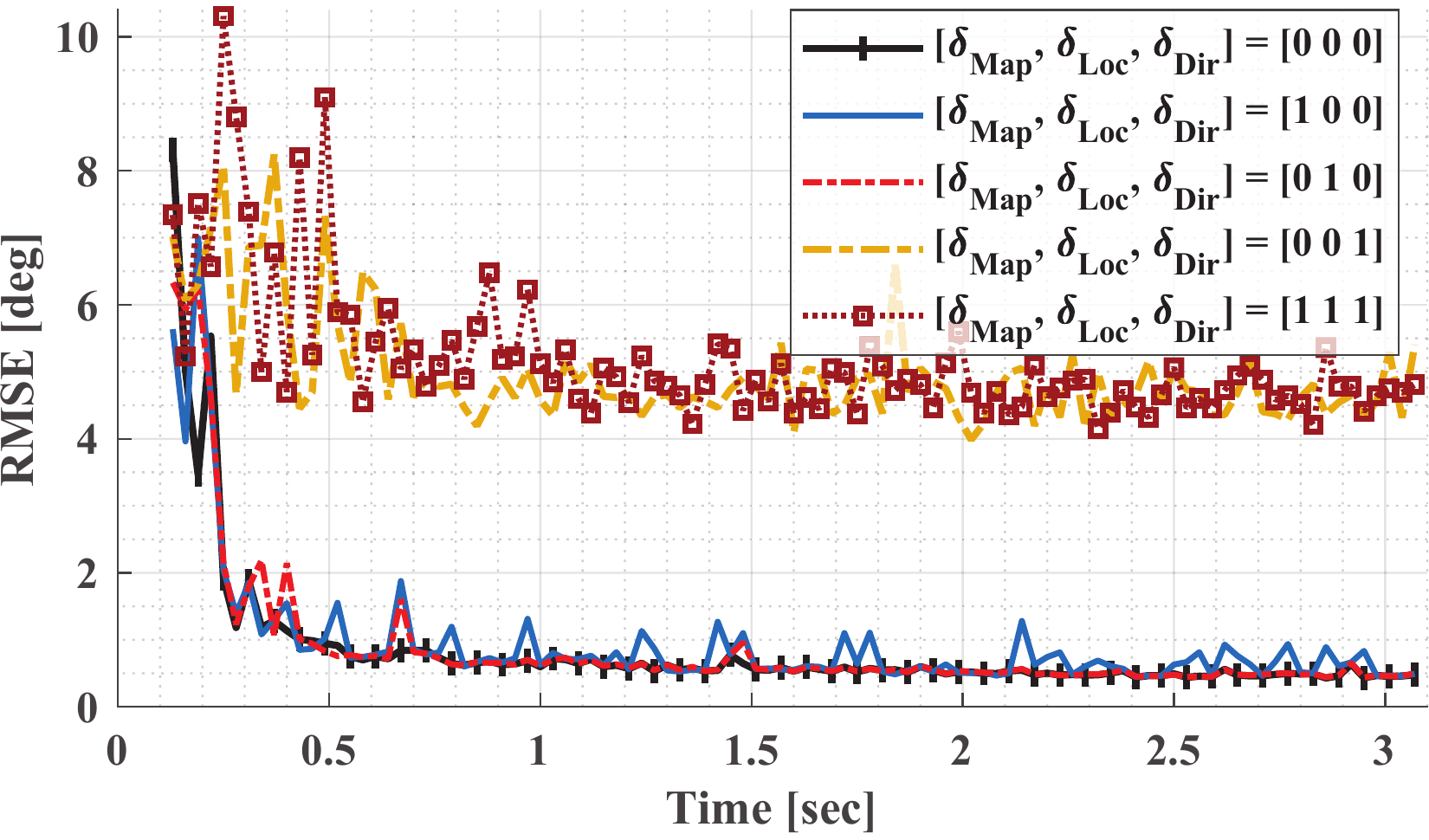}
\caption{Total RMSE of Azimuth AOA Estimation}
\label{fig:2_Error_totAzimuthRMSE}
\end{subfigure}%
\hfill
\begin{subfigure}{\columnwidth}
\centering
\includegraphics[width=0.9\columnwidth]{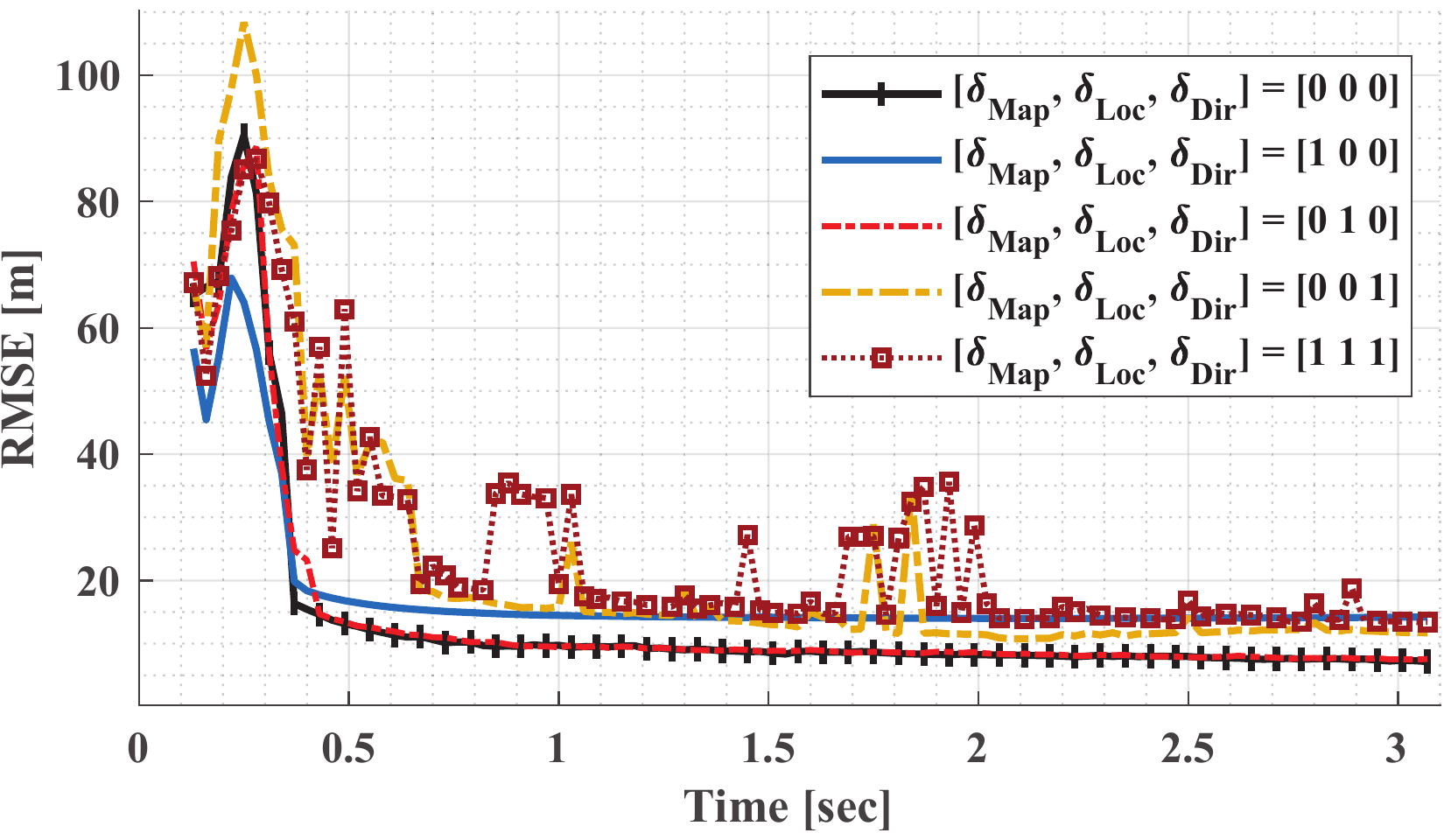}
\caption{Total RMSE of Localization}
\label{fig:2_Error_totAbsRMSE}
\end{subfigure}%
\caption{\label{fig:step6_Errors}Assessment of the proposed method's performance in the presence of diverse imperfections.}
\end{figure*}
In Fig.~\ref{fig:step6_Errors}, we observe that the azimuth AOA estimation remains robust despite errors in array location and map unavailability. However, a directional error ($\delta_\text{Dir}=1$) leads to a worsened RMSE of estimated azimuth AOAs by approximately $3.5^\circ$, which results in higher localization error. The location estimator module is robust when $\delta_\text{Loc}=1$, but its performance slightly deteriorates by around $5$ meters when $\delta_\text{Map}=1$, mainly due to a position error in the $z$ dimension. 
It is important to note that the elevation estimation error almost remains consistent across all four scenarios, converging approximately to $0.6^\circ$.

\section{Conclusion}\label{sec:con}
In this paper, we presented a novel approach for 3D AOA localization that leverages a moving array with a limited number of antennas to accurately locate multiple stationary sources. Our method effectively handles scenarios where the number of sources exceeds the number of sensors in the array. 
The proposed approach comprises several innovative components: an Energy Detector, an AOA Refiner, and a Location Estimator.
The Energy Detector algorithm filters the noisy samples by exploiting the sparsity of the received signals and the continuity of the transmitted spikes. 
Next, the initial array manifold is estimated using the previously estimated source locations and the rough 2D-AOAs obtained from the 2D-MUSIC algorithm.
Subsequently, a novel closed-loop subsystem, which includes novel sparse recovery and phase smoothing algorithms, refines this estimate to derive more accurate 2D-AOA values. Techniques such as K-SVD are integral to this subsystem, ensuring precise AOA refinement.
Finally, the 3D locations of sources are efficiently estimated by the proposed location estimator algorithm.
The simulation findings underscore the efficiency of our innovative approach in multi-source conditions, and its robustness against various signal imperfections.
Potential extensions of this work include incorporating more realistic channel models by accounting for non-line-of-sight paths and analyzing the impact of hardware impairments such as carrier frequency offset, I/Q imbalance, and other nonlinearities.

\appendices
\section{Estimating $f(N,\gamma)$}\label{app:f-est}
For a fixed number of sources $N$, we generate $T$ samples of the transmitted signal vector from a complex standard normal distribution. These samples exhibits sparsity levels of $[0, 1, 2, 3]$ with corresponding probabilities of occurrence of $[0.1, 0.65, 0.2, 0.05]$. In our simulations, we set $T=1000$. These signal vectors are then organized into columns of the matrix $\bm{S}\in\mathbb{C}^{N\times T}$.
Let $\bm{A}$ denote the corresponding array manifold, as defined in \eqref{eq:array-manifold}. By considering $\bm{\Psi}$ as a diagonal matrix, we can determine the random attenuation values due to signal propagation. Consequently, the noise-free received signal $\bm{X}$ is obtained as $\bm{X}=\bm{A}\bm{\Psi}\bm{S}$, where the elements have an average energy of $E_{s,\text{avg}}=\frac{1}{MT}\|\bm{X}\|_F^2$.
It should be noted that in the paper's formulations concerning sparse recovery, we use the term \emph{energy} to highlight the algorithm's dependence on the energy of the desired signal rather than its variance.
Next, for each SNR $\gamma$, we generate $10$ random realizations of the white complex Gaussian noise matrix $\bm{V}\in\mathbb{C}^{M\times T}$ with variance $\sigma^2_v=E_{s,\text{avg}}/\gamma$. These realizations are added to the matrix $\bm{X}$ to obtain the noisy received signal $\bm{Y}$.
For each realization, we sweep $\epsilon$ and apply Sparse Recovery Algorithm~\ref{alg:sparse-recovery} with $\hat{\epsilon}_{\text{opt}} = \epsilon$ to obtain the corresponding matrix $\widehat{\bm{S}}$. Subsequently, we calculate the average estimation error $\|\widehat{\bm{S}}-\bm{S}\|_F$ across all realizations.

\begin{figure}[h]
\centering
\begin{subfigure}{\columnwidth}
\includegraphics[width=\textwidth]{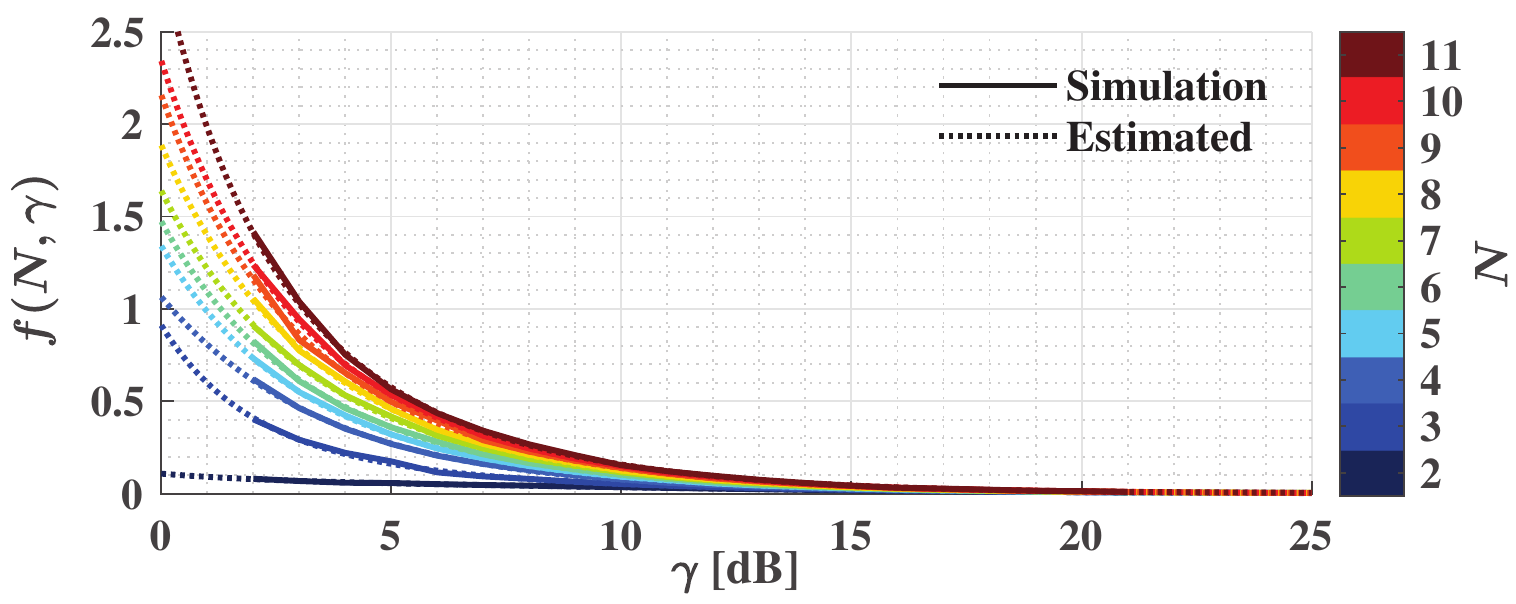}
\caption{$f(N,\gamma)$ versus $\gamma$
}
\label{fig:f_perSNR}
\end{subfigure}%
\hfill
\begin{subfigure}{\columnwidth}
\includegraphics[width=\textwidth]{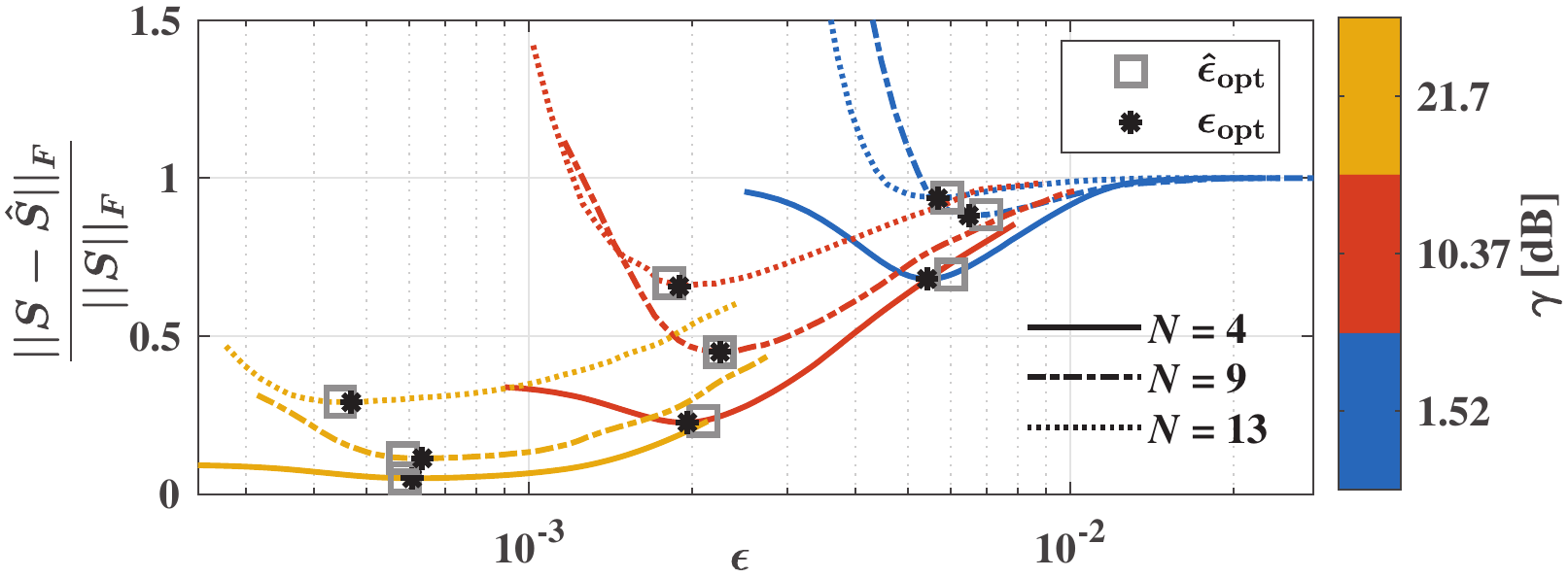}
\caption{Relative Estimation Error versus $\epsilon$}
\label{fig:f_perN}
\end{subfigure}%
\caption{
Average value of $f(N, \gamma)$ across simulation trials and its estimation for calculating $\hat{\epsilon}_{\text{opt}}$. Additionally, a comparison between the optimal value of epsilon and its estimation ($\hat{\epsilon}_\text{opt}$).
}
\label{fig:app_fig}
\end{figure}
We repeat this procedure for $N\in\{2,3,\dots,11\}$ and $\gamma\in\{2\,\mathrm{dB},3\,\mathrm{dB},\dots,21\,\mathrm{dB}\}$ to find $\epsilon_{\text{opt}}$, which attains the minimum average estimation error.
According to \eqref{eq:epsilon-opt} and using the corresponding values $E_{s,\text{avg}}$ and $\sigma^2_v$ for each obtained $\epsilon_{\text{opt}}$, we evaluate the function $f(N,\gamma)$ as follows
\begin{equation}
f(N,\gamma)=\dfrac{\epsilon_{\text{opt}}^2-M\sigma^2_v}{M E_{s,\text{avg}}}.
\end{equation}

The above procedure is repeated for $500$ Monte-Carlo trials. 
After removing outliers, $f(N,\gamma)$ is estimated by taking average over the remaining values. 
For $N\in\{2,3,\dots ,11\}$ and unseen $\gamma$, we estimate $f(N,\gamma)$ with $10^{g(N, \gamma)}$, where for a fixed $N$, $g(N,\gamma)$ indicates a polynomial with degree $4$,
\begin{equation}\label{eq:g}
g(N,\gamma)=P_{4,N}\gamma_{\mathrm{dB}}^4+P_{3,N}\gamma_{\mathrm{dB}}^3+P_{2,N}\gamma_{\mathrm{dB}}^2+P_{1,N}\gamma_{\mathrm{dB}}+P_{0,N},
\end{equation}
where $\gamma_{\mathrm{dB}}=10\log_{10}\gamma$ and $\{P_{j,N}\}_{j=0}^{4}$ denote the polynomial coefficients, which are obtained by using polynomial interpolation over the generated data. The results are shown in Fig.~\ref{fig:f_perSNR}. 
For a fixed $\gamma$ and for $N\ge 12$ or $N=1$, we estimate $f(N,\gamma)$ using linear extrapolation. 
Hence, the final formula for $f(N,\gamma)$ can be represented as
\begin{equation}\label{eq:f}
\begin{cases*}
10^{g(N,\gamma)},&if  $2\le N\le 11$,  \\
(N-10)f(11,\gamma)-(N-11)f(10,\gamma),&$\text{if}\ N>11$,\\
2f(2,\gamma)-f(3,\gamma),&if $N=1$.
 \end{cases*}
\end{equation}

Fig.~\ref{fig:f_perN}, demonstrates the effectiveness of the proposed estimator for $\epsilon_{\text{opt}}$ using the equations $\hat{\epsilon}_{\text{opt}}=\sqrt{M E_{s,\text{avg}} f(N,\gamma)+M\sigma^2_v}$, \eqref{eq:g} and \eqref{eq:f}.

\section{Spatial Conditions for Maximizing Localization RMSE}\label{app:Max_RMSE}
In noise-free conditions, the exact solution is within the search area of Algorithm~\ref{alg:opt-GP}. However, the algorithm may converge to a local optimum.
In this case, according to equations $\bm{w}'=\bm{C}^{\dagger}\left(\bm{h}-\bm{b}z\right)$ and $z=\mathcal{M}(\bm{w}')$ for the search area and also
${\Delta z}_{\text{max}}
\stackrel{\scriptscriptstyle\text{def}}{=}
\max \mathcal{M}(\bm{w}')-\min \mathcal{M}(\bm{w}')$
as the maximum height difference in the city map, the maximum possible localization RMSE is:
\begin{equation}
\label{eq:max_RMSE}
e_\text{max} = {\Delta z}_{\text{max}} \sqrt{\max\norm{\bm{C}^{\dagger}\bm{b}}^2+1}.
\end{equation}

In the following, we investigate the spatial conditions for maximizing $\norm{\bm{C}^{\dagger}\bm{b}}^2$.
Assume 2 anchors, each providing 2D-AOA estimations of a source. Let the elevation angle be constant with the value $\theta$.
Referring to \eqref{opt:lemma-map}, for $j$-th anchor we have
$\bm{C}_j=\bm{I}_2-\sin^2\theta\tilde{\bm{u}}_j\tilde{\bm{u}}_j^T$ and $\bm{b}_j=-\frac{1}{2}\sin(2\theta)\tilde{\bm{u}}_j$, where $\tilde{\bm{u}}_j = [\cos\phi_j , \sin\phi_j]^T$ and $\phi_j$ is the corresponding azimuth AOA.
According to \eqref{eq:defs}, we derive $\bm{C}=\sum_{j=1}^{2}\bm{C}_j$ and $\bm{b}=\sum_{j=1}^{2}\bm{b}_j$. After straightforward mathematical manipulations, we obtain:
\begin{equation}
\label{eq:norm2_Cb}
\norm{\bm{C}^{\dagger}\bm{b}}^2 = \dfrac{\sin^2(2\theta)\left(4-2\alpha-\sin^2\theta\sin^2(\Delta\phi)\right)}{\left(\sin^4\theta\sin^2(\Delta\phi)+4\cos^2\theta\right)^2},
\end{equation}
where $\alpha=(1-\frac{1}{2}\sin^2\theta\sin(\Delta\phi))(1-\cos(\Delta\phi))$ and $\Delta\phi=\phi_2-\phi_1$. Given $\alpha\geq0$ with its minimum value at $\Delta\phi=0$, the maximum value of $\norm{\bm{C}^{\dagger}\bm{b}}^2$ is:
\begin{equation}
\label{eq:max_norm2_Cb}
\max_{\Delta\phi}\norm{\bm{C}^{\dagger}\bm{b}}^2
\mathrel{\overset{\scriptstyle{\Delta\phi=0}}{\resizebox{\widthof{\kern12pt=}}{\heightof{$=$}}{$=$}}}
\tan^2\theta.
\end{equation}

Based on \eqref{eq:max_RMSE} and \eqref{eq:max_norm2_Cb}, as $\theta$ approaches $90^\circ$, the maximum possible RMSE increases. This means that a source positioned farther from the array's $xy$-plane projection is expected to yield a higher localization RMSE.
Additionally, if the source's azimuth AOA changes the least during the array movement ($\Delta\phi$ is small), the RMSE will rise, though the impact is less significant compared to changes in the elevation angle.



\bibliographystyle{IEEEtran}
\bibliography{Bib}

\begin{IEEEbiography}[{\includegraphics[width=1in,height=1.25in,clip,keepaspectratio]{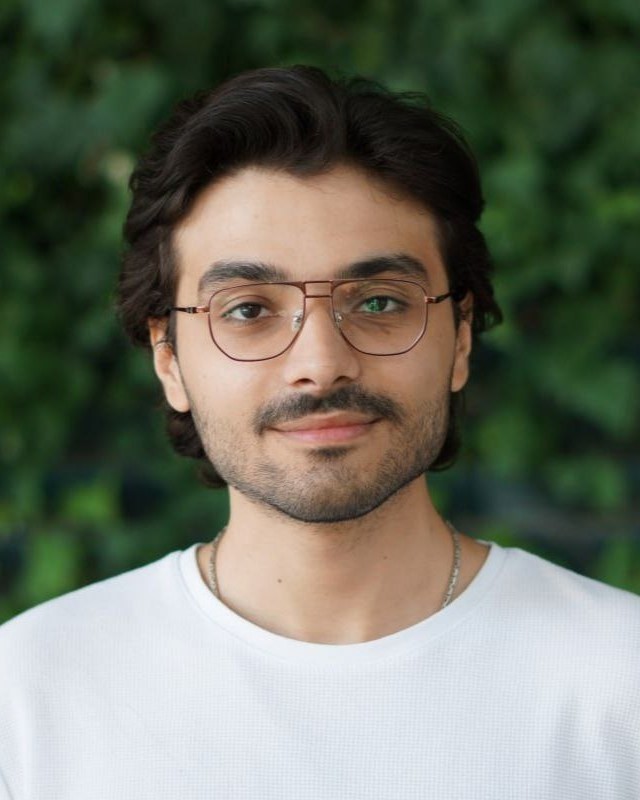}}]{Amir Mansourian }
received the B.Sc. and M.Sc. degrees in electrical engineering from the University of Tehran, Tehran, Iran, in September 2020 and February 2024, respectively. He is currently a research assistant at the University of Tehran. His research interests include wireless communication, wireless positioning, 5G and beyond technologies, physical-layer abstraction techniques, signal processing, distributed optimization, and the application of machine learning in these areas.
\end{IEEEbiography}

\begin{IEEEbiography}[{\includegraphics[width=1in,height=1.25in,clip,keepaspectratio]{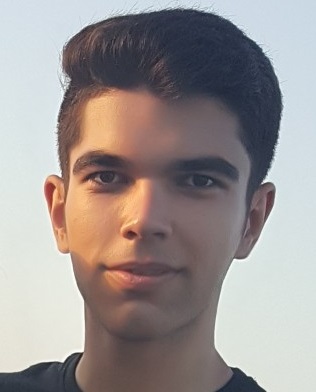}}]{Alireza Fadakar }(S'25) received the B.Sc. and M.Sc. degrees in electrical engineering from the University of Tehran, Tehran, Iran, in 2020 and 2023, respectively. In January 2025, he joined the Ming Hsieh Department of Electrical and Computer Engineering at the University of Southern California as a Ph.D. student. He is a recipient of the USC Annenberg Fellowship. In 2015, he was awarded a silver medal in the Iranian National Mathematics Olympiad. His research interests include wireless communications, signal processing, localization, channel estimation, MIMO networks, RIS-assisted communications, digital twins, and optimization, as well as the application of machine learning in these areas.
\end{IEEEbiography}

\begin{IEEEbiography}[{\includegraphics[width=1in,height=1.25in,clip,keepaspectratio]{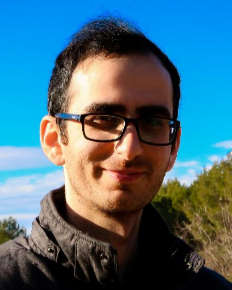}}]{Saeed Akhavan }
He completed his undergraduate, master's, and doctoral studies in electrical engineering at the University of Tehran, graduating in 2010, 2012, and 2018, respectively. From 2016 to 2017, he was affiliated with the Grenoble Polytechnic Institute in France. Presently, he serves as an assistant professor at the University of Tehran, concentrating his research on biomedical signal processing, compressed sensing, probabilistic machine learning, and both local and global optimization methodologies.
\end{IEEEbiography}

\begin{IEEEbiography}[{\includegraphics[width=1in,height=1.25in,clip,keepaspectratio]{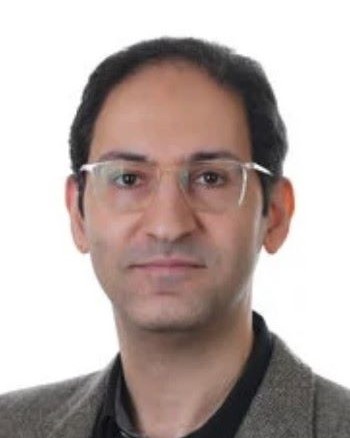}}]{Behrouz Maham }(S'07, M'10. SM'15) received the B.Sc. and M.Sc. degrees in electrical engineering from the University of Tehran, Iran, in 2005 and 2007, respectively, and the Ph.D. degree from the University of Oslo, Norway, in 2010. From September 2008 to August 2009, he was with the Department of Electrical Engineering, Stanford University, Stanford, CA, USA. He is currently an Associate Professor of the ECE Department, School of Engineering, Nazarbayev University (NU). He was an Assistant Professor with the School of Electrical and Computer Engineering, University of Tehran, from Sep. 2011 to Sep. 2015. Dr. Maham is UNESCO-TWAS-affiliated, a Senior Member of IEEE, and has more than 180 publications in major technical journals and conferences. 
He has been an editorial member of IEEE Transactions on Communications, Elsevier's Physical Communication, and John Wiley \& Sons Transactions on Emerging Telecommunications Technologies. He is also recipient of Scopus Award 2022 for outstanding research in the field of Computer Science.  His fields of interest include Wireless Communication and Networking, Signal Processing for Communications, Internet of Things and Beyond 5G. 
\end{IEEEbiography}

\end{document}